\documentclass[11pt]{article}
\pdfoutput=1

\usepackage{amsmath}
\usepackage{amssymb}
\usepackage{enumerate}
\usepackage{graphicx}
\usepackage{pifont}
\usepackage{epsfig}
\usepackage{theorem}
\usepackage{xcolor}
\usepackage{algorithmic, algorithm}
\usepackage{float}
\usepackage{tabularx}
\usepackage{hyperref}

\newcommand{\RR}{\ensuremath{\mathbb R}}

\def\Argmin{\ensuremath{\mbox{Argmin}}}

\begin{document}
%
\title{2-D Prony--Huang Transform: A New Tool for 2-D Spectral Analysis}
%
%
%

\author{J\'er\'emy Schmitt, Nelly Pustelnik, Pierre Borgnat, Patrick Flandrin, Laurent Condat}
\maketitle

\begin{abstract}
This work proposes an extension of the 1-D Hilbert Huang transform for the analysis of images. The proposed method consists in (i) adaptively decomposing an image into oscillating parts called intrinsic mode functions (IMFs) using a mode decomposition procedure, and (ii) providing a local spectral analysis of the obtained IMFs in order to get the local amplitudes, frequencies, and orientations. For the decomposition step, we propose two robust 2-D mode decompositions based on non-smooth convex optimization: a ``Genuine 2-D" approach, that constrains the local extrema of the IMFs, and a ``Pseudo 2-D" approach, which constrains separately the extrema of lines, columns, and diagonals. The spectral analysis step is based on Prony annihilation property that is applied on small square patches of the IMFs. The resulting 2-D Prony--Huang transform is validated on simulated and real data.
\end{abstract}


%

\section{Introduction}
%
%
%
%
An important challenge in image processing is the retrieval of  the local frequencies, amplitudes, and orientations of a nonstationary image. This subject presents several interesting applications for, e.g., texture classification~\cite{Ojansivu_2008_inbook_blur_itc, Guo_Z_2010_j-patt-rec_rotation_itc}, fingerprint analysis~\cite{Park_C_2005_j-patt-rec_figerprint_cuf, Yoon_S_j-ieee-tpami_altered_fad}, ocean wave characterization~\cite{Polo_I_2008_j-geo-res_oceanic_kwt, Alnajjab_B_2013_p-mets_ocean_wpf}.  To be specific, most of these images can be expressed as a sum of a trend and one (or several) \textit{amplitude modulation - frequency modulation} (AM-FM) component(s). It results that usual spectral analysis techniques (designed for analyzing one AM-FM component \cite{Huang_N_1998_p-r-soc-a_empirical_mdh}) lead to poor performance when such nonstationary images have to be analyzed. In the context of 1-D nonstationary analysis, an efficient strategy known as 1-D--Hilbert Huang Transform (HHT) has been proposed in \cite{Huang_N_1998_p-r-soc-a_empirical_mdh}. The goal of this paper is to propose the counterpart of the 1-D--HHT for nonstationary image analysis.


The 1-D--HHT is an empirical method for data analysis that favours adaptivity.  The objective of 1-D--HHT is to extract the instan\-taneous amplitudes and frequencies from a signal built as a sum of a trend and \textit{intrinsic mode functions} (IMFs). We recall that an IMF is loosely defined as a function oscillating around zero and having symmetric oscillations. To achieve this goal, the 1-D--HHT consists in a two-step procedure:
\begin{itemize}
\item a \textbf{decomposition} step, whose objective is to extract the IMFs from the data,
\item \textbf{spectral analysis} of each extracted IMF in order to estimate the instantaneous amplitudes and frequencies of each component.

\end{itemize} 
Regarding the first step, an efficient decomposition procedure known as \textit{empirical mode decomposition} (EMD) has been proposed in \cite{Huang_N_1998_p-r-soc-a_empirical_mdh}. It aims at sequentially extracting the IMF through a sifting process that is based on maxima (resp. minima) cubic spline interpolation. The second step aims at computing the analytic signal of each extracted IMF in order to access the instantaneous amplitude and phase (that leads to frequency) of each IMF. Consequently, the 2-D counterpart of the 1-D--HHT principle requires (i) a bi-dimensional mode decomposition step providing the 2-D--IMFs and (ii) a 2-D spectral analysis step allowing to extract the instantaneous amplitude, frequency, and orientation of each bi-dimensional IMF. 

A generalization of the 1-D--HHT for arbitrary space dimensions has already been proposed in~\cite{Jager_G_2010_j-aada_fast_emd}. This method combines a multidimensional extension of an EMD based on the computation of local means~\cite{Chen_Q_2010_j-acm_a_bsa} and a multidimensional generalization of analytic signal defined with the Riesz transform that is called monogenic signal~\cite{Felsberg_M_2000_p-iwafpac_multidimensional_igq, Felsberg_M_2004_j-math-imaging-vis_monogenic_ssu}. However, this method, as well as the other EMD based on a sifting procedure and interpolation steps, lacks of robustness as will be discussed further. Another 2-D spectral analysis method whose goal is close to a 2-D--HHT is the Riesz-Laplace transform proposed by Unser et al.~\cite{Unser_M_2009_j-ieee-tip_multiresolution_msa}. It combines a two-dimensional wavelet transform with a monogenic analysis~\cite{Felsberg_M_2000_p-iwafpac_multidimensional_igq, Felsberg_M_2004_j-math-imaging-vis_monogenic_ssu}. The counterpart of using a wavelet framework is the lack of adaptivity and consequently this method is less suited than EMD for analyzing nonstationary signals such as AM--FM signals. Moreover, both methods use a monogenic analysis in 2-D~\cite{Unser_M_2009_j-ieee-tip_multiresolution_msa} or n-D~\cite{Jager_G_2010_j-aada_fast_emd} for the spectral estimation step, which proved to be efficient for amplitude, phase, and orientation estimation but not so for frequency estimation as it will be seen in Section~\ref{sec:exp}.

The first contribution of this paper concerns a new robust 2-D mode decomposition procedure based on convex optimization. Indeed, the  existing 2-D--EMD methods are based on the sifting procedure whose main drawback is the lack of a rigorous mathematical definition, and consequently of convergence properties \cite{Wu_Z_2009_j-aada_multi_dee, Linderhed_A_2009_j-aada_image_emd, Nunes_J_2003_j-ima-vis-comp_Image_abe, Damerval_C_2005_j-ieee-spl_fast_afb, Xu_Y_2006_p-r-soc-a_two_dem, Huang_B_2012_j-comput-appl-math_optimization_bem, Koh_M_2013_p-icassp_perfect_rdt}, while efficient 1-D mode decomposition procedures based on convex optimization have been recently proposed in order to get stronger mathematical guarantees~\cite{Oberlin_T_2012_j-ieee-tsp_alternative_fem, Hou_T_2009_j-aada_variant_emd,Pustelnik_N_2014_j-sp_empirical_mdr}. For instance,  \cite{Pustelnik_N_2014_j-sp_empirical_mdr} proposed a mathematical formalism for 1-D--EMD based on a multicomponent proximal algorithm that combines the principle of texture-geometry decomposition \cite{Aujol_J_2006_j-ijcv_structure_tid, Gilles_J_2011_ucla-cam_Bregman_img, Gilles_J_2012_j-siam-mms_multiscale_ts} with some specific features of the usual EMD: constraints on extrema in order to extract IMFs oscillating around zero, sequential formulation of the usual EMD, or IMFs quasi-orthogonality. This method appears to have better performance (in terms of extraction or convergence guarantees) than the other convex optimization procedures as discussed in \cite{Pustelnik_N_2014_j-sp_empirical_mdr}. For this reason, we propose to extend this method to a 2-D mode decomposition formalism.

Our second contribution concerns an alternative approach to monogenic analysis, based on the \emph{annihilation} property \cite{Kay_S_1981_j-pieee_spectrum_amp, Prony_G_1795_j-polytech_essai_eea,Stoica_S_1987_book_int_sa,Kay_S_1988_book_modern_set}. This property, which was highlighted in Prony work in the eighteenth century~\cite{Prony_G_1795_j-polytech_essai_eea}, is particularly interesting for estimating sinusoids.
This annihilation technique has been adapted  to the finite rate of innovation problems in \cite{Blu_T_2008_j-ieee-spm_sparse_ssi}. When noise is involved in the data, the procedure is modified in order to incorporate a low rank constraint. This method is known as the Cadzow algorithm \cite{Blu_T_2008_j-ieee-spm_sparse_ssi, Cadzow_J_1988_j-ieee-tassp_signal_ecp}.  An improved version of Cadzow algorithm has been proposed in~\cite{Condat_L_2012_cadzow_dun} and then extended for 2-D spectral analysis in~\cite{Condat_L_2013_2D_sam}, in order to estimate the modulation parameters in structured illumination microscopy images. In this paper, the objective is to adapt this 2-D spectrum analysis technique in order to estimate the local amplitude, frequency, and orientation of an AM--FM image.

Finally, by combining the proposed variational bi-dimensional EMD with the spectral estimation based on the annihilation property, we propose an efficient adaptive 2-D spectral analysis that we call 2-D \textit{Prony--Huang Transform} (PHT), whose performances are evaluated on simulated and real data. Section~\ref{sec:emd} is focused on the 2-D--EMD step, while Section~\ref{sec:spec} presents the spectral analysis step. The experimental results and comparisons with the state-of-the art methods are presented in Section~\ref{sec:exp}.\\

\noindent \textbf{Notations} We denote by $\mathbf{y} = (\mathbf{y} [n,m])_{1\leq n\leq N_1, 1\leq m\leq N_2} \in \mathbb{R}^{N_1 \times N_2}$ the matrix expression of an image whose size is $N_1\times N_2$, the $n$-th row of the image $\mathbf{y}$ is denoted $\mathbf{y}[n,\bullet]\in \RR^{N_2}$, and $y = (y[n])_{1\leq n\leq N}\in \mathbb{R}^N$ is the vector expression of $\mathbf{y} $, such that $N = N_1 \times N_2$.

\vspace{-0.1cm}

\section{Variational 2-D--EMD}
\label{sec:emd}

\subsection{Classical 2-D--EMD}

We consider an image $\mathbf{x}\in \RR^{N_1\times N_2}$ built as a sum of bidimensional IMFs $(\mathbf{d}^{(k)})_{1\leq k\leq K}$, and a trend $\mathbf{a}^{(K)}\in  \RR^{N_1\times N_2}$, i.e.,
\begin{equation}
\mathbf{x} = \mathbf{a}^{(K)} + \sum_{k=1}^K\mathbf{d}^{(k)}. 
\end{equation}

The 2-D--EMD methods~\cite{Wu_Z_2009_j-aada_multi_dee, Linderhed_A_2009_j-aada_image_emd, Nunes_J_2003_j-ima-vis-comp_Image_abe, Damerval_C_2005_j-ieee-spl_fast_afb, Xu_Y_2006_p-r-soc-a_two_dem, Huang_B_2012_j-comput-appl-math_optimization_bem, Koh_M_2013_p-icassp_perfect_rdt} aim at sequentially extracting the IMFs $(\mathbf{d}^{(k)})_{1\leq k \leq K}$ from the data $\mathbf{x}$. The usual decomposition process is summarized in Algorithm~\ref{algo:emdbasic}.\\
%
\begin{algorithm}
\vspace{0.2cm}

Initialisation :  Set $\mathbf{a}^{(0)} = \mathbf{x}$\\
For every $k\in \{1,\ldots,K\}$
\begin{enumerate}
\item Set i=1
\item Set $\mathbf{t}^{[i]} = \mathbf{a}^{(k-1)}$
\item Compute the mean envelope of $\mathbf{t}^{[i]}$ denoted $\mathbf{m}^{[i]}$
\item $\mathbf{t}^{[i+1]} = \mathbf{t}^{[i]} - \mathbf{m}^{[i]}$
\item Iterate (i.e., $i \leftarrow i+1$) steps 3)-4) until $\mathbf{m}^{[i]}\equiv 0$
\item Set $\mathbf{d}^{(k)} = \mathbf{t}^{[i+1]}$
\item Set $\mathbf{a}^{(k)} = \mathbf{a}^{(k-1)} - \mathbf{d}^{(k)}$\\
\end{enumerate}

\caption{2-D--EMD \cite{Linderhed_A_2009_j-aada_image_emd}\label{algo:emdbasic}}
\label{alg:emd}
\end{algorithm}
%
One can easily remark that this mode decomposition procedure splits up the trend $\mathbf{a}^{(k-1)}$ into a component having IMF properties, denoted $\mathbf{d}^{(k)}$, and a residual component, denoted $\mathbf{a}^{(k)}$. This decomposition is based on the sifting process that consists in iterating the mean envelope removal to $\mathbf{t}^{[i]}$. Note that in step 3), the computation of the mean envelope $\mathbf{m}^{[i]}$ can be obtained through several procedures. For instance, it may denote the mean of the upper and lower envelopes obtained by interpolating the maxima, resp. minima, of $\mathbf{t}^{[i]}$ as proposed in Linderhed \textit{image empirical mode decompostion} (IEMD)~\cite{Linderhed_A_2009_j-aada_image_emd} or the work by Nunes et al. called \textit{bidimensional empirical mode decompostion} (BEMD)~\cite{Nunes_J_2003_j-ima-vis-comp_Image_abe}. A faster method to compute the envelopes, based on a Delaunay triangulation of the extrema, is proposed in~\cite{Damerval_C_2005_j-ieee-spl_fast_afb}. Another fast solution based on triangulation is presented in~\cite{Xu_Y_2006_p-r-soc-a_two_dem}, its main difference is that it does not compute envelopes but it directly computes the mean surface from the characteristic points of the image (maxima, minima, and saddle points). 
In \cite{Huang_B_2012_j-comput-appl-math_optimization_bem} the authors propose to estimate the upper and lower envelopes through a convex optimization procedure in order to avoid over/under shooting problems. Finally, in \cite{Liu_Z_2005_j-ieee-spl_boundary_pbe}, a tensor-product based method is provided to build the envelopes: interpolation is done separately on rows and columns of the image. Some of these methods are consequently faster and may lead to better performance but they are all based on the sifting principle, which does not have convergence guarantees.

\vspace{-0.1cm}

\subsection{Proposed 2-D--EMD}
As mentioned above, the main limitation of the existing 2-D--EMD approaches is the sifting process. In order to avoid this empirical process, we propose  to replace steps 1)-7) in Algorithm~\ref{alg:emd} by the resolution of a variational approach that extracts the trend and the IMF of order $k$ from the trend of order $k-1$. 
The proposed criterion is the following, for every $k\in\{1,\ldots,K\}$,
\begin{equation}
\label{eq:crit_gen}
(a^{(k)},d^{(k)}) \in \underset{a\in \RR^N,d\in \RR^N}{\Argmin} \,  \phi_k(a) + \psi_k(d) + \varphi_k(a,d;a^{(k-1)})
\end{equation}
where $\phi_k$ and $\psi_k$  denote convex, lower semi-continuous, and proper functions from $\RR^N$ to $]-\infty,+\infty]$ that impose the trend and IMF behaviors to the components $a^{(k)}$ and $d^{(k)}$ respectively, while $\varphi_k(\cdot,\cdot ;a^{(k-1)})$ denotes a convex, lower semi-continuous, proper function from $\RR^N\times \RR^N$ to $]-\infty,+\infty]$ that aims at modeling that $a^{(k-1)}$ is close to $a^{(k)} + d^{(k)}$. 

As proposed in \cite{Pustelnik_N_2014_j-sp_empirical_mdr} for 1-D--EMD, the coupling term is chosen quadratic:
\begin{equation}
(\forall (a,d) \in \RR^N \times \RR^N) \quad \varphi_k(a,d;a^{(k-1)}) = \Vert a+d -a^{(k-1)}\Vert^2.
\end{equation}
Such a coupling term makes the method robust to sampling artefacts.

The smoothness of the $k$-order trend $a^{(k)}$ is obtained by imposing a constraint on its isotropic total variation, i.e., $\phi_k$ is defined as follows, for every $\mathbf{a} \in \RR^{N_1\times N_2}$,
\begin{equation}
\label{}
\small{\rho^{(k)} \!\!\sum_{n = 1}^{N_1}\!\sum_{m = 1}^{N_2}\!\!\sqrt{\vert \mathbf{a}[n-1,m] - \mathbf{a}[n,m]\vert^2 \!+ \!\vert \mathbf{a}[n,m-1] - \mathbf{a}[n,m]\vert^2}}
\end{equation}
with a regularization parameter $\rho^{(k)}> 0$. 

At this stage, one can notice the similarities with the texture-geometry decomposition strategies \cite{Aujol_J_2006_j-ijcv_structure_tid, Gilles_J_2011_ucla-cam_Bregman_img, Gilles_J_2012_j-siam-mms_multiscale_ts} when K=1. For this class of methods the function $\psi_k$ is chosen to model oscillating signals, for example it can be a $\ell_1$-norm (TV-$\ell_1$)~\cite{Aujol_J_2006_j-ijcv_structure_tid}, or the $G$-norm that is associated to the Banach space of signals with large oscillations (TV-$G$)~\cite{Meyer_Y_2001_book_oscillating_pip}, \cite{Gilles_J_2011_ucla-cam_Bregman_img}, ~\cite{Gilles_J_2012_j-siam-mms_multiscale_ts}. However, such strategies lack of adaptivity. For this reason we propose to integrate the IMF properties in the function $\psi_k$. To achieve this goal, we extend the 1-D solution proposed in \cite{Pustelnik_N_2014_j-sp_empirical_mdr} to the bi-dimensional problem. We describe two solutions that are the \textit{genuine 2-D} (G2D) approach, based on 2-D local extrema, and the \textit{pseudo 2-D} (P2D) approach, where  lines, columns, and diagonals extrema are separately constrained (see \cite{Wu_Z_2009_j-aada_multi_dee} for a comparison between G2D and P2D approaches in the usual sifting-based EMD procedure).\\

\noindent \textbf{G2D approach} \quad For every $k\in\{1,\ldots,K\}$, we identify the $P_k$ extrema of $\mathbf{a}^{(k-1)}$ whose locations are denoted by $\underline{i}^{(k)}[\ell] \in \{1,\ldots, N_1\} \times \{1,\ldots,N_2\}$. 
For every $\ell\in\{1,\ldots,P_k\}$ such that $\underline{i}^{(k)}[\ell]$ denotes a maxima (resp. a minima), we denote $(\underline{i}^{(k)}_{1}[\ell],\underline{i}^{(k)}_{2}[\ell],\underline{i}^{(k)}_{3}[\ell]\big)$ the locations of the three closest minima (resp. maxima) in the sense of Euclidean distance. We want to impose that $\mathbf{d}(\underline{i}^{(k)}[\ell])$ is approximatively symmetric with respect to its mirror-point that would be on the minima (resp. maxima) envelope. This condition can be obtained by imposing a constraint on the extrema of $\mathbf{d}$:

{\small{\begin{equation}
\label{eq:conG2D}
\Bigg|  \mathbf{d}\big[\underline{i}^{(k)}[\ell]\big] +  \frac{ \alpha_1^{(k)}[\ell] \mathbf{d}\big[\underline{i}^{(k)}_1[\ell]\big]  + \alpha^{(k)}_2[\ell] \mathbf{d}\big[\underline{i}^{(k)}_2[\ell]\big]  +\alpha_3^{(k)}[\ell] \mathbf{d}\big[\underline{i}^{(k)}_3[\ell]\big]  } {\alpha_1^{(k)}[\ell] +\alpha_2^{(k)}[\ell]  +\alpha_3^{(k)}[\ell]   }  \Bigg|,
\end{equation}}}

\noindent where $\big({\alpha}^{(k)}_{j}[\ell]\big)_{1\leq j \leq 3}$ are computed so that $\underline{i}^{(k)}[\ell]$ is the barycenter of the locations $\big(\underline{i}^{(k)}_{j}[\ell]\big)_{1\leq j \leq 3}$ weighted by the $\big({\alpha}^{(k)}_{j}[\ell]\big)_{1\leq j \leq 3}$.
This penalization can be globally rewritten as:
\begin{equation}
\label{ }
(\forall d\in \RR^N)\quad \psi_k(d) = \nu^{(k)} \| \textbf{M}_{G2D}^{(k)} d \|_1,
\end{equation}
where $\textbf{M}_{G2D}^{(k)} \in \mathbb{R}^{P_k \times N}$ is a sparse matrix modelling the constraint imposed on $d$, i.e., Eq.~\eqref{eq:conG2D} can be written $\vert \textbf{M}^{(k)}_{G2D}[\ell,\bullet] d \vert$, where $\textbf{M}^{(k)}_{G2D}[\ell,\bullet]$ denotes the $\ell$-th row of $\textbf{M}_{G2D}^{(k)} $.

The main difficulty of this strategy lies in the detection of extrema locations. Several strategies have been proposed in the literature, for instance in \cite{Linderhed_A_2009_j-aada_image_emd} a pixel is considered as a local extremum if its value is maximum/minimum in a $3 \times 3$ neighbourhood. On the other hand, there are directional strategies that designate a pixel as a maximum (resp. minimum) when its value is greater (resp. lower) than the two closest pixels in any of the 4 principal directions of the image (horizontal, vertical, diagonal, and anti-diagonal). This method makes easier the extraction of oriented textures, that is the reason why we have retained this approach in the present paper. The problem of this second approach lies in the handling of saddle points, which are minima in one direction and maxima in another direction. In this work, we choose not to take these points into account.\\

\pagebreak

\noindent \textbf{P2D approach} \quad This solution constrains extrema of each line, column, diagonal and anti-diagonal rather than dealing with local 2-D extrema. The proposed strategy is described for the constraint applied on the $n$-th row. We denote $\big(n,{i}^{(k)}[\ell]\big)_{1\leq \ell \leq P_{k,n}}$ the locations of local maxima/minima in the $n$-th row of $\mathbf{a}^{(k-1)}$. The condition that imposes a zero mean envelope is
\begin{equation}
\label{eq:conP2D}
\Bigg|  \mathbf{d}[n,i^{(k)}[\ell]] +  \frac{ \alpha^{(k)}_1[\ell] \mathbf{d}[n,i^{(k)}[\ell-1]]  +  \alpha^{(k)}_2[\ell]  \mathbf{d}[n,i^{(k)}[\ell+1]]  } {\alpha^{(k)}_1[\ell] + \alpha^{(k)}_2[\ell] }  \Bigg|,
\end{equation}
where $({\alpha}^{(k)}_{1}[\ell],{\alpha}^{(k)}_{2}[\ell] )$ are computed so that ${i}^{(k)}[\ell]$ is the barycenter of the locations $({i}^{(k)}[\ell-1], {i}^{(k)}[\ell+1])$ weighted by the $({\alpha}^{(k)}_{1}[\ell],{\alpha}^{(k)}_{2}[\ell] )$. The extrema-based constraint can be written for each row $n\in \{1,\ldots,N_1\}$,  $\vert\textbf R_{n}^{(k)}\mathbf{d}[n,\bullet]^\top\vert$, where $\textbf{R}_{n}^{(k)} \in \mathbb{R}^{P_{k,n} \times N_2}$ denotes the linear combination of some elements of the $n$-th row $\mathbf{d}[n,\bullet]$ creating a constraint of a zero mean envelope for the component $d^{(k)}$. 

Considering the whole image, the constraint can be written  $\| \textbf R^{(k)} d \|_1$ where  $\textbf R^{(k)} = \text{diag} (\textbf R_{1}^{(k)}, \ldots, \textbf R_{N_1}^{(k)})$ is a block diagonal matrix, which is highly sparse. 
We apply the same type of constraint to the columns ($\textbf C^{(k)}$), the diagonals ($\textbf D^{(k)}$), and the anti-diagonals ($\textbf A^{(k)}$) of the image, leading to the penalization:
\begin{equation}
(\forall d \in \RR^N) \quad \psi_k(d) = \sum_{l=1}^4\nu_{l}^{(k)}\| \textbf M_{l}^{(k)} d \|_1
\end{equation}
where $\textbf M_{1}^{(k)} = \textbf R^{(k)}$, $\textbf M_{2}^{(k)} = \textbf C^{(k)}$, $\textbf M_{3}^{(k)} =\textbf  D^{(k)}$, $\textbf M_{4}^{(k)} = \textbf A^{(k)}$ denote matrices in $\RR^{N\times N}$. In this paper, we used the same regularization parameters for the four directions, i.e., for every $l \in \{1,\ldots,4 \}$, $\nu_{l}^{(k)} \equiv \nu^{(k)}$.

\subsection{Algorithm}
For both proposed solutions (G2D--EMD or P2D--EMD), the resulting criteria are convex, non-smooth, and involve sparse (but non-Toeplitz) linear operators. According to the recent literature in convex optimization, we propose to adapt the primal-dual splitting algorithm proposed in~\cite{Condat_L_2012_j-ota_primal_dsm} for solving~\eqref{eq:crit_gen}.
Other efficient primal-dual proximal algorithms such as the one proposed in \cite{Combettes_P_2012_j-svva_pri_dsa, Briceno_L_2011_j-siam-opt_mon_ssm, Vu_B_2013_j-nfao_variable_meo} could have been employed. In this paper, we will not discuss and compare the performance of these algorithms in order to focus on the performance of the decomposition procedure. However, in our simulation, the algorithm proposed in [36] appears slightly faster in term of convergence of the iterates.
The iterations are specified in Algorithm~\ref{alggenuine} for G2D--EMD, and in Algorithm~\ref{algpseudo} for P2D--EMD. For further details on the algorithmic solution and proximal tools, one may refer to \cite{Combettes_P_2010_inbook_proximal_smsp}. In order to lighten the notations, we rewrite the total variation penalization as $\phi_k = \rho^{(k)} \| \textbf L \cdot \|_{2,1}$, with $\textbf L = [\textbf L_H^\top \textbf L_V^\top]^\top$ where $\textbf L_H$ and $\textbf L_V$ denote the operators associated to the horizontal and vertical finite differences.  For the P2D--EMD algorithm, we denote $\textbf M_{P2D}^{(k)} = \text{diag}(\textbf M_{1}^{(k)},\textbf M_{2}^{(k)}, \textbf M_{3}^{(k)}, \textbf M_{4}^{(k)})$. Parameters $\sigma$ and $\tau$ are chosen so as to ensure the convergence of the algorithm, see~\cite{Condat_L_2012_j-ota_primal_dsm} for further details.

{\small{
\begin{algorithm}[h]
\caption{G2D--EMD algorithm}
\label{alggenuine}
Set $\mathbf{a}^{(0)} = \mathbf{x}$, \\
For every $k\in \{1,\ldots,K\}$
\begin{enumerate}
\item Compute $\textbf M_{G2D}^{(k)}$ from $\mathbf a^{(k-1)}$,
\item Set $\beta = 1 + \|\textbf M_{G2D}^{(k)}\|^2$,
\item Set $\sigma >0$ and let $\tau= 0.9/(\sigma \beta + 2)$,
\item Initialize $a^{[0]}$ and $d^{[0]}$ in $\RR^N$, 
\item Initialize $y_0^{[0]}$ in $\RR^{2N}$ and $y_1^{[0]} \in \RR^N$ 
\item For $i=0,1,\cdots$\\
{\small{$\left \lfloor \begin{array}{l}
\hspace{-0.3cm}\mbox{ $a^{[i+1]} = a^{[i]} - 2 \tau (a^{[i]}+d^{[i]}-a^{(k-1)} ) - \tau \textbf L^{\top}y_{0}^{[i]}$}\\
\hspace{-0.3cm}\mbox{ $d^{[i+1]} = d^{[i]} - 2 \tau (a^{[i]}+d^{[i]} -a^{(k-1)} ) - \tau (\textbf M_{G2D}^{(k)})^{\top} y_{1}^{[i]}$}\\
\hspace{-0.3cm}\mbox{ $y_{0}^{[i+1]} = \text{prox}_{\sigma (\rho^{(k)} \| \cdot \|_{1,2})^{\ast}} (y_{0}^{[i]} + \sigma \mathbf L (2a^{[i+1]} - a^{[i]}))$}\\
\hspace{-0.3cm}\mbox{ $y_{1}^{[i+1]} = \text{prox}_{\sigma (\nu^{(k)} \| \cdot \|_{1})^{\ast}} (y_{1}^{[i]} + \sigma \mathbf M_{G2D}^{(k)} (2a^{[i+1]} - a^{[i]}))$}\\ 
\end{array} \right.$}}
\item Set $d^{(k)} = \lim_{i\to \infty} d^{[i]}$ and $a^{(k)} = \lim_{i\to \infty} a^{[i]}$.
\end{enumerate}
\end{algorithm}}}


{\small{\begin{algorithm}[h!]
\caption{P2D--EMD algorithm}
\label{algpseudo}
Set $\mathbf{a}^{(0)} = \mathbf{x}$, \\
For every $k\in \{1,\ldots,K\}$
\begin{enumerate}
\item Compute $(\textbf M_l^{(k)})_{1\leq l \leq 4}$ from $\mathbf a^{(k-1)}$,
\item Set $\beta = 1 + \|\textbf M_{P2D}^{(k)}\|^2$,
\item Set $\sigma >0$ and let $\tau= 0.9/(\sigma \beta + 2)$,
\item Initialize $a^{[0]}$ and $d^{[0]}$ in $\RR^N$, 
\item Initialize $y_0^{[0]}$ in $\RR^{2N}$ and $y_l^{[0]} \in \RR^N$ for $l=1,\cdots, 4$.
\item For $i=0,1,\cdots$\\
  {\small{$\left \lfloor \begin{array}{l} 
\hspace{-0.3cm}\mbox{ $a^{[i+1]} = a^{[i]} - 2 \tau (a^{[i]}+d^{[i]}-a^{(k-1)} ) - \tau \textbf L^{\top}y_{0}^{[i]}$}\\
\hspace{-0.3cm}\mbox{ $d^{[i+1]} = d^{[i]} - 2 \tau (a^{[i]}+d^{[i]} -a^{(k-1)} ) - \tau \sum_{l=1}^{4} (\textbf M_{l}^{(k)})^{\top} y_{l}^{[i]}$}\\
\hspace{-0.3cm}\mbox{ $y_{0}^{[i+1]} = \text{prox}_{\sigma (\rho^{(k)} \| \cdot \|_{1,2})^{\ast}} (y_{0}^{[i]} + \sigma \textbf L (2a^{[i+1]} - a^{[i]}))$}\\
\hspace{-0.3cm}\mbox{ $\text{For} \ l=1,\cdots,4$ }\\
\hspace{-0.3cm} \; \; \lfloor \mbox{ $y_{l}^{[i+1]} = \text{prox}_{\sigma (\nu_{l}^{(k)} \| \cdot \|_{1})^{\top}} (y_{l}^{[i]} + \sigma \textbf M_{l}^{(k)} (2a^{[i+1]} - a^{[i]}))$}\\ \end{array} \right.$}}
\item Set $d^{(k)} = \lim_{i\to \infty} d^{[i]}$ and $a^{(k)} = \lim_{i\to \infty} a^{[i]}$.
\end{enumerate}
\end{algorithm}}}

\vspace{-0.3cm}

\section{Spectral analysis}
\label{sec:spec}

The previous section was dedicated to methods to extract 2-D IMFs. In this section, we now focus on the estimation of the instantaneous frequency, amplitude, and orientation of each IMF.  After a short review of monogenic analysis, usually employed for analysing 2-D IMFs \cite{Jager_G_2010_j-aada_fast_emd}, we propose a new 2-D spectral analysis method based on Prony annihilation property.

\vspace{-0.2cm}
\subsection{2-D spectral estimation based on monogenic signal}

We first recall that for a given real-valued 1-D signal $d \in \RR^N$, the associated analytic signal $d_{\text{a}} \in \mathbb{C}^N$, which by definition involves the signal itself and its Hilbert transform, can also be written under a polar form involving instantaneous phase $\chi \in \RR^N$ and amplitude $\alpha \in \RR^N$ such as:
\begin{equation}
d_{\text{a}}   = d + j \mathcal{H}(d) =\alpha e^{j\chi},
\end{equation}
where $\mathcal{H}(d)$ is the Hilbert transform of $d$, which consists in a convolution by an all pass filter $ h$ characterized by its transfer function $H_{\omega} = -j \omega / | \omega | $.
%
These two formulations make easy the computation of the instantaneous amplitude and the instantaneous phase as the absolute value of the analytic signal and its argument.

The Riesz transform is the natural 2-D extension of the Hilbert transform \cite{Felsberg_M_2004_j-math-imaging-vis_monogenic_ssu}. 
The Riesz transform of a 2-D signal $\mathbf{d}$ can be expressed as $\mathbf{d}_{\text{r}} = (\mathbf{d}_1, \mathbf{d}_2) = (\mathbf h_1 \ast \mathbf{y} , \mathbf h_2 \ast \mathbf{y})$,
where the filters $(\mathbf h_l)_{1\leq l \leq 2}$ are characterized by their 2-D transfer functions $(\mathbf H_l)_{\underline{ \omega}} = -j \omega_l / \|\underline{ \omega}\| $ with $\underline{ \omega} = (\omega_1 , \omega_2)$.  Based on the Riesz transform, the monogenic signal is the counterpart in 2-D of the analytic signal defined as a three-component signal $\mathbf{d}_{\text{m}} = (\mathbf{d},\mathbf{d}_1,\mathbf{d}_2)$~\cite{Felsberg_M_2004_j-math-imaging-vis_monogenic_ssu}.
Similarly to the analytic signal, the monogenic signal enables to compute easily the local amplitude, phase, and orientation at each pixel through the relations, for every $(n,m)\in \{1,\ldots, N_1\}\times \{1,\ldots, N_2\}$, 
\begin{eqnarray}
\label{eqalpha}
\boldsymbol{\alpha}[n,m] & = &   \sqrt{\big(\mathbf{d}[n,m]\big)^2 + \big(\mathbf{d}_1[n,m]\big)^2 +\big( \mathbf{d}_2[n,m]\big)^2} \nonumber\\
\label{eqxi}
\boldsymbol{\chi}[n,m] & = & \arctan \Bigg(\frac{\sqrt{\big(\mathbf{d}_1[n,m]\big)^2 + \big(\mathbf{d}_2[n,m]\big)^2}}{\mathbf{d}[n,m]}\Bigg) \nonumber\\
\label{eqtheta}
\boldsymbol{\theta}[n,m] & = & \arctan (\mathbf{d}_2[n,m]/\mathbf{d}_1[n,m]).
\end{eqnarray}
The local frequency $\boldsymbol{\eta}[n,m]$ is then obtained by differentiating the local phase $\boldsymbol{\chi}[n,m]$ along the direction given by the orientation $\boldsymbol{\theta}[n,m]$, see~\cite{Unser_M_2009_j-ieee-tip_multiresolution_msa} for further details.

The estimation of the orientation proposed in \eqref{eqtheta} lacks of robustness because it does not take into account the orientation of neighbouring pixels. Unser \emph{et al.} \cite{Unser_M_2009_j-ieee-tip_multiresolution_msa} derived an improved estimation based on a minimization procedure including a smoothness neighbourhood constraint. This improved orientation estimation also gives a coherency map $\boldsymbol{\lambda}[n,m]$, which models the degree of directionality of the local neighborhood and gives a general reliability index of the estimation.

\vspace{-0.2cm}
\subsection{2-D spectral analysis based on the annihilation property}

The monogenic analysis is efficient for the analysis of instantaneous amplitudes and orientations but unfortunately fails for the frequency estimation, as we will see in section~\ref{sec:exp}. In this section, we propose an alternative method, based on the annihilation property of a discrete cosine function. While several papers deal with this problem in 1-D \cite{Blu_T_2009_p-sta_generalized_apt}, \cite{Blu_T_2008_j-ieee-spm_sparse_ssi, Condat_L_2012_cadzow_dun}, the bi-dimensional cosine estimation is still challenging. Recently, in the context of structured illumination microscopy, an efficient 2-D spectral estimation strategy has been proposed in order to estimate the global modulation parameters \cite{Condat_L_2013_2D_sam}. While in structured illumination microscopy the modulation is uniform through the image, in the present context the modulation may vary from a location to another. For this reason, we propose to adapt the spectral strategy proposed in~\cite{Condat_L_2013_2D_sam} in order to locally estimate the amplitudes, phases, and orientations of the IMFs.\\%

\noindent \textbf{An IMF is locally a cosine function} \quad  For every $k\in \{1, \ldots,K\}$, we divide the $k$-th estimated IMF $\textrm{d}^{(k)}$ into square patches of size $\overline{N}^{(k)}$, i.e., 
for every $(\overline{n}, \overline{m}) \in \big\{1,\ldots, \Big\lfloor \frac{N_1}{\overline{N}^{(k)}} \Big\rfloor\big\}\times \big\{1,\ldots, \Big\lfloor\frac{N_2}{\overline{N}^{(k)}} \Big\rfloor\big\},$
\begin{equation}
\mathbf{p}^{(k)}_{\overline{n},\overline{m}} = (\mathbf{d}^{(k)}[n,m])_{(n,m)\in \mathcal{N}^{(k)}_{\overline{n},\overline{m}}}
\end{equation}

\noindent where 
\begin{equation}
\mathcal{N}^{(k)}_{\overline{n},\overline{m}} = \{(n,m)\in \{(\overline{n}-1)\overline{N}^{(k)} + 1,\ldots,\overline{n}\overline{N}^{(k)}\}\times \{(\overline{m}-1)\overline{N}^{(k)} + 1,\ldots,\overline{m}\overline{N}^{(k)}\} \}
\end{equation}
and  we locally model it by a discrete cosine function that is, for every $(n',m')\in\{1,\ldots,\overline{N}^{(k)}\}\times\{1,\ldots,\overline{N}^{(k)}\}$,
\begin{equation}
\mathbf{p}^{(k)}_{\overline{n},\overline{m}}[n',m'] \approx \boldsymbol{\alpha}^{(k)}_{\overline{n},\overline{m}} \cos\big(2\pi \boldsymbol{\xi}^{(k)}_{\overline{n},\overline{m}} n'  + 2\pi \boldsymbol{\zeta}^{(k)}_{\overline{n},\overline{m}} m' + \boldsymbol{\chi}^{(k)}_{\overline{n},\overline{m}}\big)
\end{equation}
where $\boldsymbol{\alpha}^{(k)}_{\overline{n},\overline{m}}$ models the local amplitude for the patch $(\overline{n},\overline{m})$ of the $k$-th IMF, while $\boldsymbol{\xi}^{(k)}_{\overline{n},\overline{m}}$ and $\boldsymbol{\zeta}^{(k)}_{\overline{n},\overline{m}}$ are respectively the local frequencies toward lines and columns. Then, the local frequency and orientation are respectively given by:
\begin{equation}
\boldsymbol{\eta}^{(k)}_{\overline{n},\overline{m}}  =  \sqrt{(\boldsymbol{\xi}^{(k)}_{\overline{n},\overline{m}})^2 + (\boldsymbol{\zeta}^{(k)}_{\overline{n},\overline{m}})^2}, \\
\end{equation}
and
\begin{equation}
\boldsymbol{\theta}^{(k)}_{\overline{n},\overline{m}}  =  \arctan (\boldsymbol{\xi}^{(k)}_{\overline{n},\overline{m}} / \boldsymbol{\zeta}^{(k)}_{\overline{n},\overline{m}}).
\end{equation}

The problem is then to estimate the parameters $\boldsymbol \alpha^{(k)}$, $\boldsymbol\xi^{(k)}$, $\boldsymbol\zeta^{(k)}$, $\boldsymbol\chi^{(k)}$ which best fit the data $\mathbf{p}^{(k)}$.\\

\noindent \textbf{Principle of annihilation property in 1-D} \quad For every $n'\in \{1,\ldots,\overline{N}\}$ with $\overline{N} \geqslant 3$, a 1-D discrete cosine function $p[n'] = \alpha \cos (2 \pi \xi n' + \chi)$ can be written as a sum of two complex exponentials. Thus, according to the annihilation property \cite{Blu_T_2008_j-ieee-spm_sparse_ssi}, the sequence $(p[n'])_{1 \leq n' \leq \overline{N}}$ admits an annihilating filter $f = (f[k])_{0\leq k \leq 2}$ which satisfies:
\begin{equation}
\big(\forall n'= \{3, \cdots,\overline{N}\}\big) \quad \sum_{k=0}^{2} f[k] p[n'-k] = 0,
\end{equation}
and has the following $Z$-transform:
\begin{align}
F(z) &=  f[0] + f[1] z^{-1} + f[2] z^{-2} \\
&=  f[0] (1 - e^{-j 2 \pi \xi} z^{-1}) (1 - e^{+j 2 \pi \xi} z^{-1}) 
\label{cosannihil}
\end{align}
that means  $f[0] = f[2]$ (i.e, $f$ is symmetric) and $f[1] = -f[2](e^{j 2\pi \xi} + e^{-j 2 \pi \xi})$. It is then straightforward to compute $\alpha$, $\xi$, and $\chi$ from the annihilating filter. Indeed, according to Eq.~(\ref{cosannihil}), the roots of the polynomial $F(z)$ are on the unit complex circle and are $e^{-j 2 \pi \xi}$ and $e^{+j 2 \pi \xi}$, which leads to the value of $\xi$. Then, by linear regression, we retrieve the complex amplitude $\alpha e^{j\chi}$. Consequently, the main difficulty consists in estimating the annihilating filter $f$.

First, it can be shown easily that, if $f = (f[0], f[1], f[2])$ is an annihilating filter of the cosine function $p$, any non trivial filter $f'=(f'[k])_{0 \leq k \leq L}$, where $2 \leqslant L \leqslant \overline{N}-3$, is also an annihilating filter of $p$ if the roots of $f$ are roots of $f'$. The inverse is true: if $f'$ is a non trivial annihilating filter of $p$, then the roots of $f$ are roots of $f'$. 
The annihilating equation can then be rewritten as:
\begin{equation}
\big(\forall n'= \{L+1, \cdots,\overline{N}\}\big) \quad \sum_{k=0}^{L} f'[k] p[n'-k] = 0,
\end{equation}
or equivalently in the matrix form, $\textrm T_L(p) f' = 0$, with
\begin{equation}
\label{ }
\textrm T_L(p) = \scriptsize\left(\begin{array}{cccc} 
p[\overline{N}-L] & \cdots & p[\overline{N}-1] & p[\overline{N}] \\
p[\overline{N}-L-1] & \cdots & p[\overline{N}-2] & p[\overline{N}-1]  \\ 
\vdots &  & \vdots & \vdots  \\
p[1] & \cdots & p[L] & p[L+1] \\ 
\hline \hline
p[L+1] & \cdots & p[2] & p[1] \\ 
\vdots &  & \vdots & \vdots  \\
p[\overline{N}]  & \cdots & p[\overline{N}-L+1] & p[\overline{N}-L] \end{array}\right)
\normalsize
\end{equation}
where $\textrm T_L(p) \in \RR^{2(\overline{N}-L) \times (L+1)}$. 
The symmetrization of $\textrm T_L(p)$ ensures the symmetry of the filter. 

Second, according to~\cite{Blu_T_2008_j-ieee-spm_sparse_ssi}, 
$p$ is a sinusoid if and only if, for any $L$, $\textrm T_L(p)$ has a rank of two. The simplest annihilating filter $h$ is obtained from $\textrm T_2(p)$ by considering its right singular vector corresponding to the 3rd singular value which is zero (due to $L=2$ and that the rank of  $\textrm T_2(p)$ is 2, the two others singular values are non-zero).

This strategy cannot be applied directly on degraded data. In~\cite{Blu_T_2008_j-ieee-spm_sparse_ssi,Condat_L_2012_cadzow_dun}, the authors propose an iterative strategy to estimate the matrix $\widehat{t} = \textrm T_L(\widehat{p})$ of rank 2, where $\widehat{p}$ denotes the denoised sinusoidal signal that is the closest from $p$. Consequently, the denoising strategy consists in solving the following structured low-rank approximation (SLRA) problem:
\begin{equation}
\label{srla1-D}
\widehat{t} \in \underset{t\in \mathcal{T} \cap \mathcal{R}_2}{\Argmin} \; \| \sqrt{ \textrm P} \circ (t - \textrm T_L(p)  \|_F^2,
\end{equation}
where $\textrm P \in \RR^{2(\overline{N}-L) \times (L+1)}$ denotes a weighting matrix whose entries are inversely equal to the number of times where the entry models the same element, $\circ$ is the entrywise product, $\| \cdot \|_F$ is the Frobenius norm. $\mathcal{T}$ is the set of matrices $\mathcal{T} = \{ (\textrm T_L(c) ) \ : \ c \in \mathbb{R}^{N} \}$, and $\mathcal{R}_2$ is the set of matrices with a maximal rank of $2$. The problem can be solved with an iterative primal-dual algorithm as proposed in~\cite{Condat_L_2012_cadzow_dun}. 
$L$ is chosen so that $\textrm T_L (p)$ is as close to a square matrix as possible, in order to improve the convergence speed of the algorithm.\\




\noindent \textbf{Spectral analysis of IMFs} \quad The procedure described previously has been extended for 2-D spectral analysis in~\cite{Condat_L_2013_2D_sam} when the modulation is uniform through the whole image. In our work, in the scope of providing an adaptive 2-D spectral analysis method designed for nonstationary images, we will estimate amplitudes, frequencies and orientations \emph{locally}. The idea is then to apply the method proposed in~\cite{Condat_L_2013_2D_sam} on the local patches $\mathbf{p}^{(k)}_{\overline{n},\overline{m}}\in \RR^{\overline{N}^{(k)} \times \overline{N}^{(k)}}$. 

In the situation where the $(k,\overline{n},\overline{m})$-th patch $\mathbf{p}^{(k)}_{\overline{n},\overline{m}}$ 
is a strict sinusoid, there exists two annihilating filters $\mathbf f_{\overline{n},\overline{m}}^{(k)}$ and  $\mathbf g_{\overline{n},\overline{m}}^{(k)}$, both symmetric and of size $3$, which annihilate respectively the rows and the columns of the patch. The roots of the $Z$-transform of $\mathbf f_{\overline{n},\overline{m}}^{(k)}$ are $e^{-j 2 \pi \boldsymbol\xi_{\overline{n},\overline{m}}^{(k)}}$ and $e^{+j 2 \pi \boldsymbol\xi_{\overline{n},\overline{m}}^{(k)}}$, which leads to $\boldsymbol\xi_{\overline{n},\overline{m}}^{(k)}$ as it is chosen positive. The same calculation with $\mathbf g_{\overline{n},\overline{m}}^{(k)}$ gives $\pm \boldsymbol\zeta_{\overline{n},\overline{m}}^{(k)}[\overline{n},\overline{m}]$, the sign of $\boldsymbol\zeta_{\overline{n},\overline{m}}^{(k)}$ has to be disambiguated in order to compute the orientation. Finally, a linear regression gives us the complex amplitude $\boldsymbol\alpha_{\overline{n},\overline{m}}^{(k)} e^{j\boldsymbol\varphi_{\overline{n},\overline{m}}^{(k)}}$ and disambiguates the sign of $\boldsymbol\zeta_{\overline{n},\overline{m}}^{(k)}$, see~\cite{Condat_L_2012_cadzow_dun} for more details.

The IMFs extracted with the EMD procedures described in Section~\ref{sec:emd} do not behave exactly like a local cosine. Consequently, a SRLA based procedure is used in order to achieve an efficient 2-D-block spectral estimation. The problem to solve is:
\begin{equation}
\underset{\substack{(\textrm u,\textrm v) \in \mathcal{T}\\\textrm u \in \mathcal{R}_2\\ \textrm v \in \mathcal{R}_2}}{\min} \| \sqrt{\textrm P} \circ (\textrm u - \textrm U_{L^{(k)}}(\mathbf{p}_{\overline{n},\overline{m}}^{(k)})) , \sqrt{\textrm P} \circ (\textrm v - \textrm V_{L^{(k)}}(\mathbf{p}_{\overline{n},\overline{m}}^{(k)}))  \|_F^2 
\label{srla1}
\end{equation}
where $\textrm U_L^{(k)}$ and $\textrm V_L^{(k)}$ map respectively the lines and columns of a patch of size $\overline{N}^{(k)} \times \overline{N}^{(k)}$ into a centro-symmetric Toeplitz matrix of size $2 \overline{N}^{(k)} (\overline{N}^{(k)} - L^{(k)}) \times (L^{(k)}+1)$: each row (resp. column) of the patch is mapped into a Toeplitz matrix of size $\overline{N}^{(k)} (\overline{N}^{(k)} - L^{(k)}) \times (L^{(k)}+1)$ and these matrices are stacked on the top of each other. The matrix is then duplicated, the order of rows and columns is reversed, and the duplicated matrix is stacked under the first one. See~\cite{Condat_L_2013_2D_sam} for more details on the construction of $\textrm U_{L^{(k)}}(\mathbf{p}_{\overline{n},\overline{m}}^{(k)})$ and $\textrm V_{L^{(k)}}(\mathbf{p}_{\overline{n},\overline{m}}^{(k)})$. 
The weighting matrix $\textrm P$ of size $2 \overline{N}^{(k)} (\overline{N}^{(k)} - L^{(k)}) \times (L^{(k)}+1)$ is defined similarly as for the original 1-D SRLA. To summarize, the goal of this minimization problem is to estimate the matrices $\widehat{\textrm u}_{\overline{n},\overline{m}}^{(k)}$ and $\widehat{\textrm v}_{\overline{n},\overline{m}}^{(k)}$ of rank $2$, so that $\widehat{\textrm u}_{\overline{n},\overline{m}}^{(k)} = \textrm U_{L^{(k)}} (\widehat{\mathbf{p}}_{\overline{n},\overline{m}}^{(k)})$ and $\widehat{\textrm v}_{\overline{n},\overline{m}}^{(k)} = \textrm V_{L^{(k)}} (\widehat{\mathbf{p}}_{\overline{n},\overline{m}}^{(k)})$, where $\widehat{\mathbf{p}}_{\overline{n},\overline{m}}^{(k)}$ is the $(k,\overline{n},\overline{m})$-th denoised sinusoidal patch. In practice, $L^{(k)}$ is chosen to be equal to $\overline{N}^{(k)} - 3$ so that the matrices are as compact as possible.  \\

\noindent \textbf{Algorithm for analysing IMFs} \quad 
Similarly as in~\cite{Condat_L_2013_2D_sam}, the SRLA problem is solved with a primal-dual algorithm, which alternates between a gradient descent with respect to the cost function (the squared Frobenius norm) with projections $\mathcal{P}_{\mathcal{T}}$ and $\mathcal{P}_{\mathcal{R}_2}$ to enforce the constraints. $\mathcal{P}_{\mathcal{T}}$ is the orthogonal projection of a pair of matrices $(\textrm u, \textrm v)$ on $\mathcal{T}$ that consists in averaging the coefficients of $\textrm u$ and $\textrm v$ corresponding to the same pixel of the image. $\mathcal{P}_{\mathcal{R}_2}$ is the orthogonal projection on the set of matrices of rank at most $2$, it is done by SVD truncation, that consists to set to zero all singular values except the two largest ones. The algorithm is described in Algorithm \ref{algsrla}. \\

\begin{algorithm}
\caption{Spectral estimation algorithm}
\label{algsrla}
\begin{algorithmic}
\footnotesize
\STATE For $\overline{n} = 1, \cdots, \lfloor N_1/\overline{N}^{(k)} \rfloor, \ \overline{m} = 1, \cdots, \lfloor N_2/\overline{N}^{(k)} \rfloor $, \\
\STATE STEP 1 -- \textbf{Initialization}\\
$\quad$$\left \lfloor \begin{array}{l}
\mbox{  Choose the parameters $\mu > 0$, $\gamma \in ]0,1[$}\\
\mbox{   Set the initial estimates $\textrm u^{(0)}$, $\textrm v^{(0)}$, $\textrm s^{(0)}$, $\textrm t^{(0)}$} \\
\end{array} \right.$
\STATE STEP 2 -- \textbf{Iteration} : for $l= 0,1, \cdots$\\
$\quad$$\left \lfloor \begin{array}{l}
\mbox{    $\textrm u^{(l+1)} = P_{\mathcal{R}_2}\Big(\textrm s^{(l)} + \mu \big(\textrm u^{(l)} - \textrm s^{(l)}\big) - \mu \textrm P \circ \big(\textrm u^{(l)} - \textrm U_{L^{(k)}}(\mathbf{p}_{\overline{n}, \overline{m}}^{(k)})\big)\Big)$}\\
\mbox{ $\textrm v^{(l+1)} = P_{\mathcal{R}_2}\Big(\textrm t^{(l)} + \mu \big(\textrm v^{(l)} - \textrm t^{(l)}\big) - \mu \textrm P \circ \big(\textrm v^{(l)} - \textrm V_{L^{(k)}}(\mathbf{p}_{\overline{n}, \overline{m}}^{(k)})\big)\Big)$}\\
\mbox{ $(\textrm s^{(l+1)}, \textrm t^{(l+1)}) = (\textrm s^{(l)}, \textrm t^{(l)}) - (\textrm u^{(l+1)}, \textrm v^{(l+1)}) $} \\
\mbox { $\; \; \; \; \; \; \; \; +  P_{\mathcal{T}} \big( 2(\textrm u^{(l+1)}, \textrm v^{(l+1)}) - (\textrm s^{(l)}, \textrm t^{(l)}) \big)$}\\
\end{array} \right.$
\STATE STEP 3 -- \textbf{Parameters estimation} :  \\
$\quad$$\left \lfloor \begin{array}{l}
\mbox{ $\widehat{\textrm u}_{\overline{n},\overline{m}}^{(k)}= \textrm u^{(l)}$, $\widehat{\textrm v}_{\overline{n},\overline{m}}^{(k)} = \textrm v^{(l)}$ }\\
\mbox{ Compute $\widehat{\mathbf{p}}_{\overline{n},\overline{m}}^{(k)}$ from $\widehat{\textrm u}_{\overline{n},\overline{m}}^{(k)}$ or $\widehat{\textrm v}_{\overline{n},\overline{m}}^{(k)}$}\\
\mbox{ Compute the annihilating filters $\widehat{\textbf{\textrm f}}_{\overline{n},\overline{m}}^{(k)}$ and $\widehat{\textbf{\textrm g}}_{\overline{n},\overline{m}}^{(k)}$ from $\widehat{\mathbf{p}}_{\overline{n},\overline{m}}^{(k)}$ } \\
\mbox { Compute $\widehat{\boldsymbol{\alpha}}_{\overline{n},\overline{m}}^{(k)}$, $\widehat{\boldsymbol{\xi}}_{\overline{n},\overline{m}}^{(k)}$, $\widehat{\boldsymbol{\zeta}}_{\overline{n},\overline{m}}^{(k)}$ from $\widehat{\textbf{\textrm f}}_{\overline{n},\overline{m}}^{(k)}$ and $\widehat{\textbf{\textrm g}}_{\overline{n},\overline{m}}^{(k)}$}\\
\mbox{ Compute $\widehat{\boldsymbol{\eta}}_{\overline{n},\overline{m}}^{(k)}$, $\widehat{\boldsymbol{\theta}}_{\overline{n},\overline{m}}^{(k)}$ from $\widehat{\boldsymbol{\xi}}_{\overline{n},\overline{m}}^{(k)}$  and $\widehat{\boldsymbol{\zeta}}_{\overline{n},\overline{m}}^{(k)}$ }
\end{array} \right.$

\normalsize

\end{algorithmic}
\end{algorithm}


\noindent \textbf{Coherency index} \quad Inspired by ~\cite{Unser_M_2009_j-ieee-tip_multiresolution_msa}, we introduce a new coherency index, defined as $\widehat{\boldsymbol{\lambda}} =  \{ \widehat{\boldsymbol{\lambda}}_{\overline{n},\overline{m}}^{(k)} \}_{k,\overline{n},\overline{m}} \in [0;1]$ in order to provide a degree of quality of the spectral estimation. For every $(k,\overline{n},\overline{m})$, $\widehat{\boldsymbol{\lambda}}_{\overline{n},\overline{m}}^{(k)}$ is given by the sum of the two higher singular values of $\widehat{\textbf{\textrm p}}_{\overline{n},\overline{m}}^{(k)}$, normalized to have an index between 0 and 1. $\widehat{\boldsymbol{\lambda}}$, which is highly linked to the amplitude $\widehat{\boldsymbol{\alpha}}$, informs us about the local oscillatory character of data:  it is higher on oscillating parts of the signal, and lower on non-oscillating parts and parts containing only noise. \\

\noindent \textbf{Size of the patches} \quad The patch size $\overline{N}^{(k)}$ should be chosen so that the IMF can be modeled as a sinusoidal function inside a patch. This means that it should be small enough so that the frequency and orientation can be considered constant inside a patch, and large enough so that each patch contains at least one period of oscillation. 

\section{Experiments}
\label{sec:exp}

\subsection{Simulations}

The first experiment is on simulated data of size $N = 512\times 512$, consisting in a sum of two localized texture components $\mathbf{x}^{(1)}$ and $\mathbf{x}^{(2)}$ and a piecewise constant background $\mathbf{x}^{(3)}$. The background is formed with two piecewise constant patches: one rectangular patch and one ellipsoidal patch. The component $\mathbf{x}^{(1)}$ (resp. $\mathbf{x}^{(2)}$) models a modulated signal of central frequency $\eta_1 = 120/512$ (resp. $\eta_2=60/512$). 

%
%
%
%
%

\subsubsection{EMD}

We compare our two variational EMD approaches (G2D--EMD and P2D--EMD) with several state-of-the-art decomposition methods that are (a) a classical 2-D--EMD method, image empirical mode decomposition (IEMD)~\cite{Linderhed_A_2009_j-aada_image_emd}, which is a natural 2-D extension of EMD based on 2-D interpolation of extrema using thin-plate spline, (b) a texture-cartoon decomposition methods based on total variation decomposition~\cite{Aujol_J_2006_j-ijcv_structure_tid} and (c) the Gilles-Osher texture-geometry decomposition~\cite{Gilles_J_2011_ucla-cam_Bregman_img}, which is an iterative procedure designed to solve the Meyer $G$-norm texture-cartoon decomposition problem (similarly to the proposed solution, for both texture-geometry decomposition, we denote $\rho^{(k)}$ the cartoon regularization parameter and $\nu^{(k)}$ the texture regularization parameter).

\begin{figure*}
\begin{center}
\footnotesize
\begin{tabular}{cccc}

Simulated data: $\mathbf{x}$ &
1st component: $\mathbf{x}^{(1)}$&
2nd component:  $\mathbf{x}^{(2)}$&
Background: $\mathbf{x}^{(3)}$ \\

\includegraphics[width=1.2in]{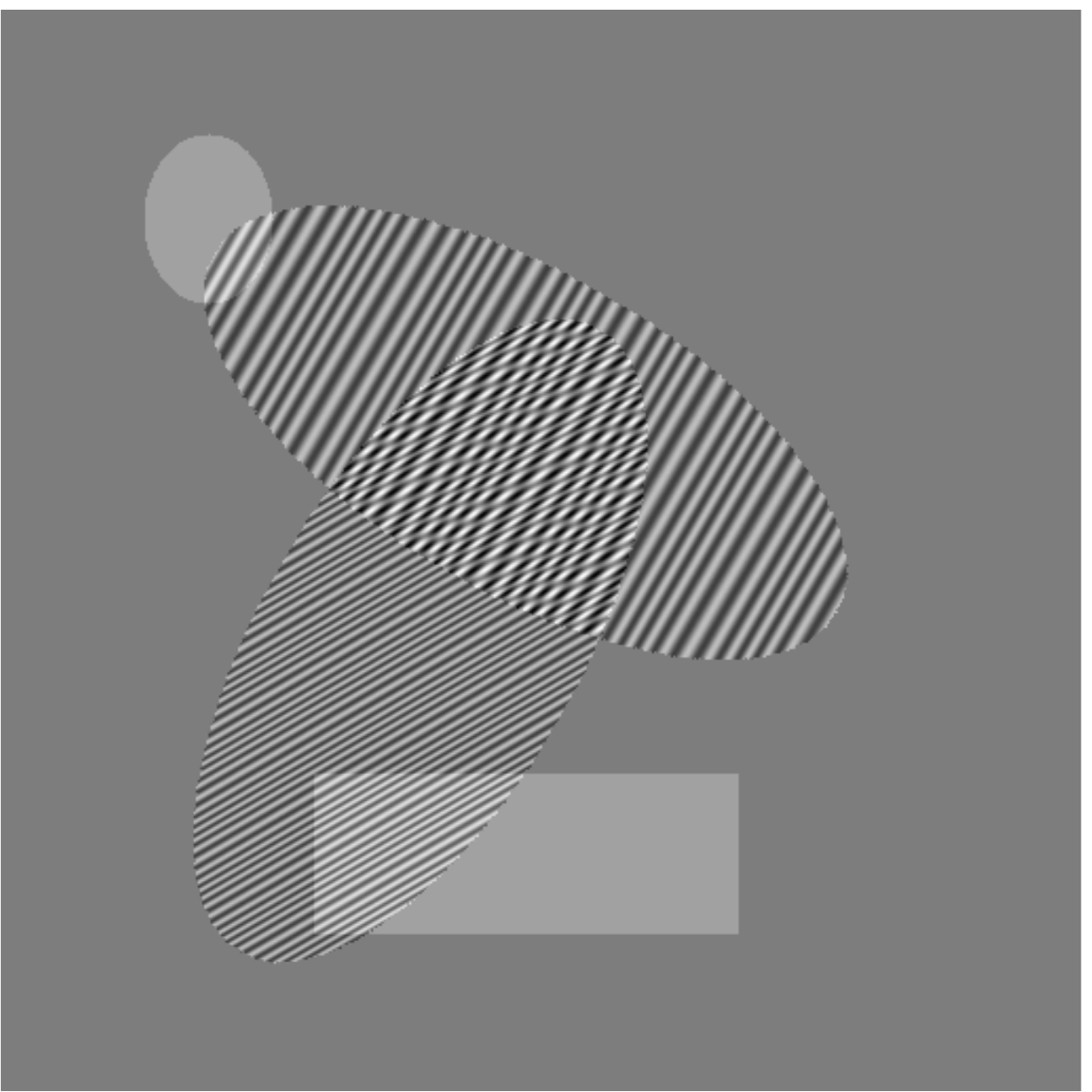} &
\includegraphics[width=1.2in]{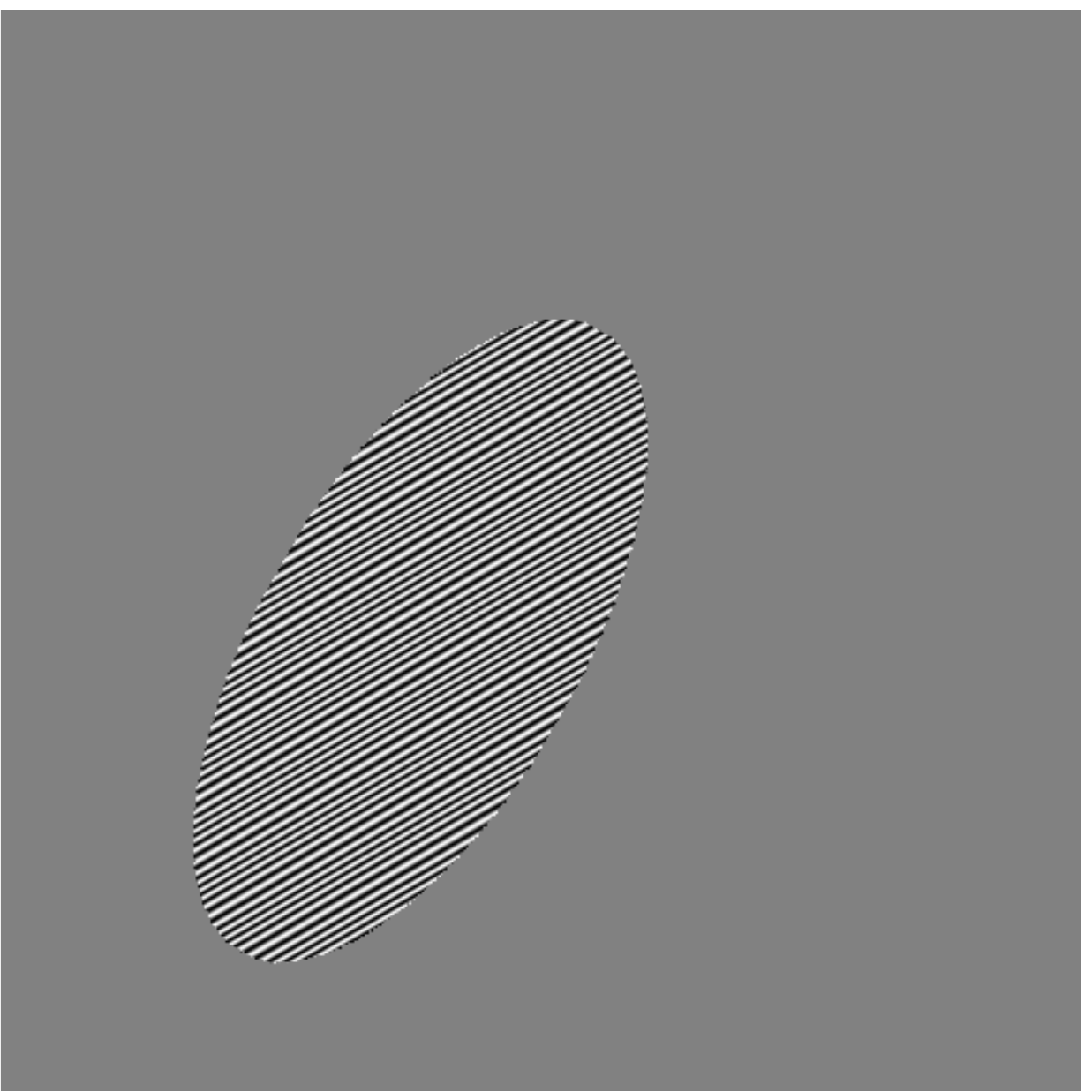} &
\includegraphics[width=1.2in]{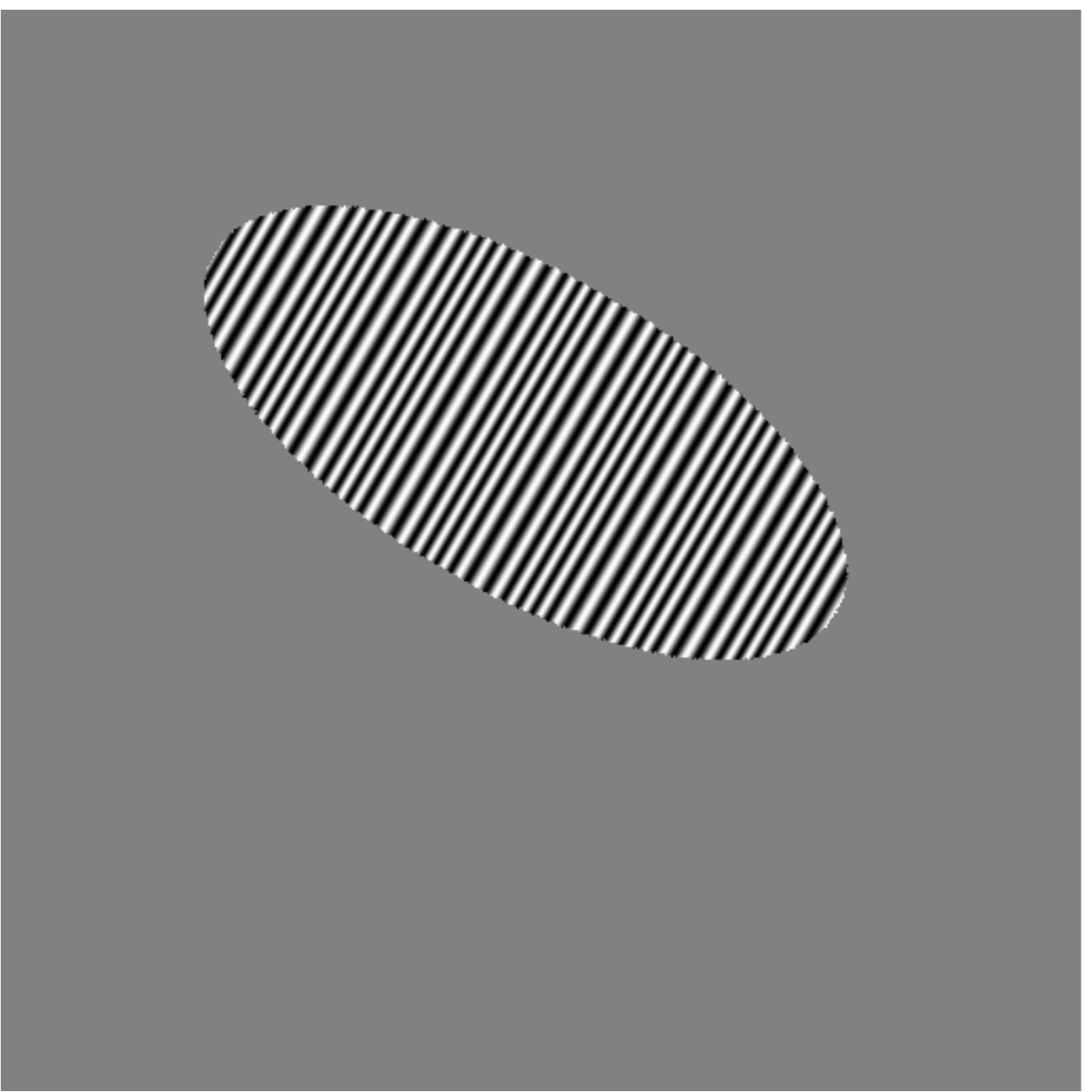} &
\includegraphics[width=1.2in]{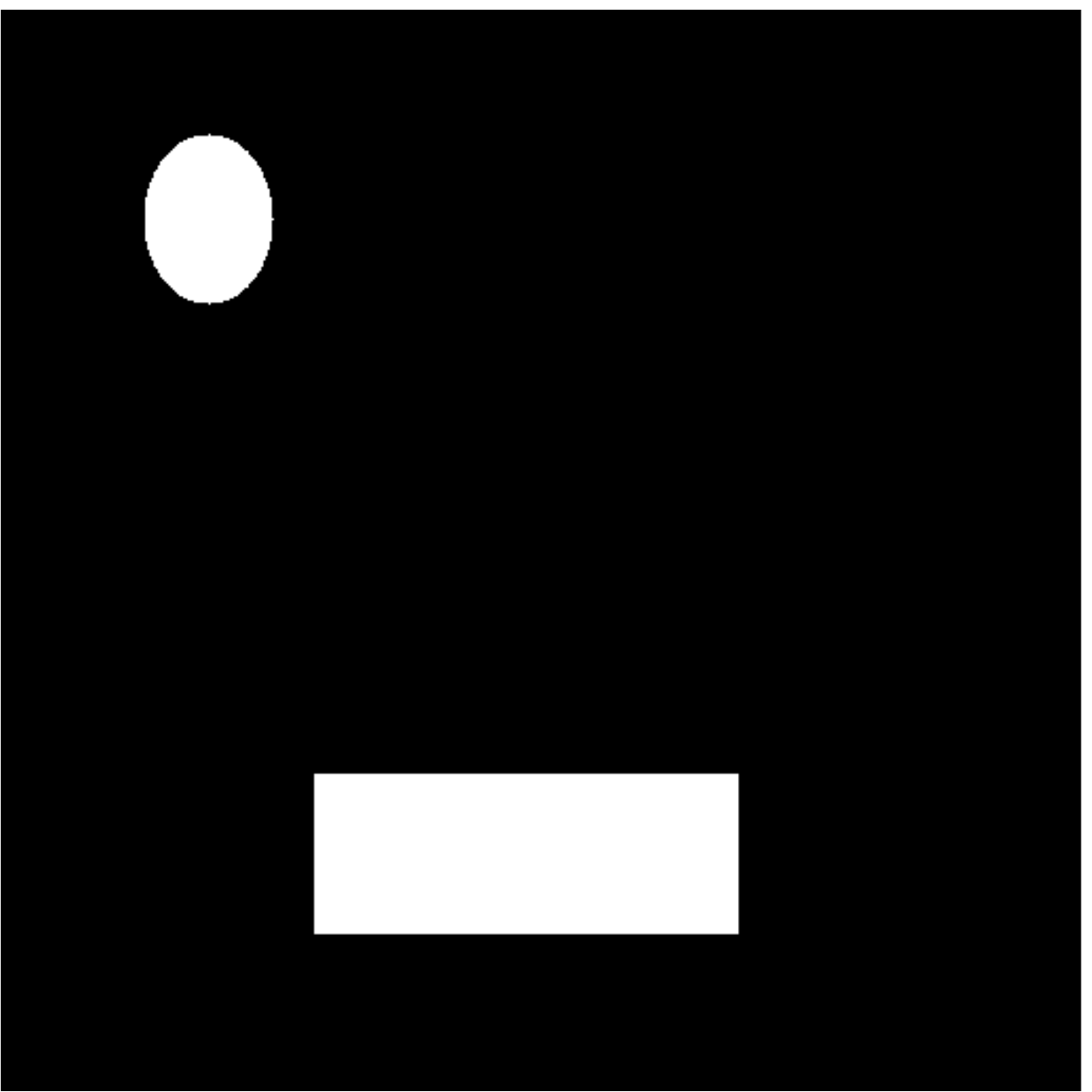} \\

&&&\\

&
1st mode: $\mathbf{d}^{(1)}$&
2nd mode:  $\mathbf{d}^{(2)}$&
Residual: $\mathbf{a}^{(2)}$ \\



\rotatebox{90}{\raisebox{-3cm}{\hspace{0.6cm} IEMD~\cite{Linderhed_A_2009_j-aada_image_emd}}}&
\includegraphics[width=1.1in]{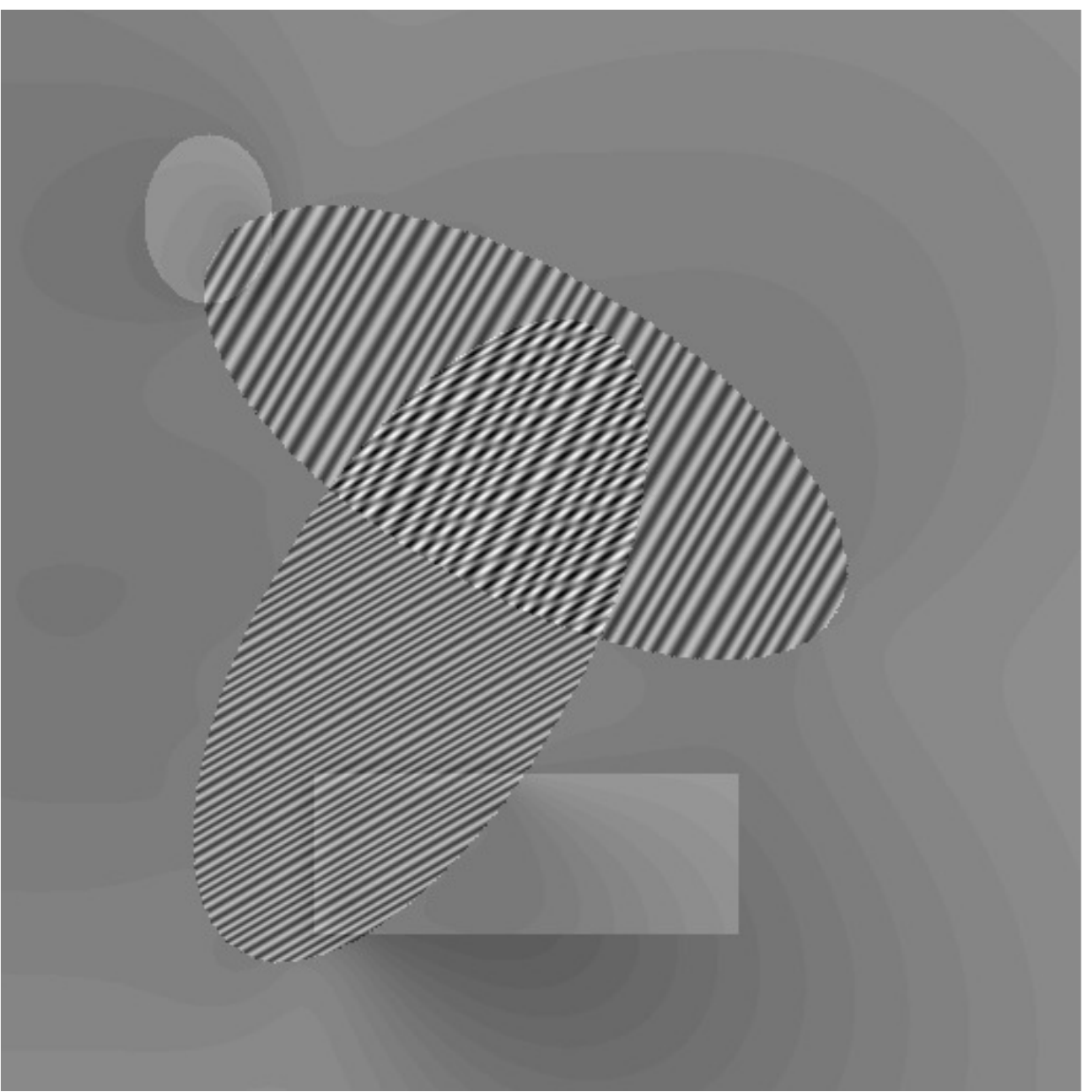} &
\includegraphics[width=1.1in]{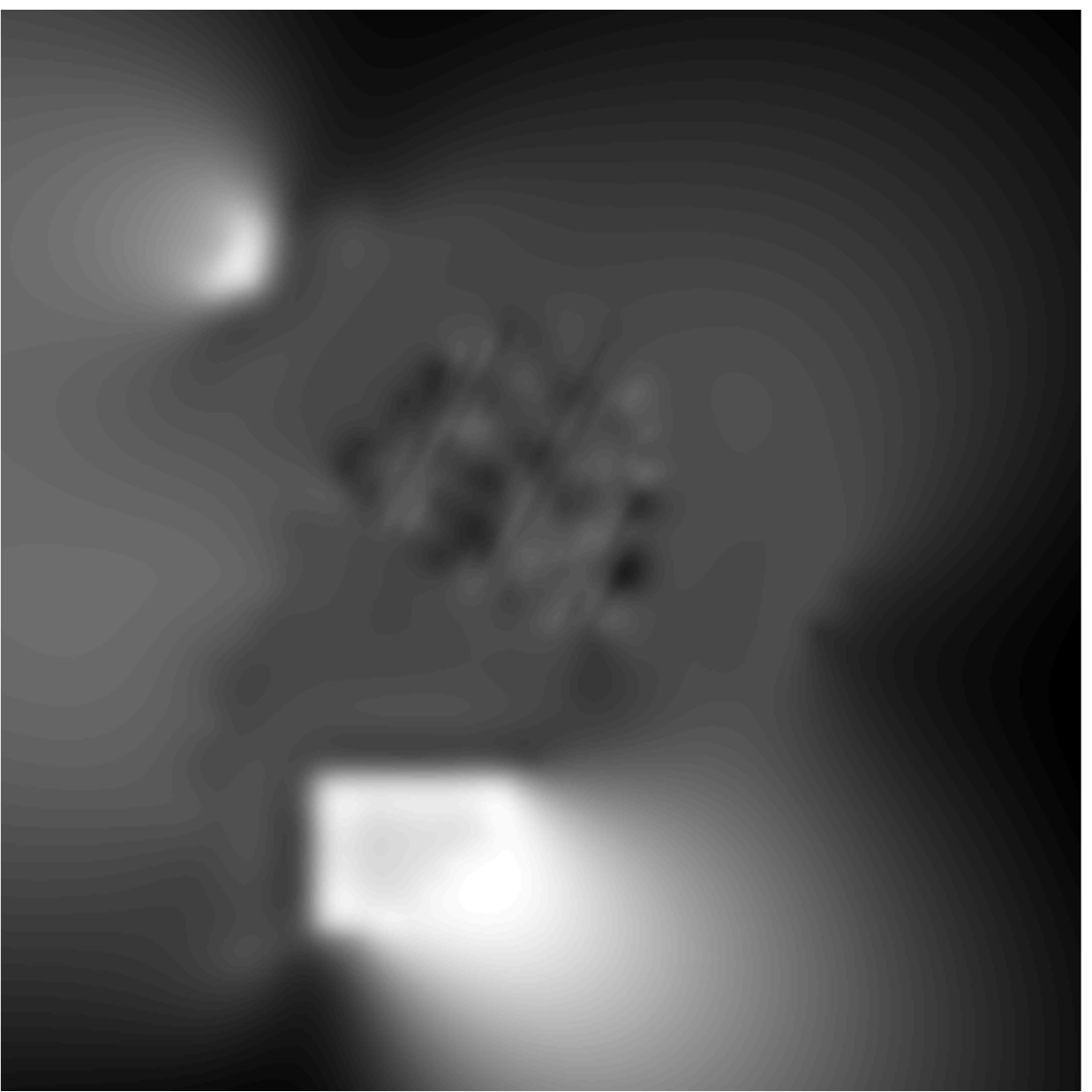} &
\includegraphics[width=1.1in]{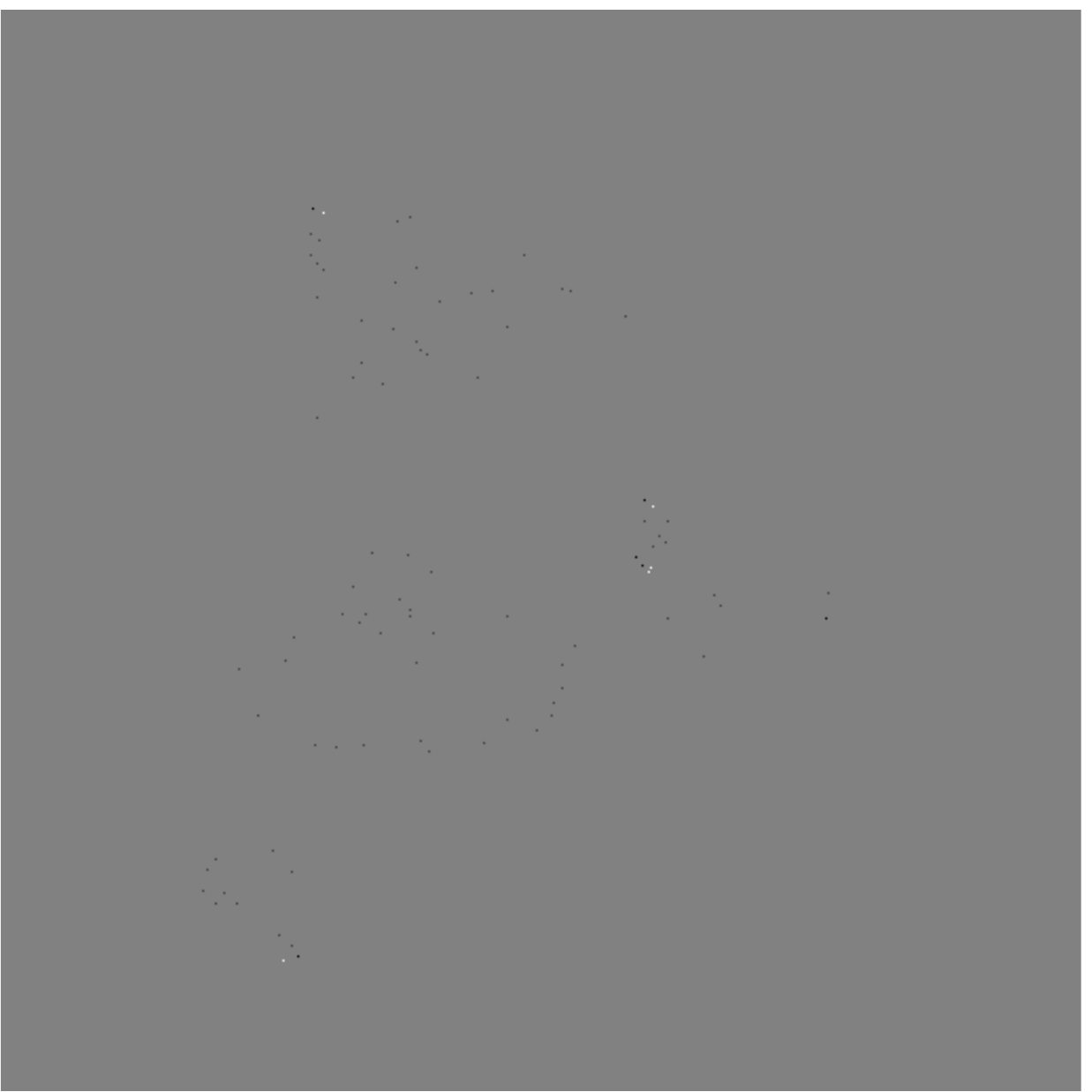} \\

\rotatebox{90}{\raisebox{-3cm}{\hspace{0.9cm} TV~\cite{Aujol_J_2006_j-ijcv_structure_tid}}}&
\includegraphics[width=1.1in]{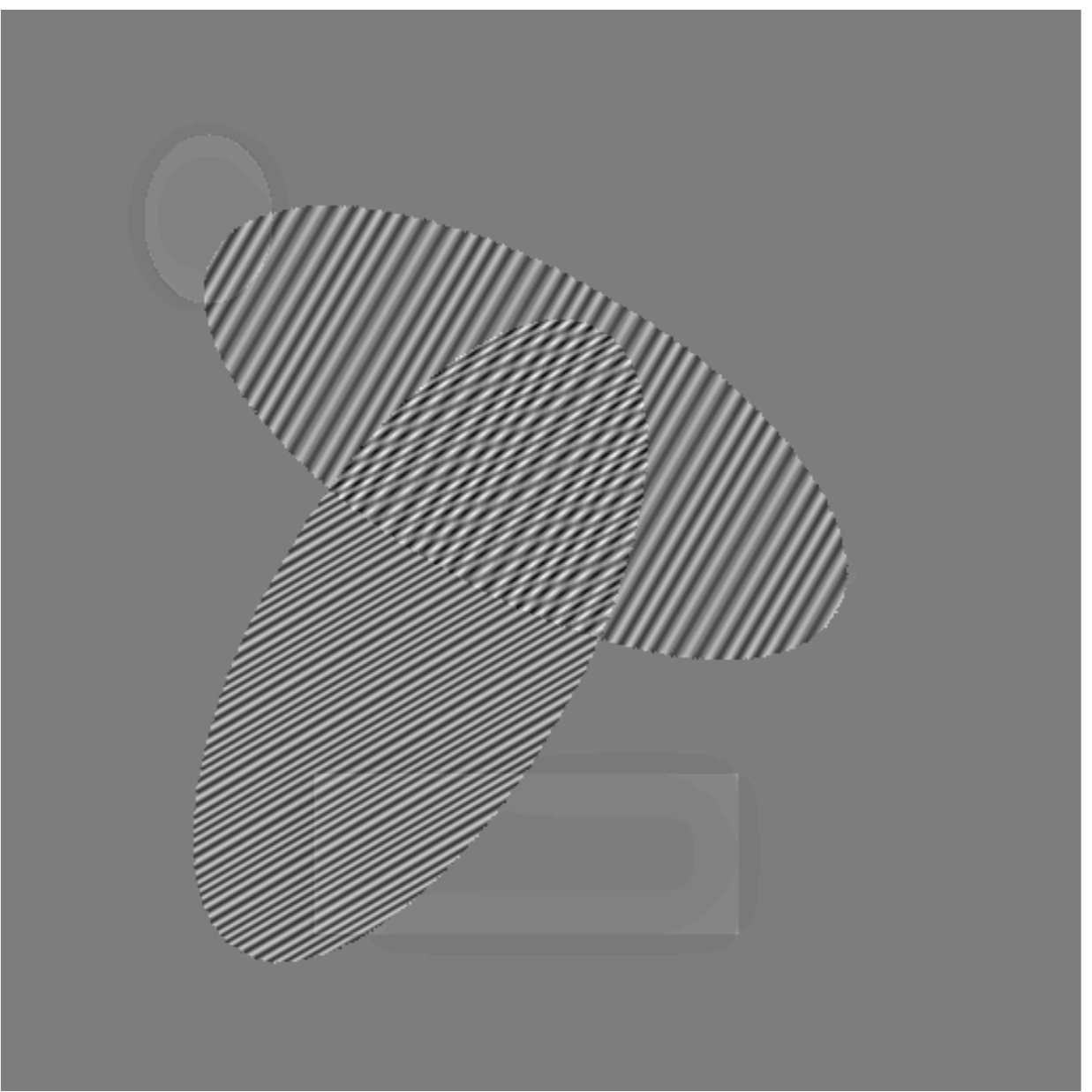} &
\includegraphics[width=1.1in]{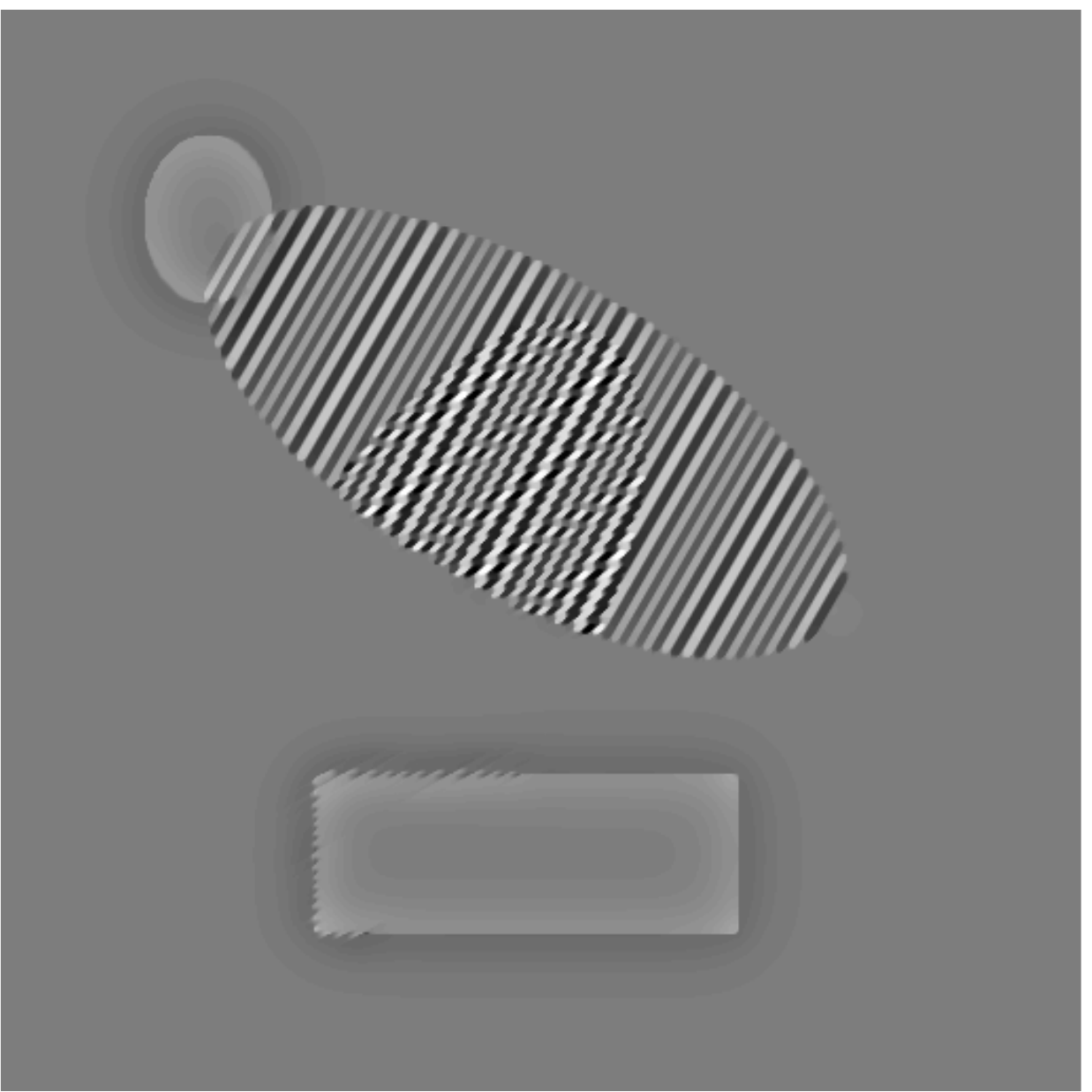} &
\includegraphics[width=1.1in]{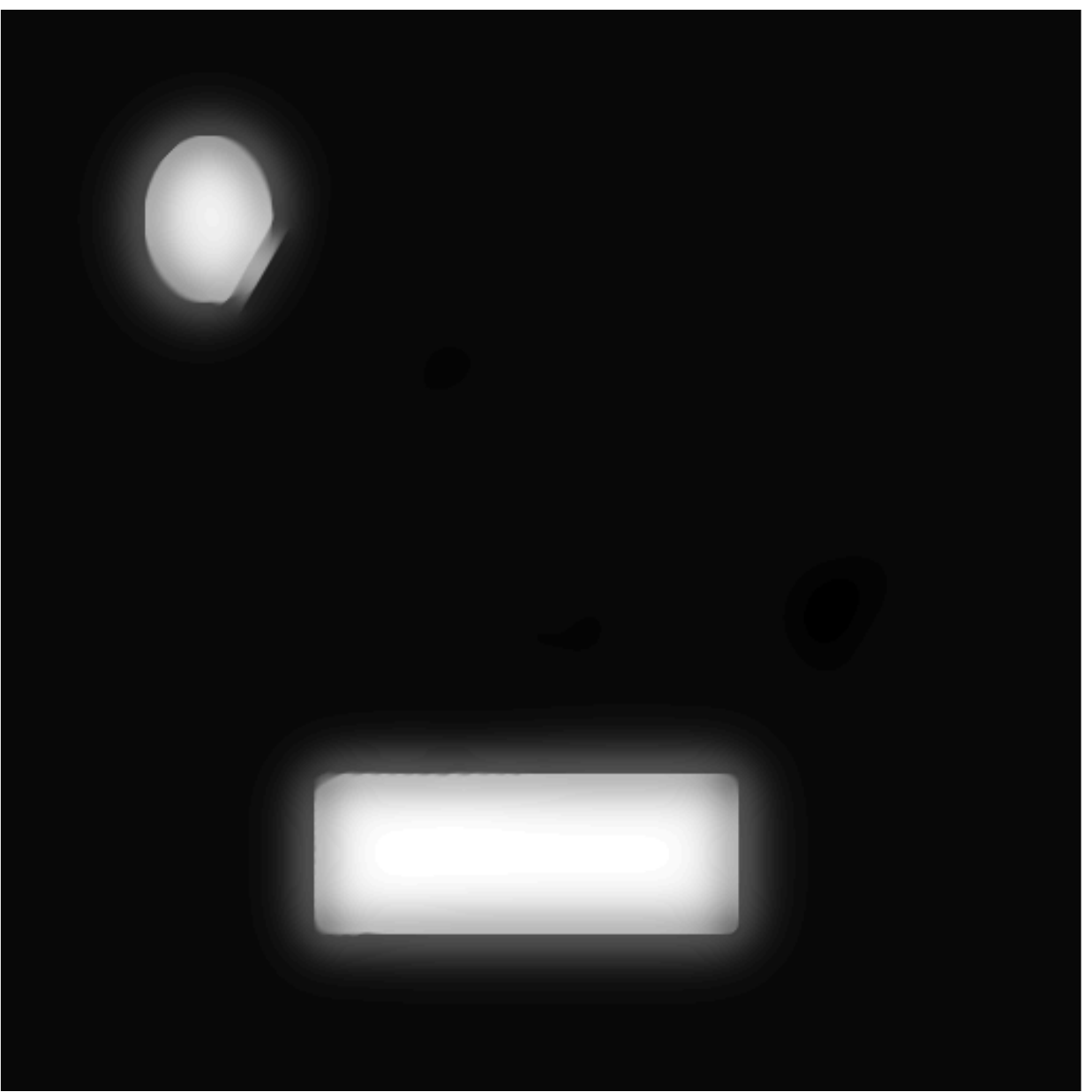} \\

\rotatebox{90}{\raisebox{-3cm}{\hspace{0.1cm} Gilles-Osher~\cite{Gilles_J_2011_ucla-cam_Bregman_img}}}&
\includegraphics[width=1.1in]{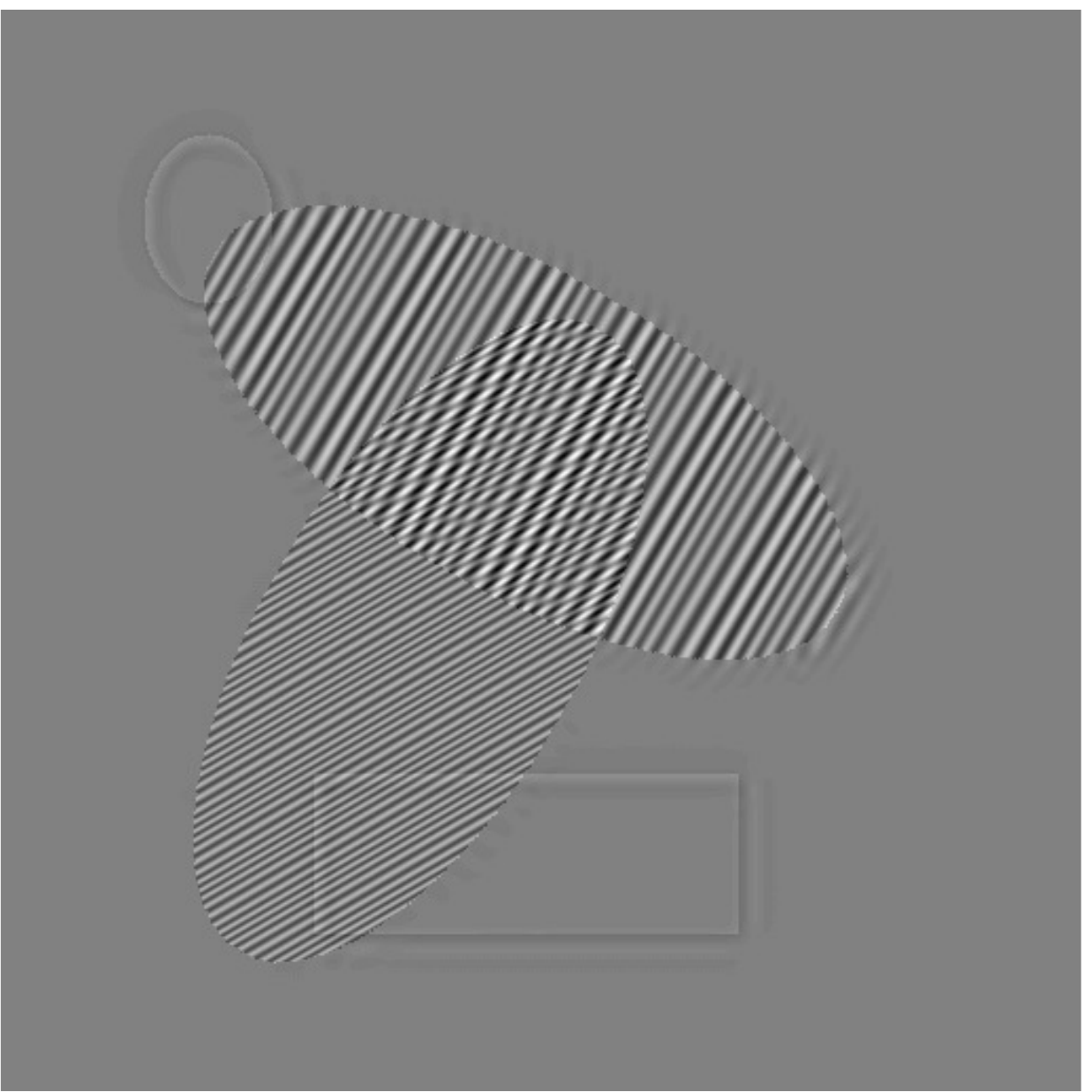} &
\includegraphics[width=1.1in]{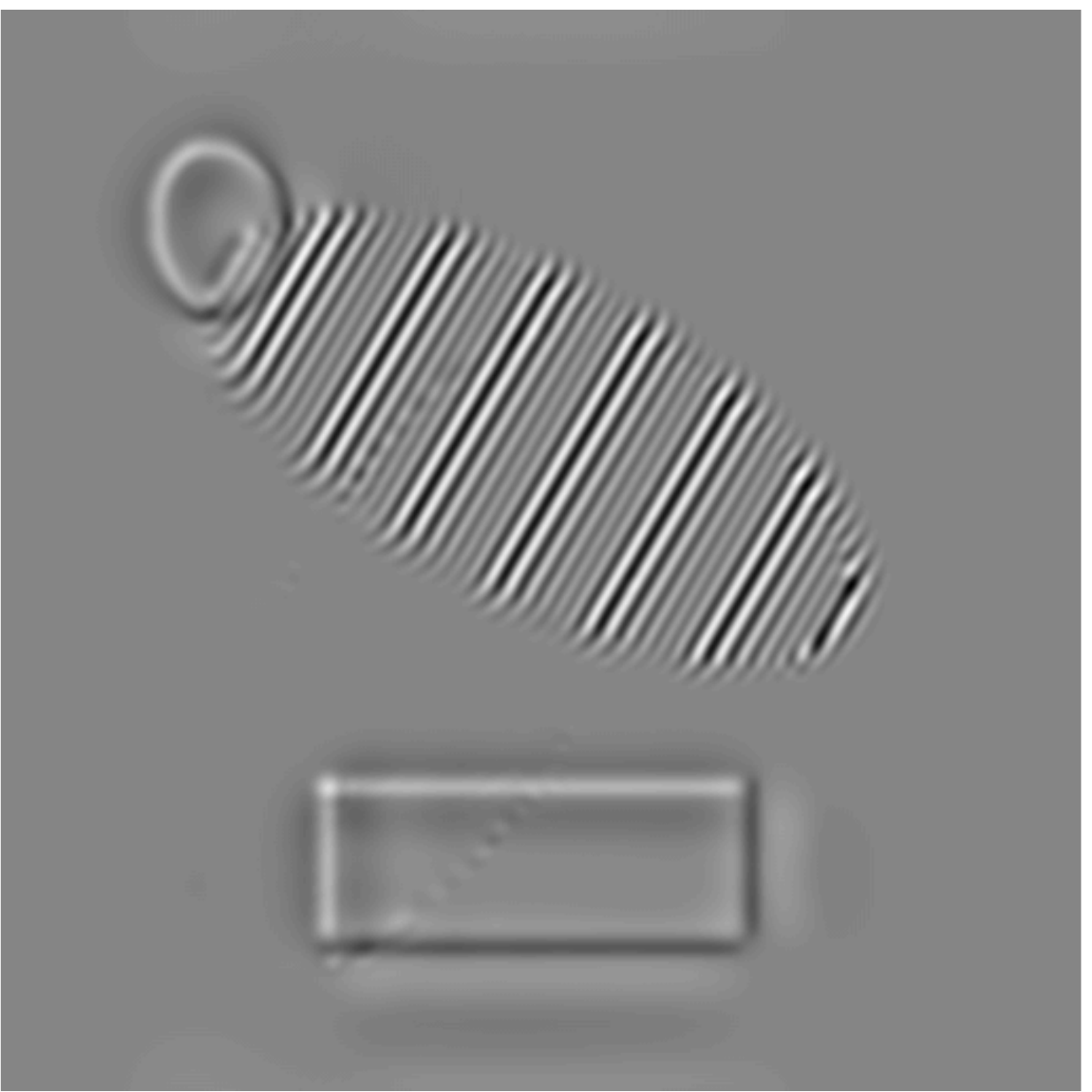} &
\includegraphics[width=1.1in]{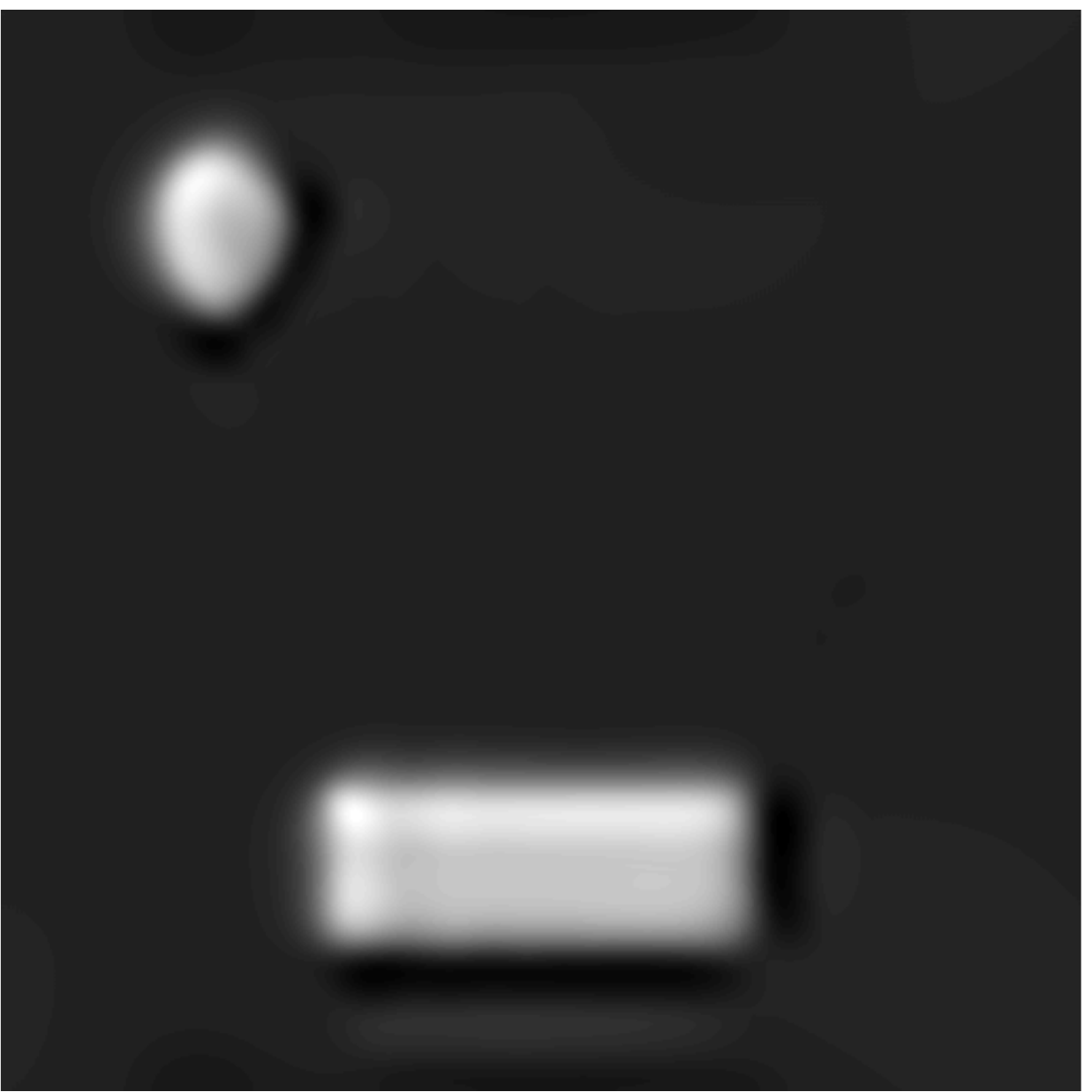} \\

\rotatebox{90}{\raisebox{-3cm}{\hspace{0.3cm} Proposed G2D}} &
\includegraphics[width=1.1in]{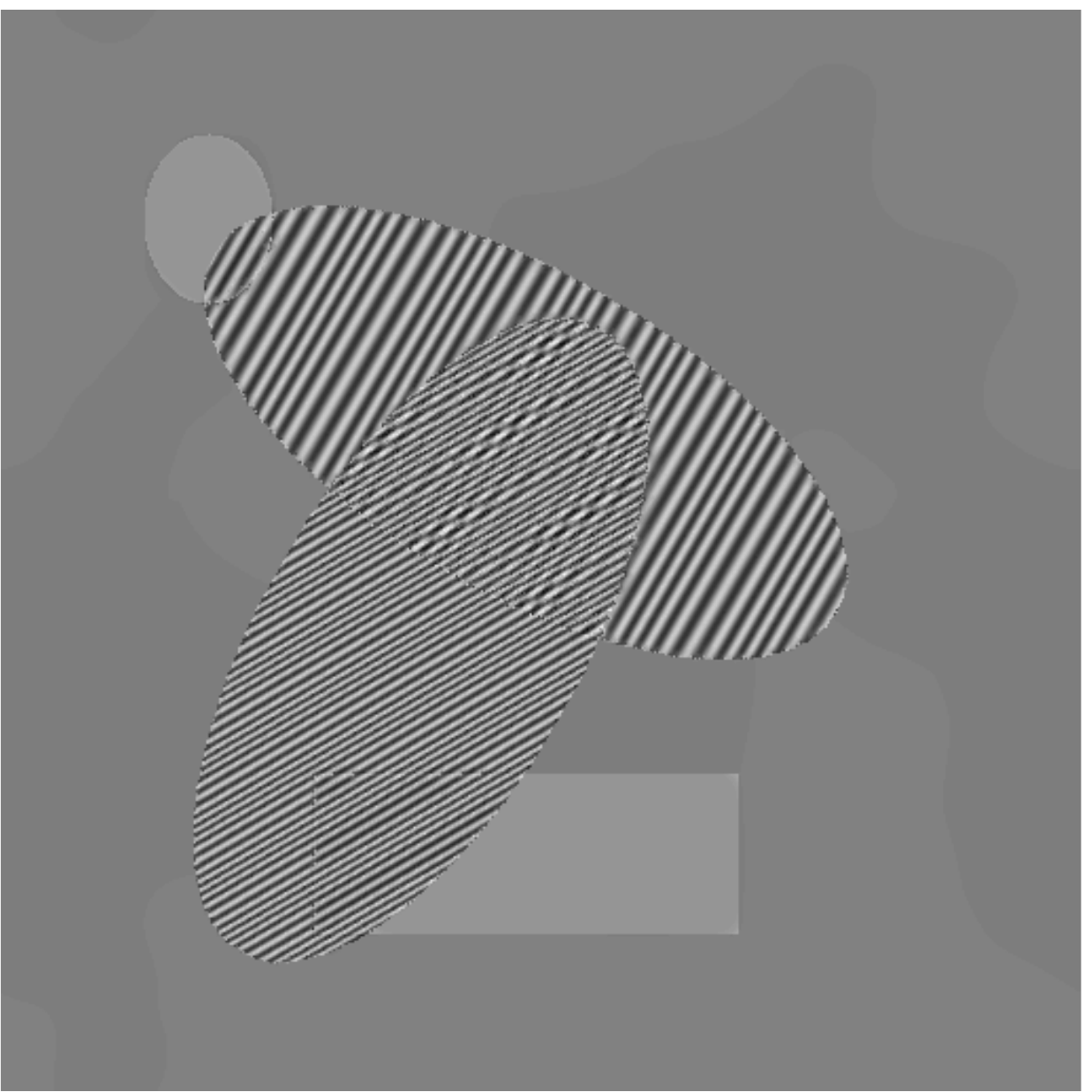} &
\includegraphics[width=1.1in]{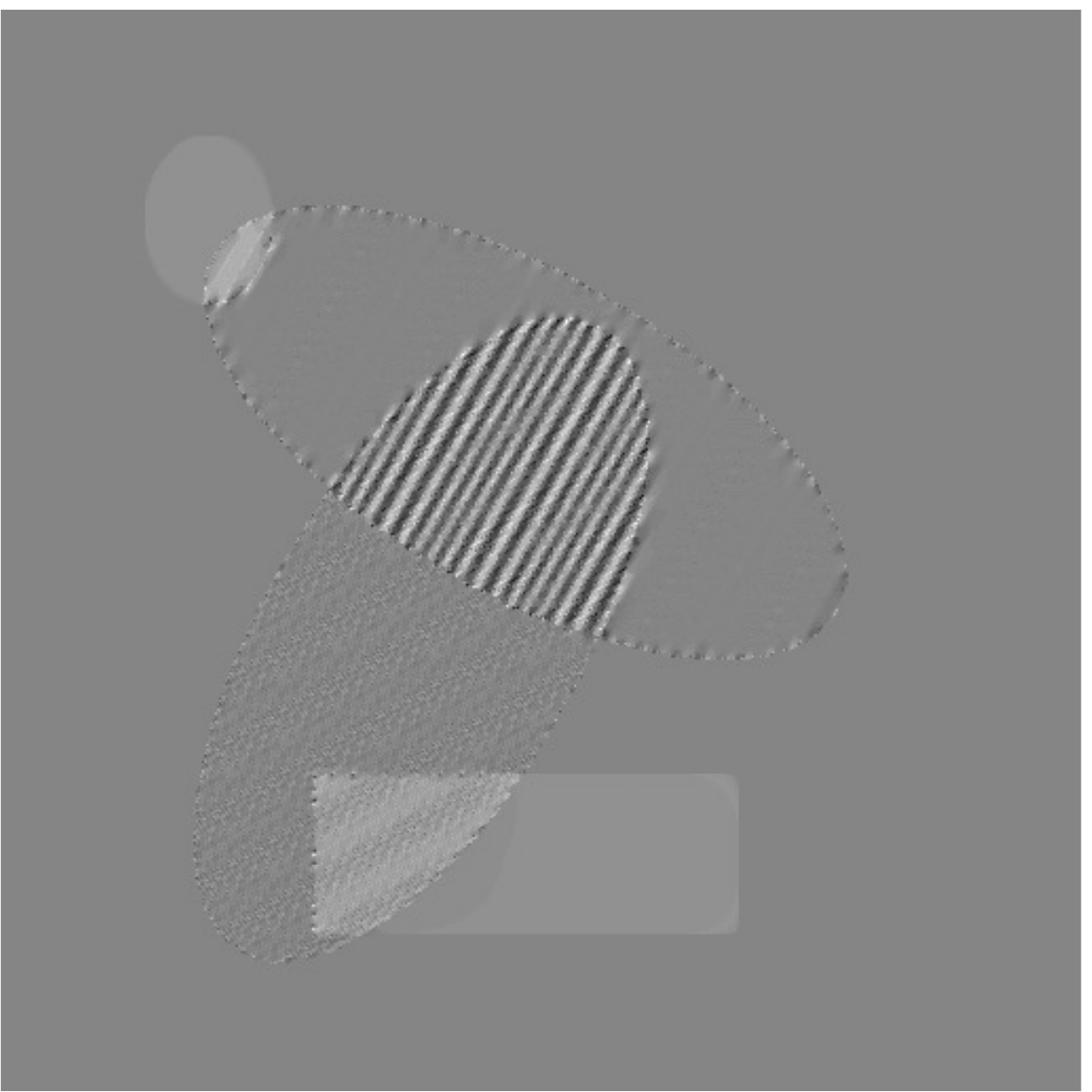} &
\includegraphics[width=1.1in]{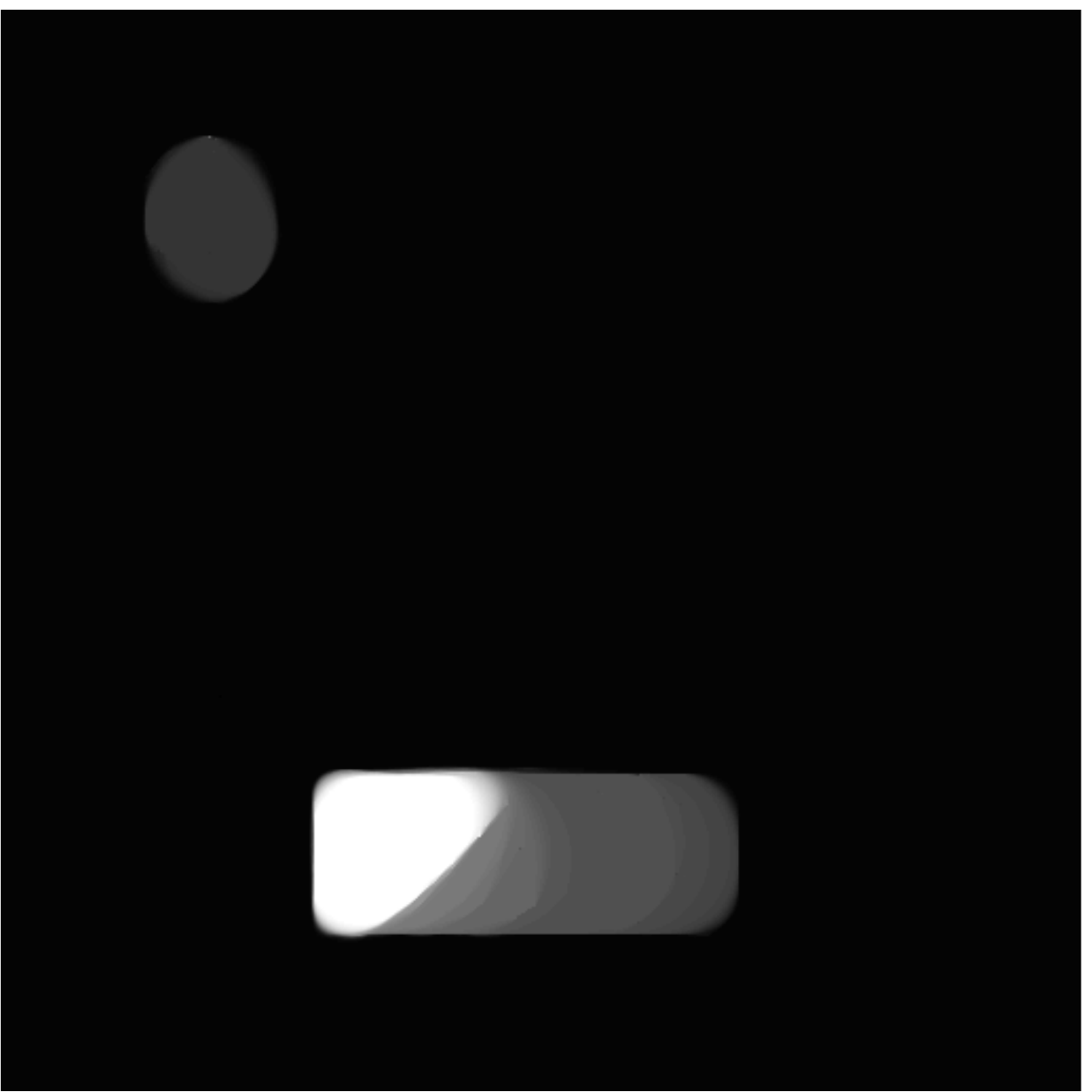}\\

\rotatebox{90}{\raisebox{-3cm}{\hspace{0.3cm} Proposed P2D}}&
\includegraphics[width=1.1in]{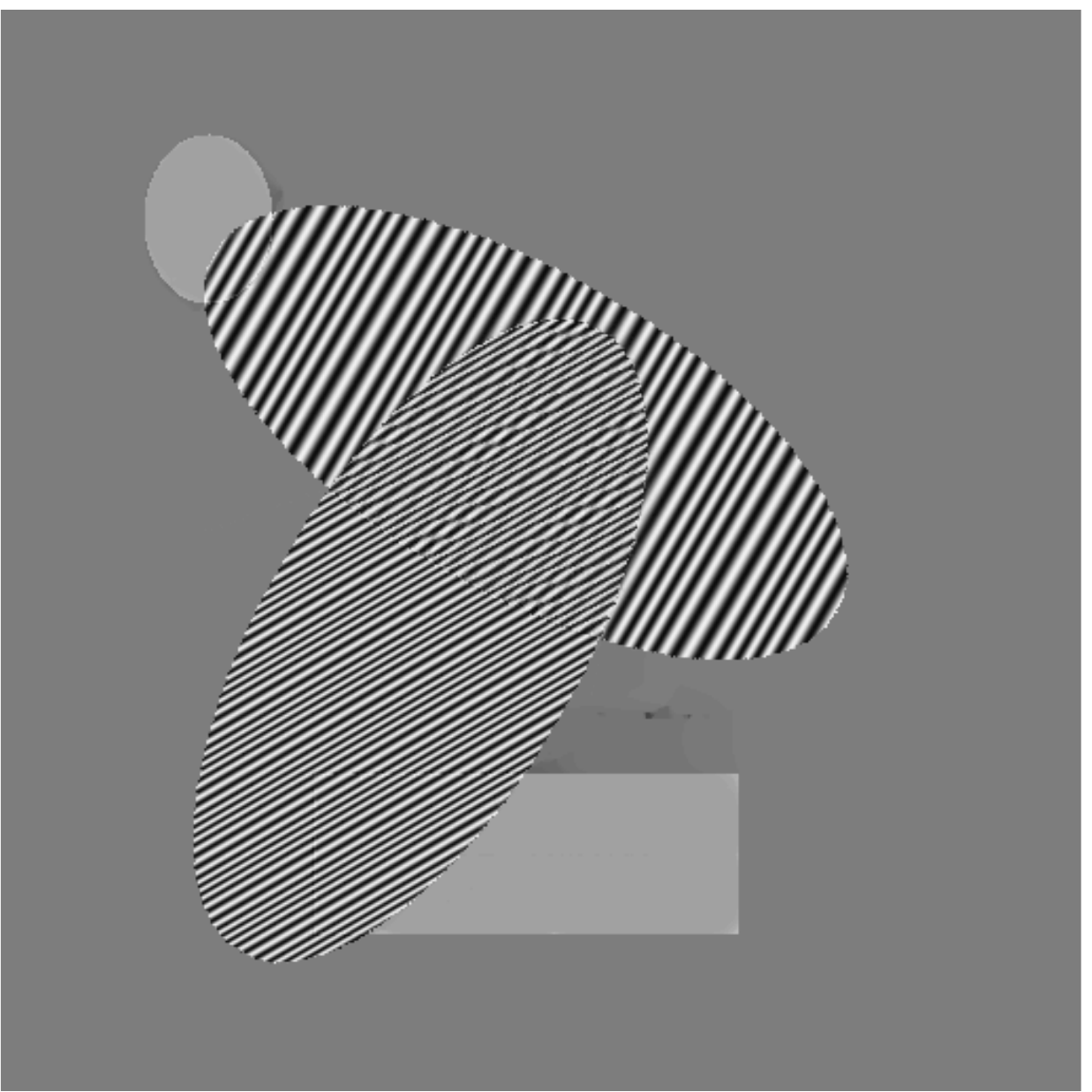} &
\includegraphics[width=1.1in]{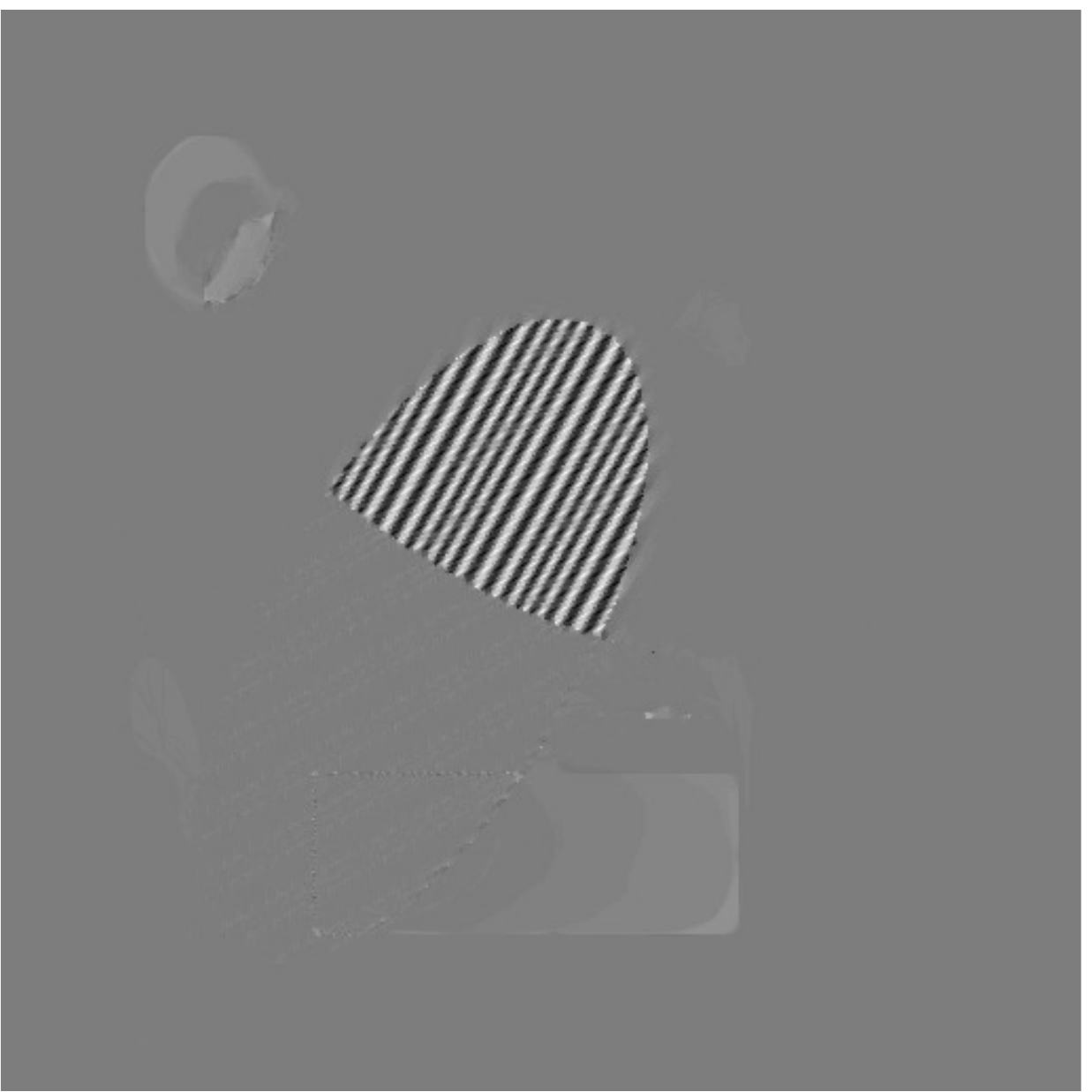} &
\includegraphics[width=1.1in]{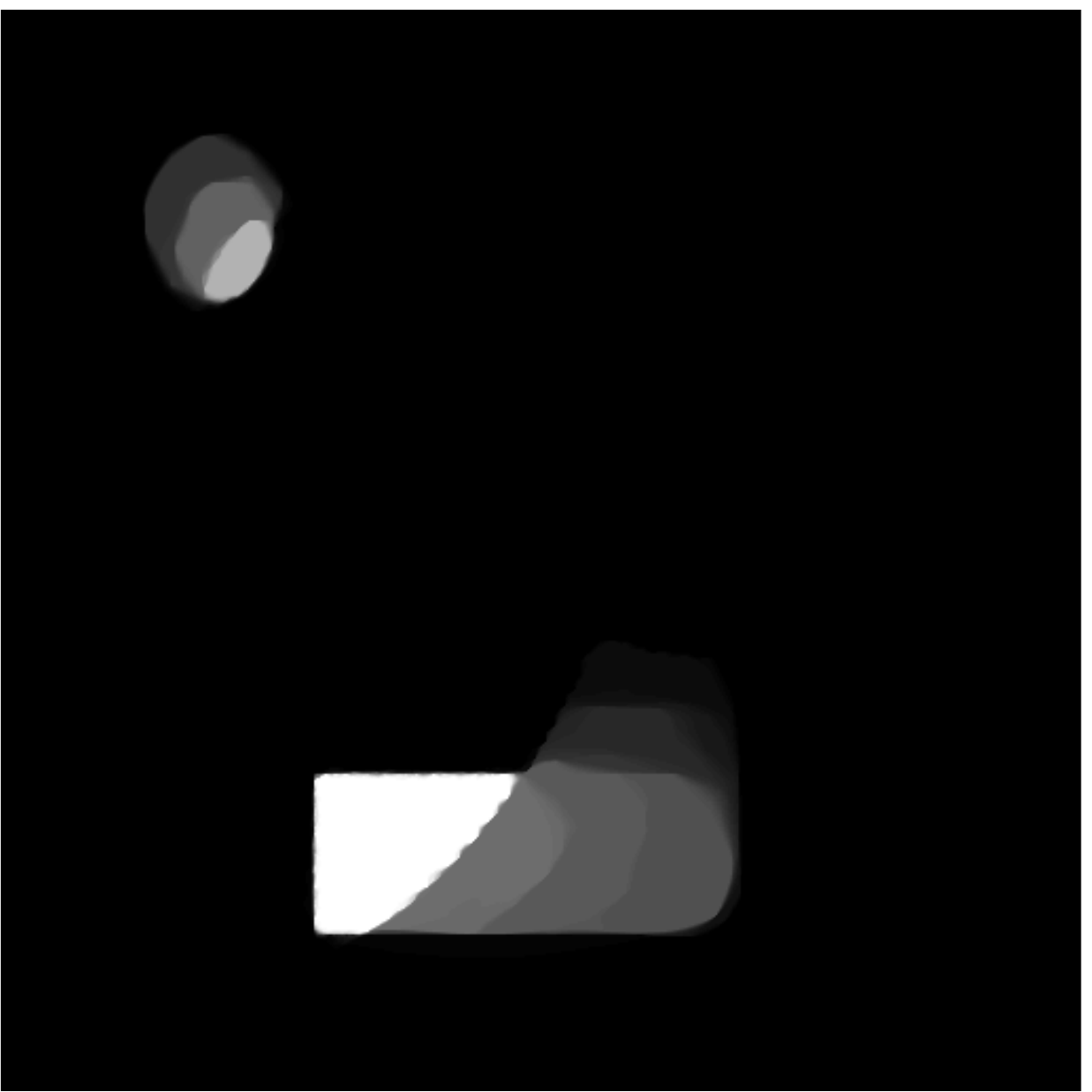}\\

\end{tabular}
\normalsize
\caption{Decomposition of the simulated data obtained with different methods. 1st row: Simulated data. 
2nd row: Image Empirical Mode Decomposition~\cite{Linderhed_A_2009_j-aada_image_emd}, 3rd row: Total Variation based decomposition~\cite{Aujol_J_2006_j-ijcv_structure_tid}, 4th row: Gilles-Osher based decomposition~\cite{Gilles_J_2011_ucla-cam_Bregman_img}, 5th row: G2D--EMD, 6th row: P2D--EMD. On 1st row, from the left to the right the columns present $\mathbf{x}$, $\mathbf{x}^{(1)}$, $\mathbf{x}^{(2)}$, $\mathbf{x}^{(3)}$. From the left to the right the columns present $\mathbf{d}^{(1)}$, $\mathbf{d}^{(2)}$ and $\mathbf{a}^{(2)}$.}
\label{fig:simuemd}

\end{center}
\end{figure*}

In our experiments, the regularization parameters are chosen as follows for G2D--EMD: 
$\rho^{(1)} = 0.02$, $\nu^{(1)} = 1000$,  $\rho^{(2)} = 0.05$, $\nu^{(2)} = 1$ while for P2D--EMD, we use $\rho^{(1)} = 0.3$, $\nu^{(1)} = 0.3 $,  $\rho^{(2)} = 1$, $\nu^{(2)} = 0.1$. For Total Variation decomposition method, we set $\rho^{(1)} = 70$ and $\rho^{(2)} = 100$. For Gilles-Osher method, we set $\rho^{(1)} = 10^4$, $\nu^{(1)} = 10^3$, $\rho^{(2)} = 10$, and $\nu^{(2)} = 10$. The results are displayed on Figure~\ref{fig:simuemd}.

First of all, our method provides a good separation of the different components. It has the expected behaviour of a 2-D--EMD, especially the P2D--EMD method: the locally fastest oscillating components are extracted at each step of the decomposition, even if their frequencies are nonstationary. The G2D--EMD approach also gives good results, but the oscillating components are not so perfectly separated. 
On contrary, the state-of-the-art IEMD does not manage to separate at all the components $\mathbf{x}_1$ and $\mathbf{x}_2$.
In comparison with other approaches like texture-cartoon decomposition, the proposed 2-D--EMD approach provides more adaptivity and a better management of nonstationary signals.
The TV based approach does not give a good separation of the three oscillating components. Gilles-Osher solution is not suited for nonstationary signals: some of the slower part of the frequency modulated component $\mathbf{x}_2$ is on the 2nd mode, while its faster part is localized on the first mode. 

To estimate the computational time of each method, we define a stopping criterion based on the norm of the difference between two successive iterates set to $10^{-6}$. The complete decomposition into two modes needs around 3 minutes with TV decomposition, around 6 minutes with Gilles-Osher decomposition, and less than 15 minutes with P2D--EMD. The decomposition using G2D--EMD is substantially longer and takes too much time to reach the stopping criterion, so we have stopped the algorithm after $10^4$ iterations. Then, the full decomposition with G2D--EMD takes a little less than 1 hour. P2D--EMD takes a few more time than other state-of-the-art methods, but it is compensated with the better separation performance.

%

\subsubsection{Spectral analysis}

\begin{figure*}
\begin{center}
\footnotesize
\begin{tabular}{p{0.2cm}ccccccc}


 &
1st mode&
 Amplitude &
 &
 Frequency &
 &
 Orientation \\

 \rotatebox{90}{Riesz-Laplace~\cite{Unser_M_2009_j-ieee-tip_multiresolution_msa}} &
 \includegraphics[width=1.1in]{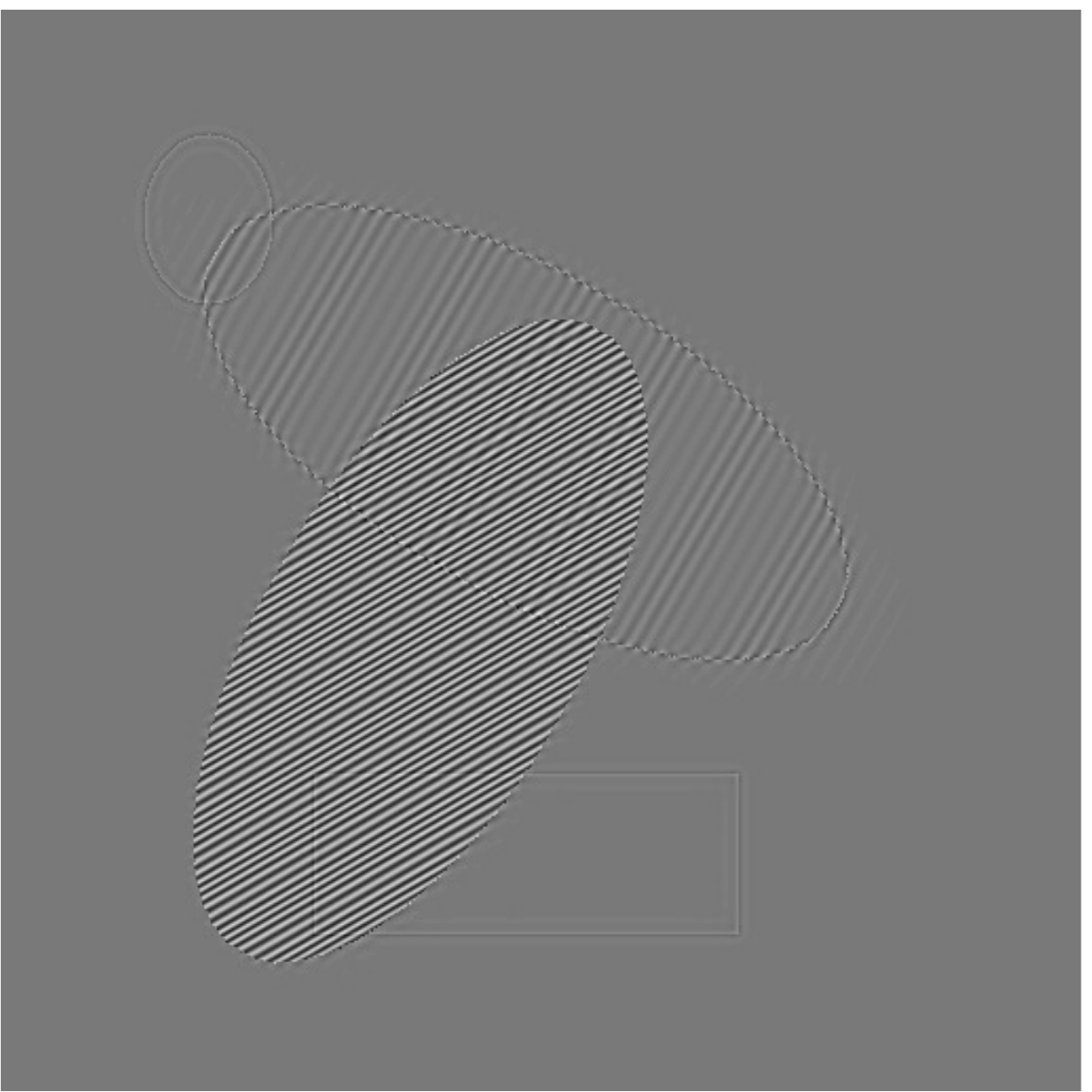} &
\includegraphics[width=1.1in]{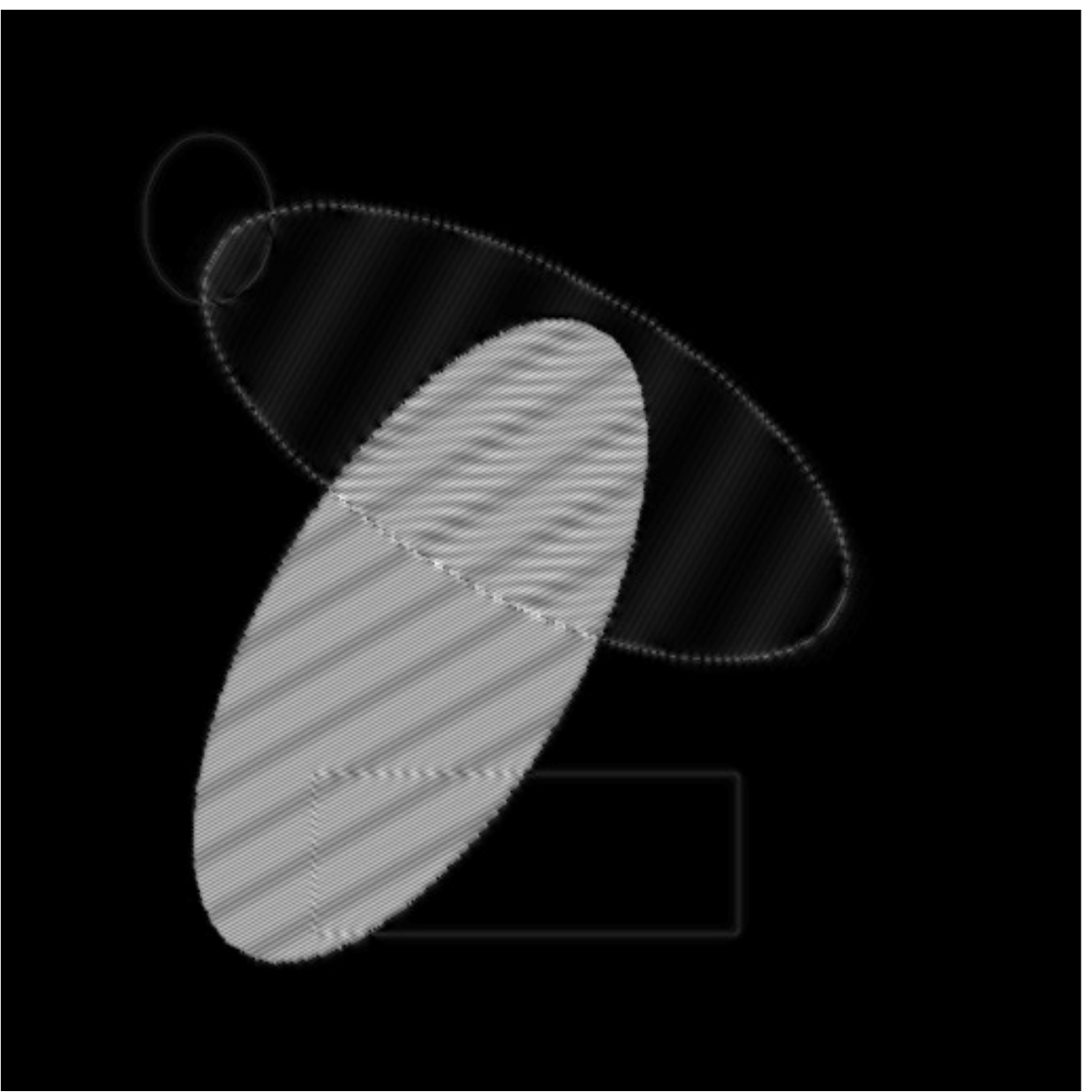} &
\includegraphics[width=0.16in]{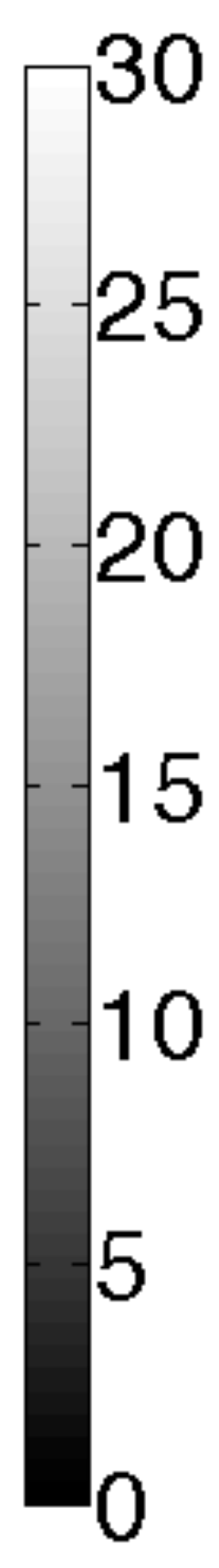} &
\includegraphics[width=1.1in]{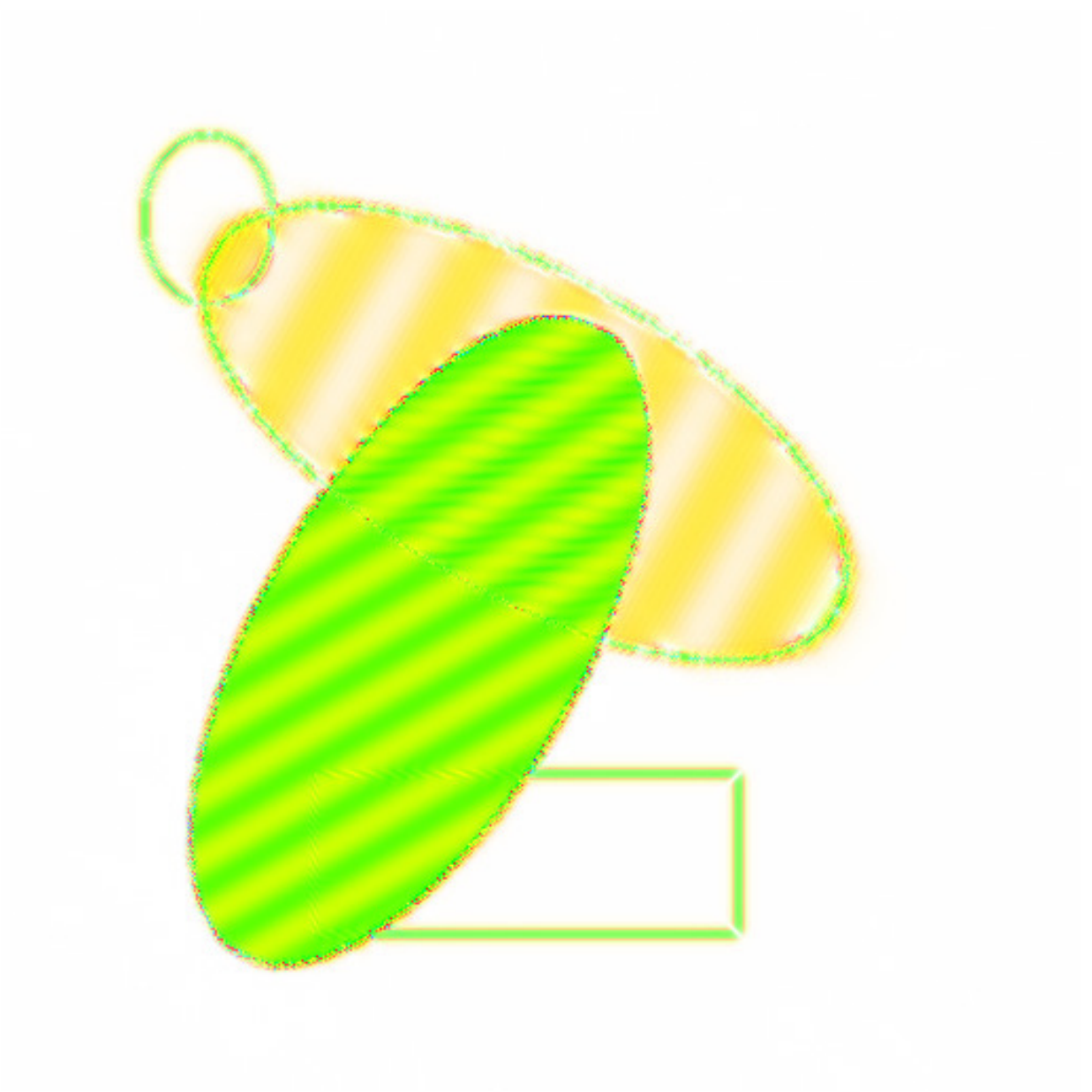} &
\includegraphics[width=0.17in]{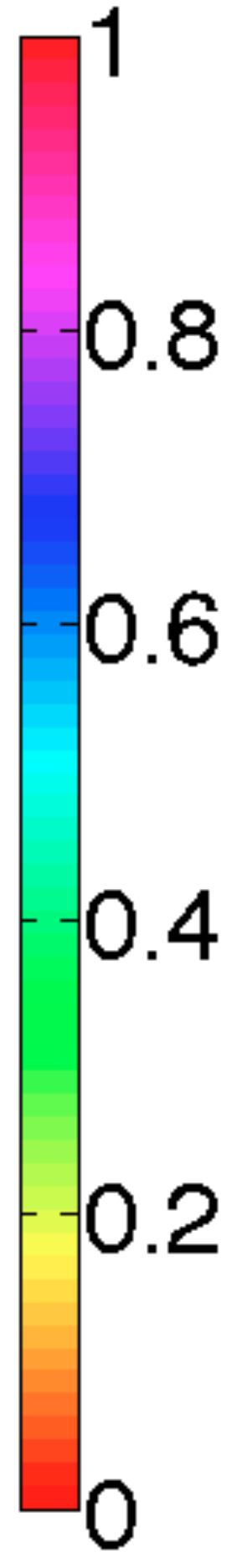} &
\includegraphics[width=1.1in]{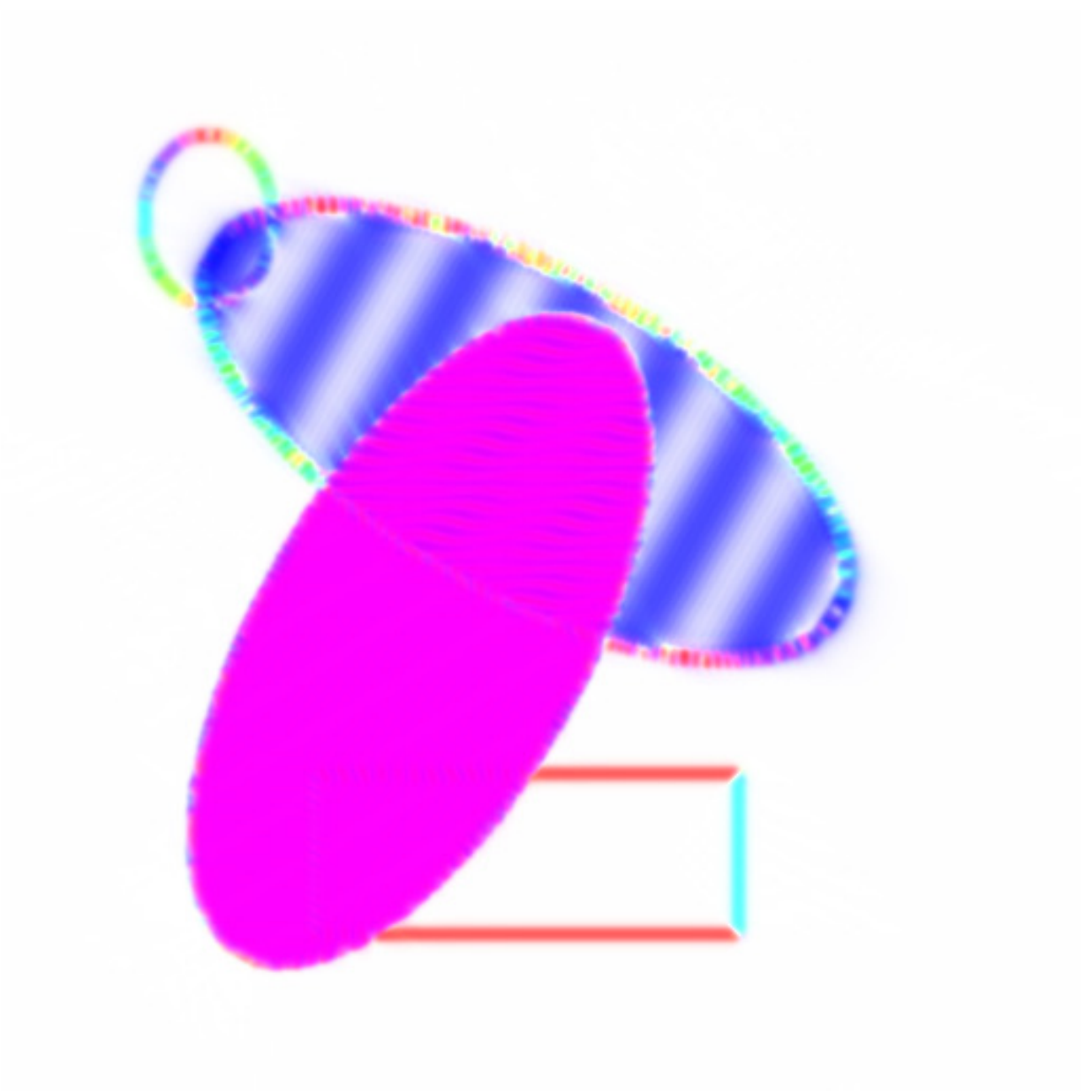} &
\includegraphics[width=0.2in]{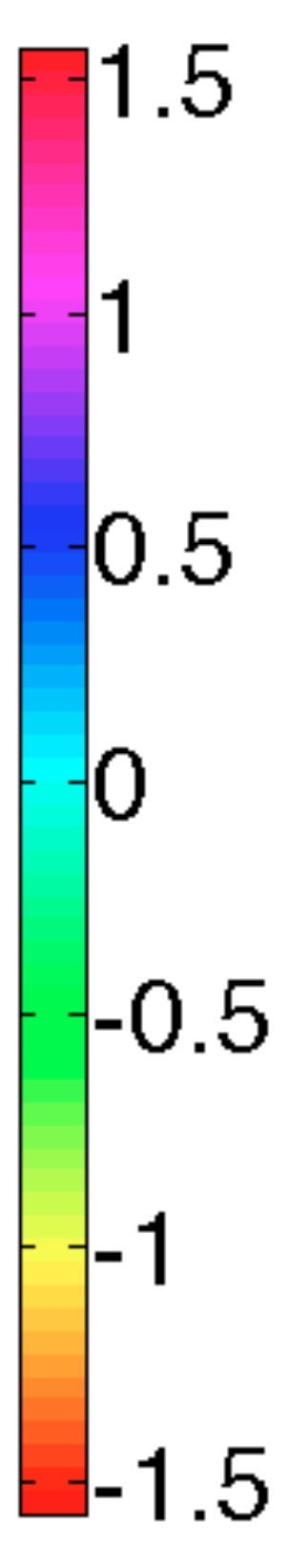}
 \\

\rotatebox{90}{\hspace{0.5cm} G2D--HHT} &
\includegraphics[width=1.1in]{figures/imf1_g2d_0p02_1000-eps-converted-to.pdf} &
\includegraphics[width=1.1in]{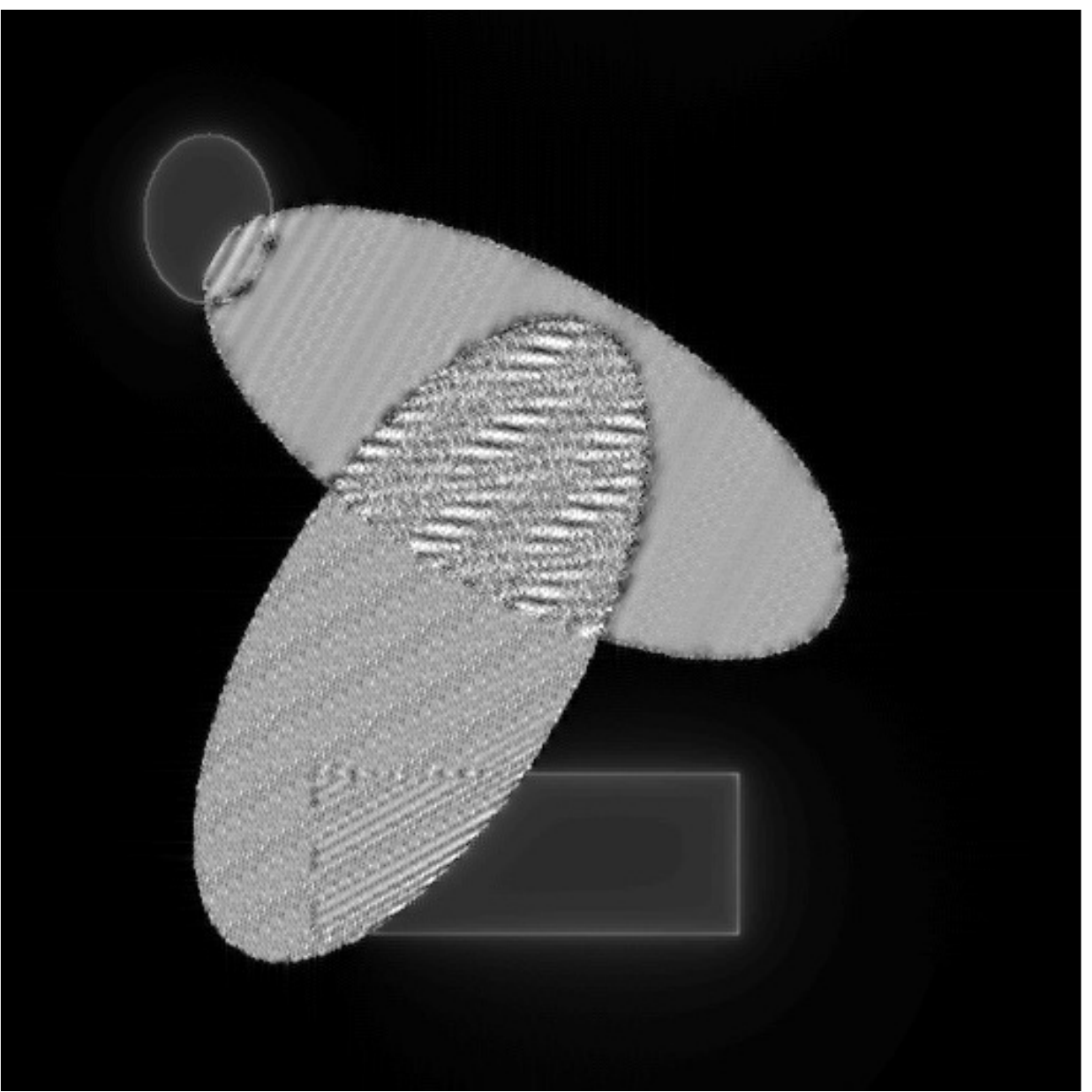} &
\includegraphics[width=0.16in]{figures/colorbar_amplitude_simu-eps-converted-to.pdf} &
\includegraphics[width=1.1in]{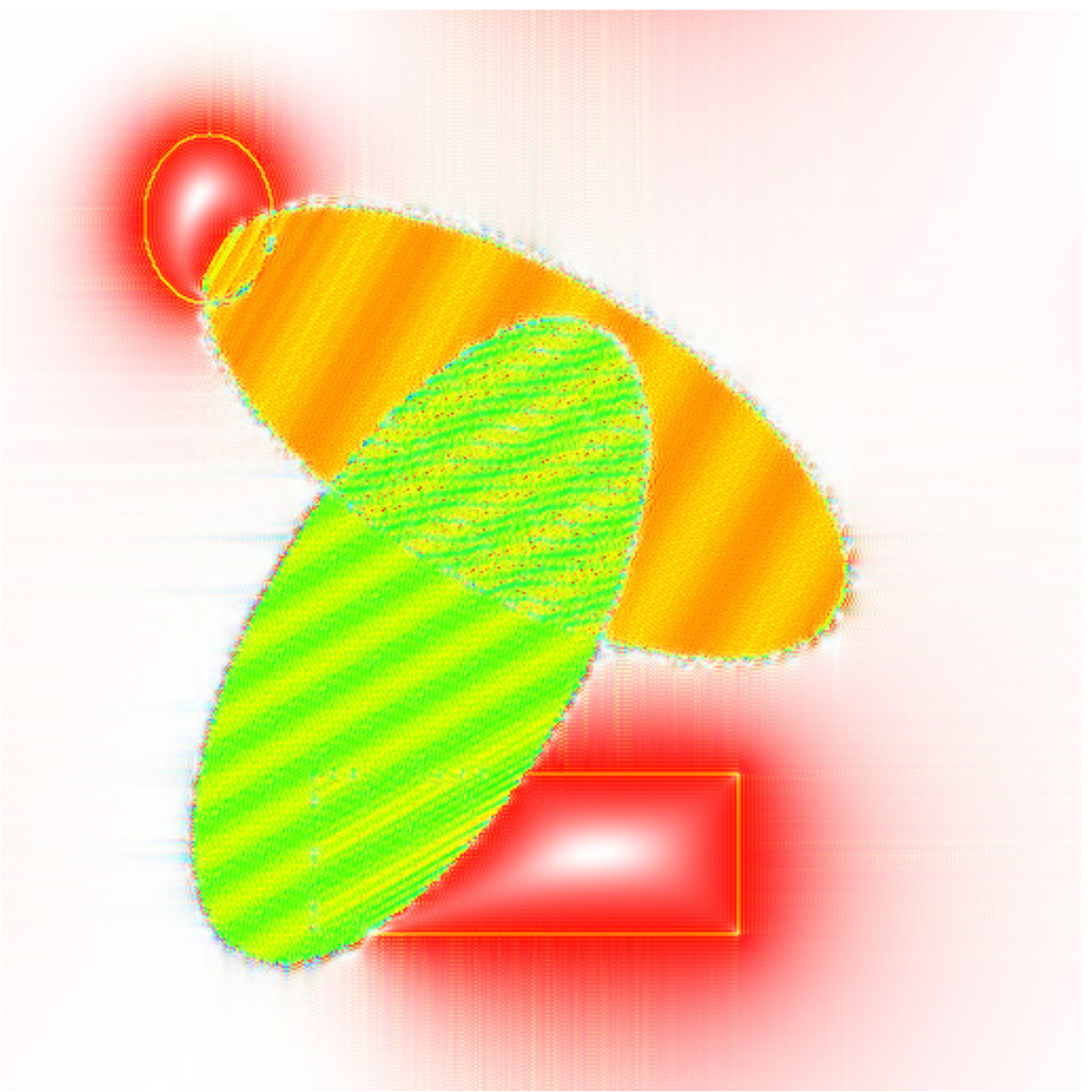} &
\includegraphics[width=0.17in]{figures/colorbar_freq-eps-converted-to.pdf} &
\includegraphics[width=1.1in]{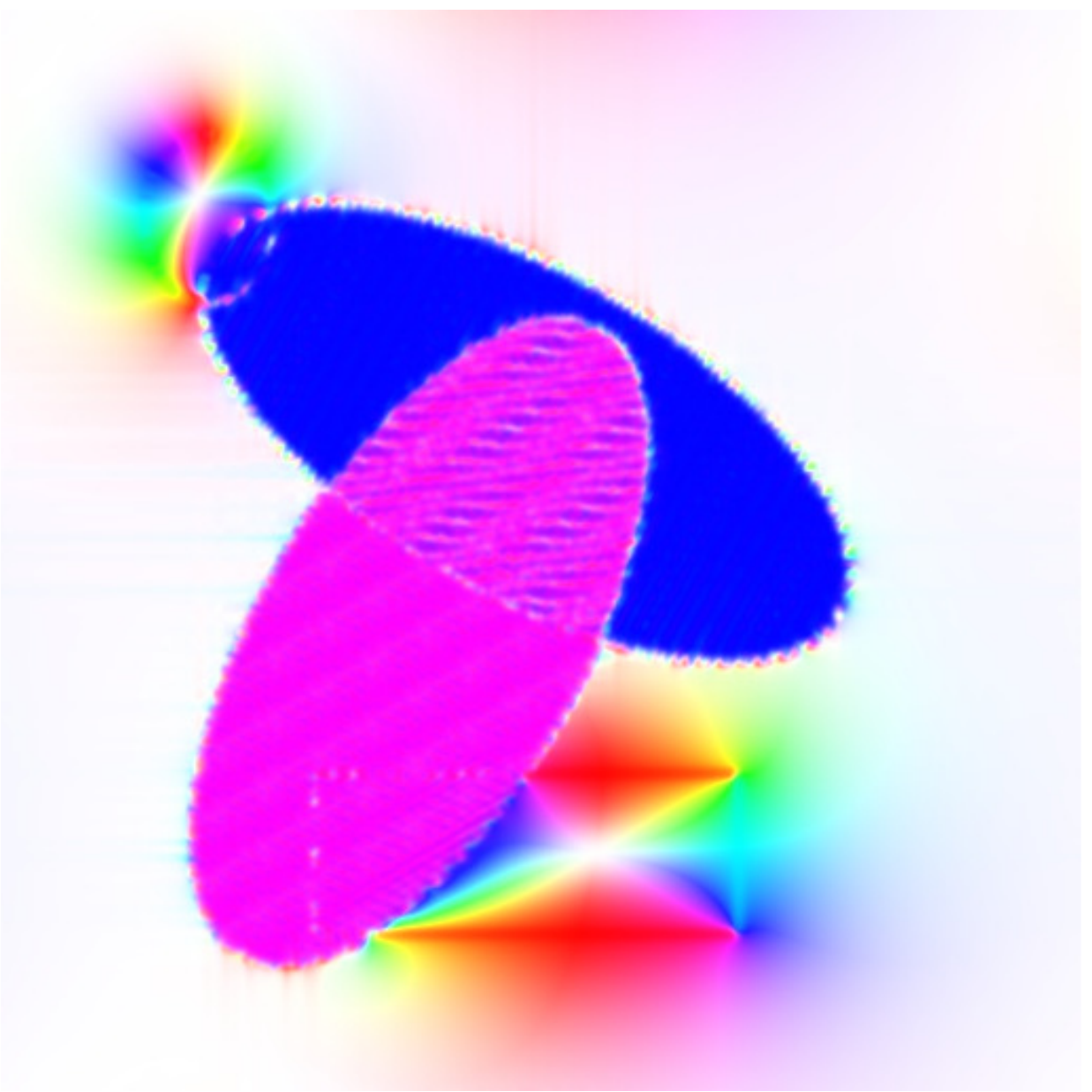} &
\includegraphics[width=0.2in]{figures/colorbar_orientation-eps-converted-to.pdf}
\\

\rotatebox{90}{Proposed G2D--PHT} &
\includegraphics[width=1.1in]{figures/imf1_g2d_0p02_1000-eps-converted-to.pdf} &
\includegraphics[width=1.1in]{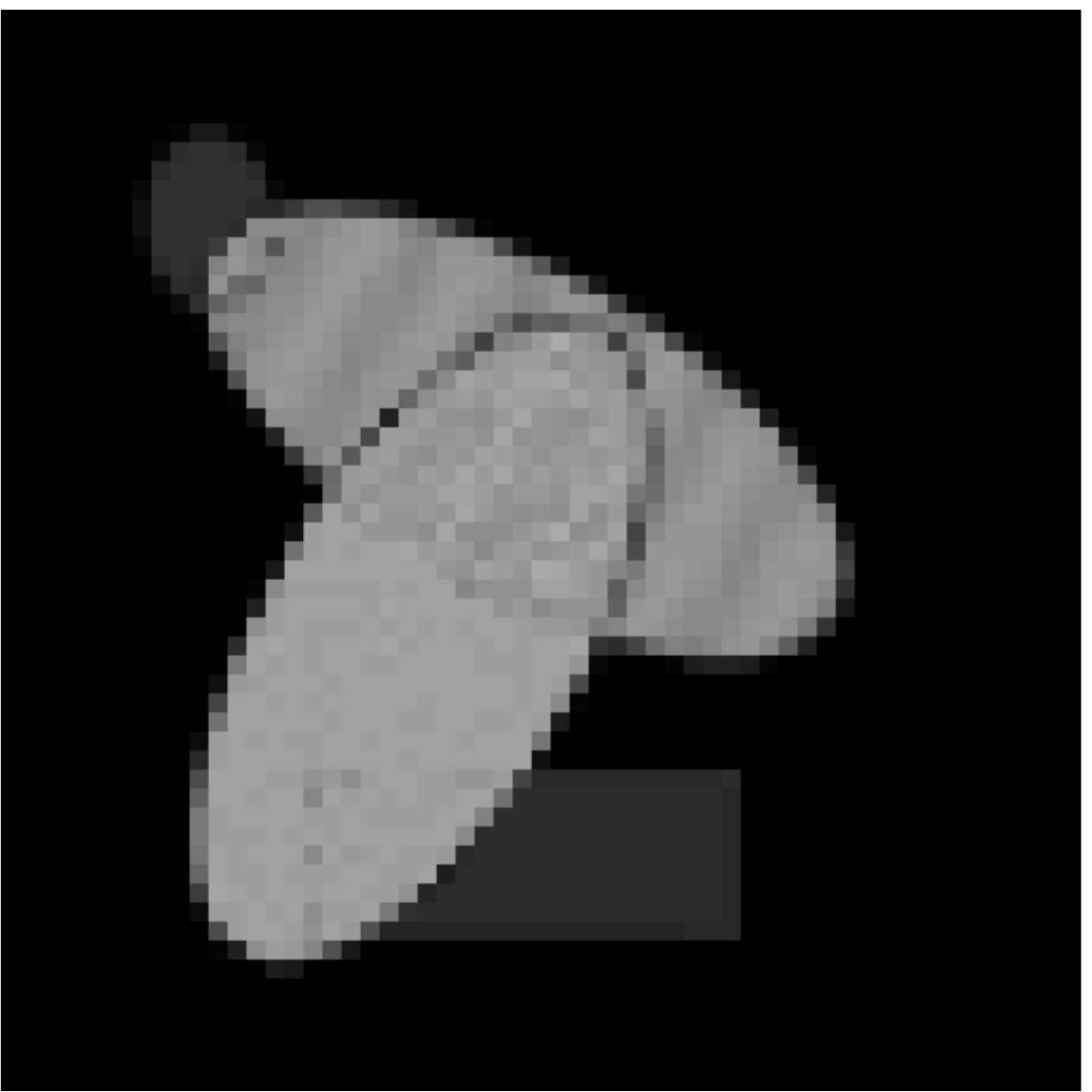} &
\includegraphics[width=0.16in]{figures/colorbar_amplitude_simu-eps-converted-to.pdf} &
\includegraphics[width=1.1in]{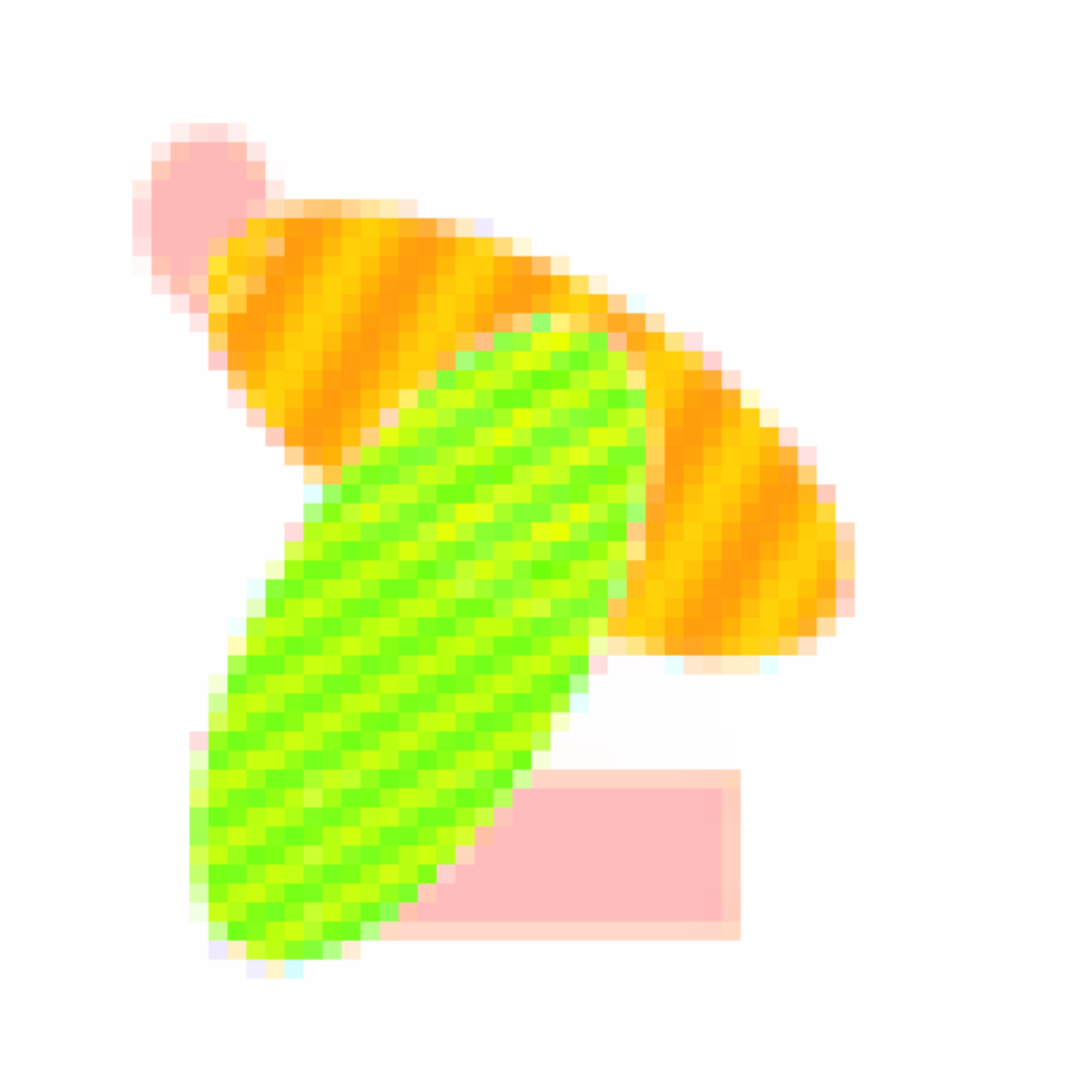} &
\includegraphics[width=0.17in]{figures/colorbar_freq-eps-converted-to.pdf} &
\includegraphics[width=1.1in]{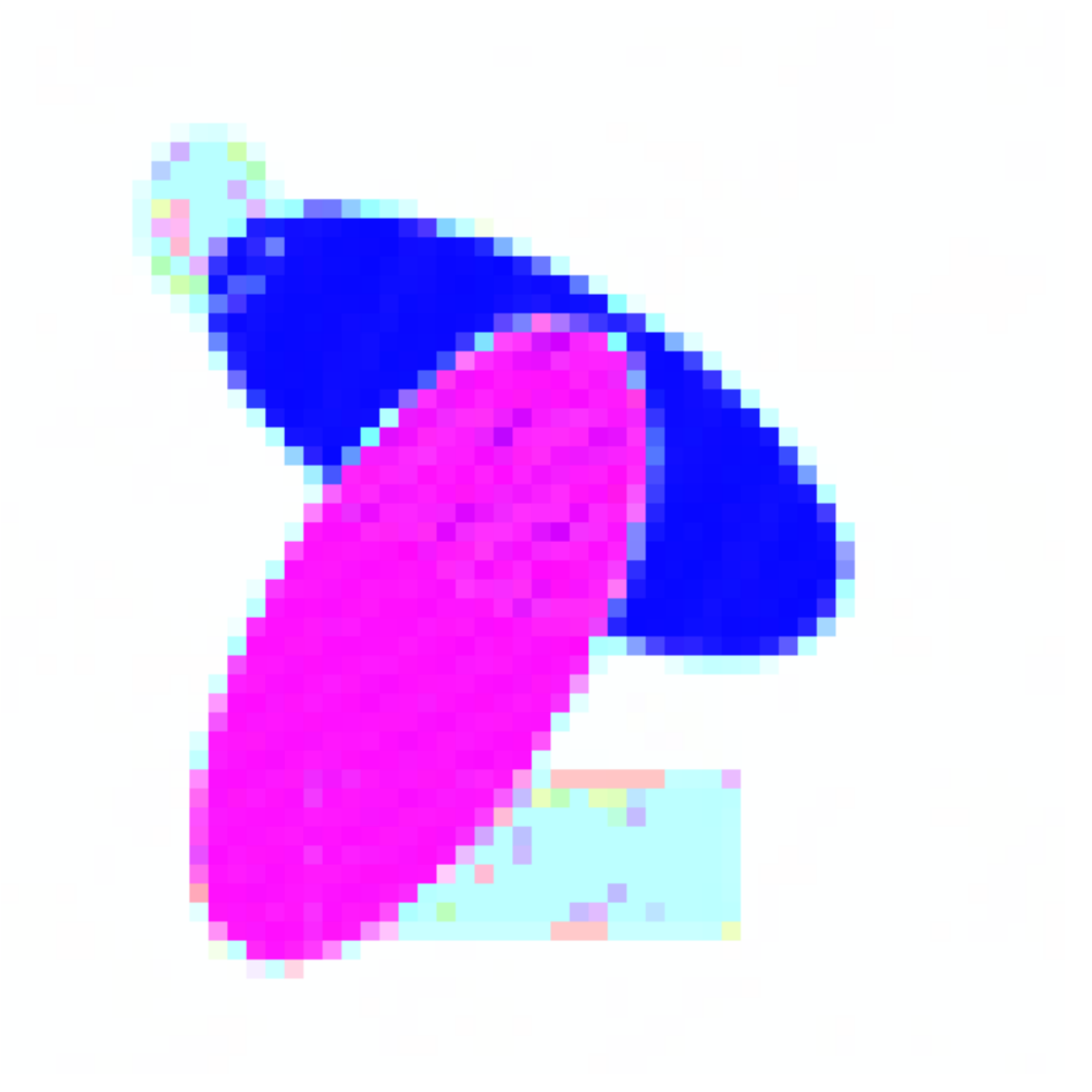} &
\includegraphics[width=0.2in]{figures/colorbar_orientation-eps-converted-to.pdf}
 \\

\rotatebox{90}{\hspace{0.2cm} P2D--HHT~\cite{Schmitt_J_2013_p-icassp_2D_HHT} }&
\includegraphics[width=1.1in]{figures/exp3_vemd_imf1_0p3_0p3-eps-converted-to.pdf} &
\includegraphics[width=1.1in]{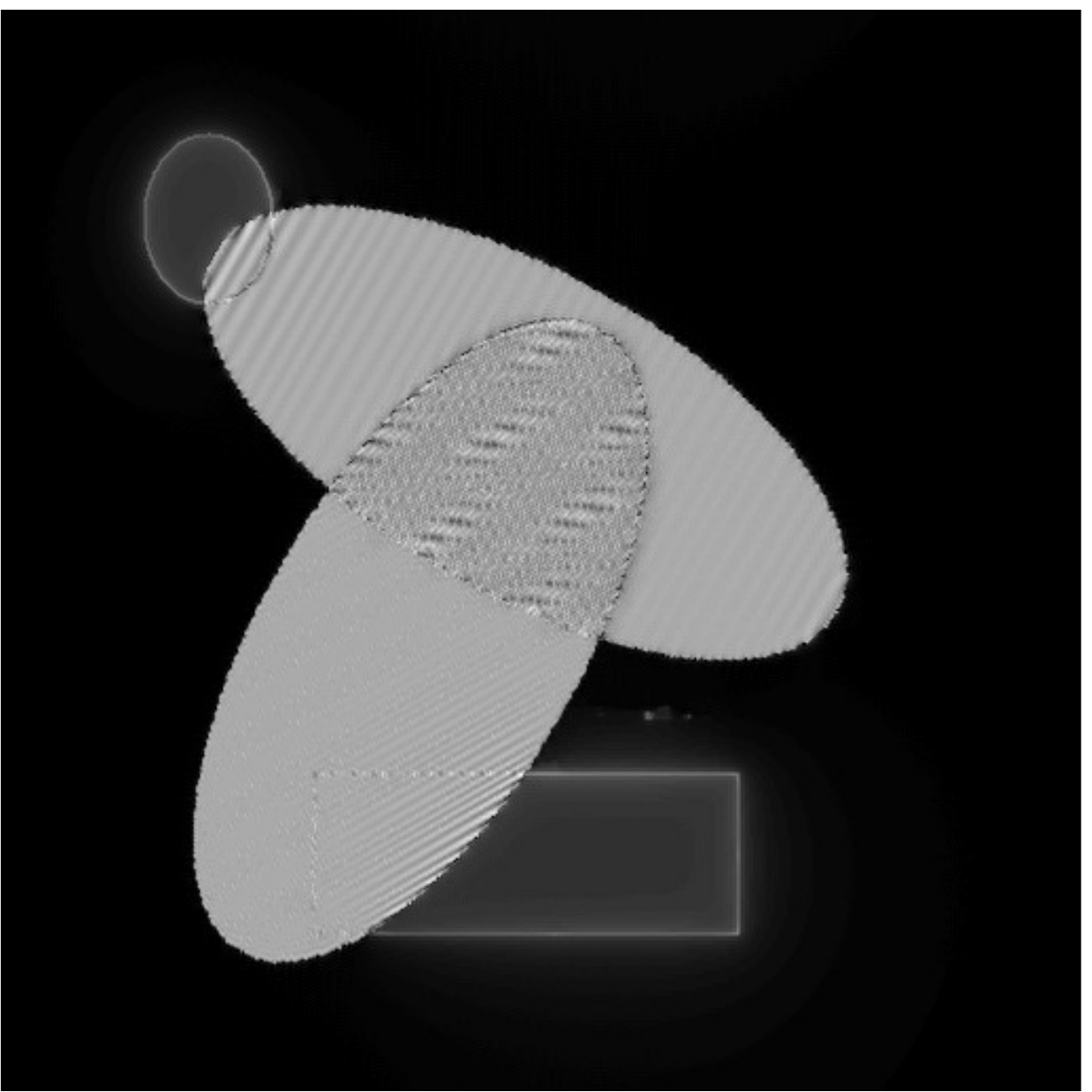} &
\includegraphics[width=0.16in]{figures/colorbar_amplitude_simu-eps-converted-to.pdf} &
\includegraphics[width=1.1in]{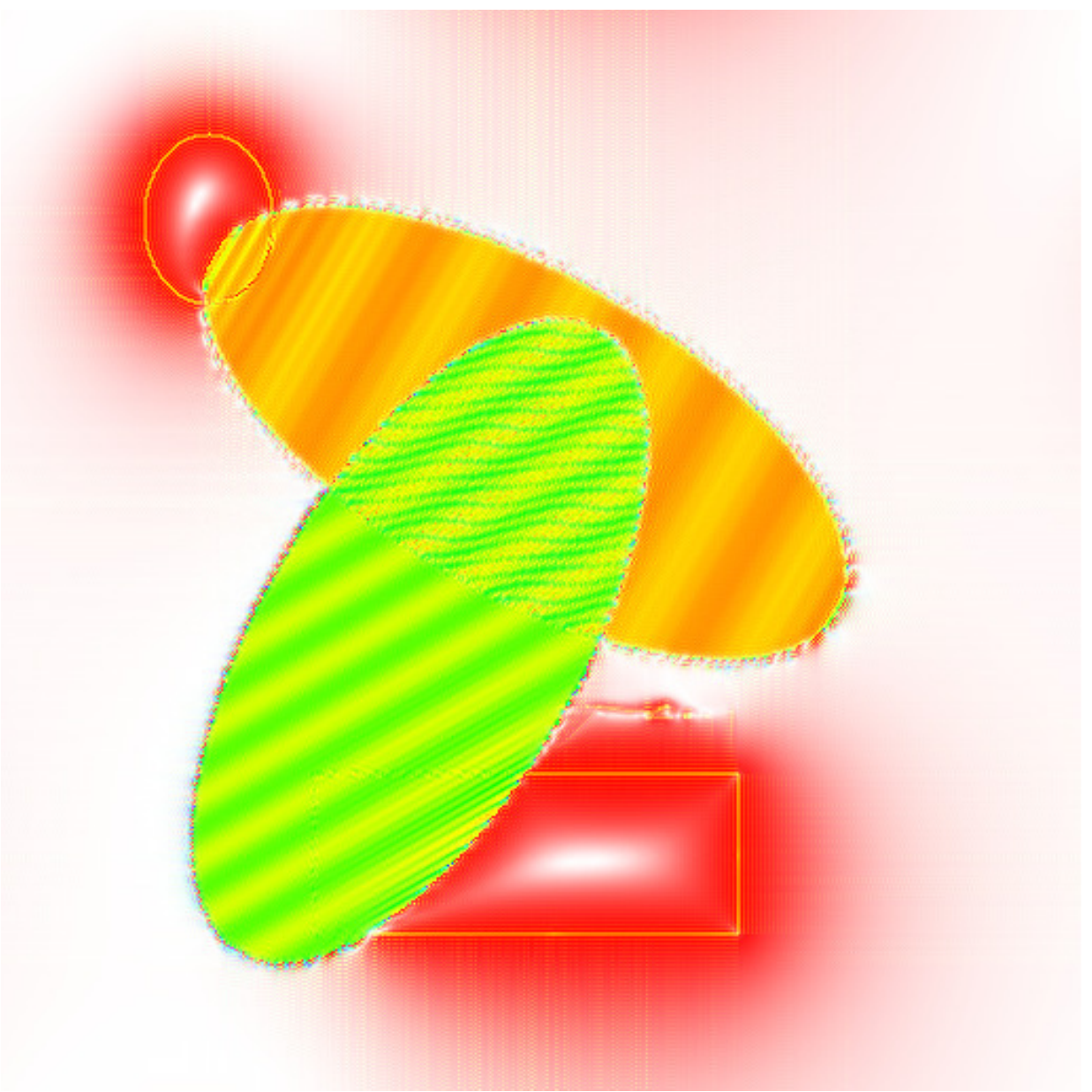} &
\includegraphics[width=0.17in]{figures/colorbar_freq-eps-converted-to.pdf} &
\includegraphics[width=1.1in]{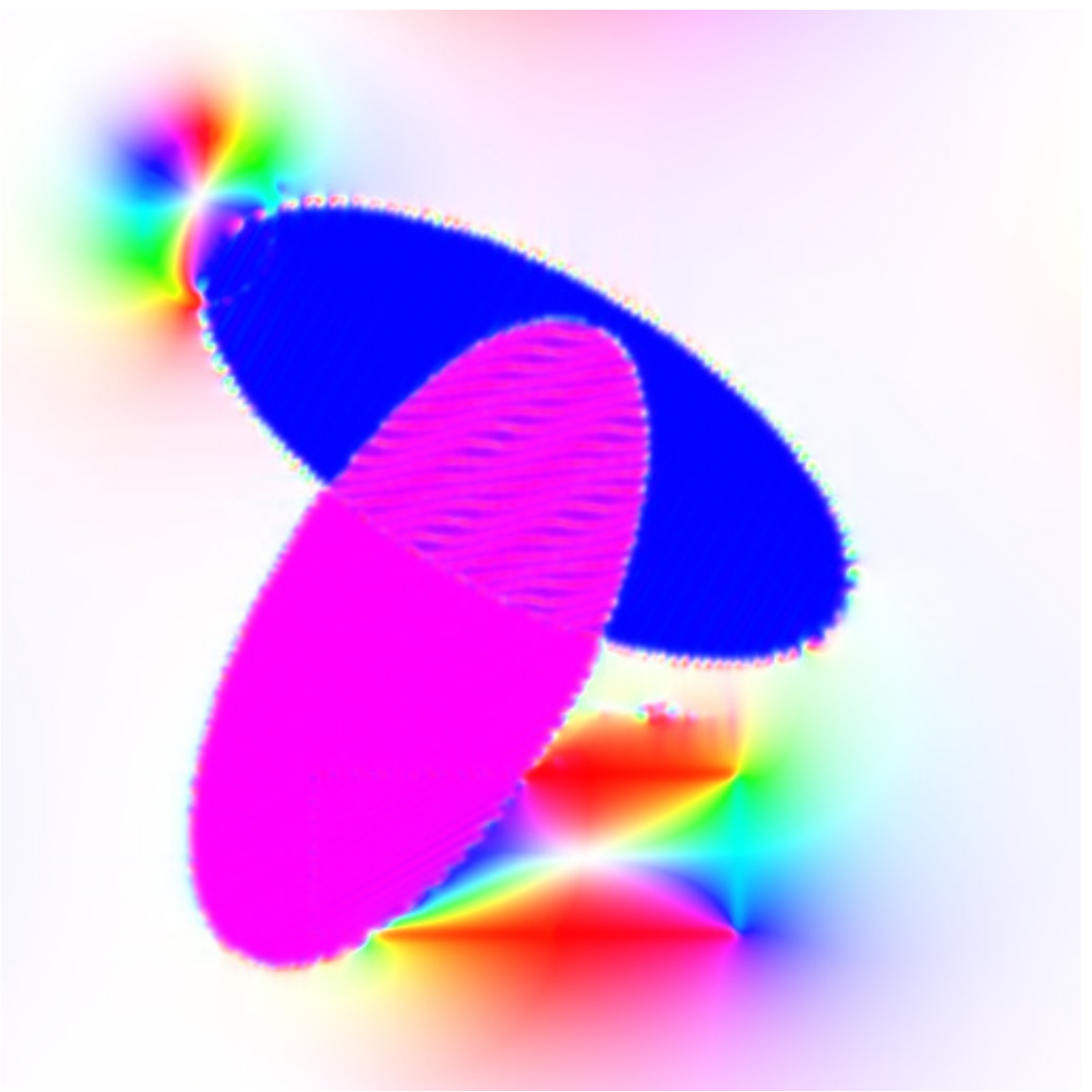} &
\includegraphics[width=0.2in]{figures/colorbar_orientation-eps-converted-to.pdf}
 \\

\rotatebox{90}{Proposed P2D--PHT }&
\includegraphics[width=1.1in]{figures/exp3_vemd_imf1_0p3_0p3-eps-converted-to.pdf} &
\includegraphics[width=1.1in]{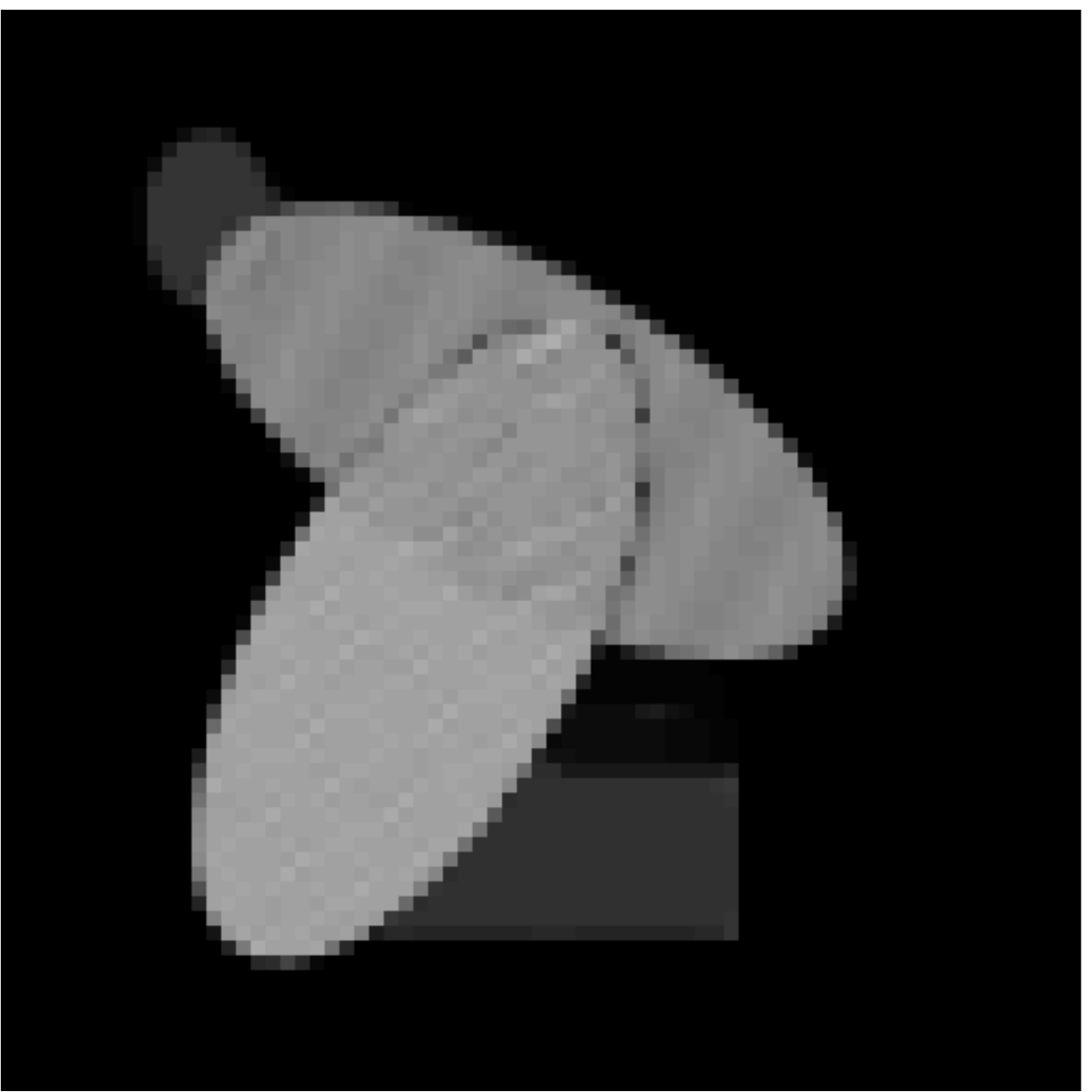} &
\includegraphics[width=0.16in]{figures/colorbar_amplitude_simu-eps-converted-to.pdf} &
\includegraphics[width=1.1in]{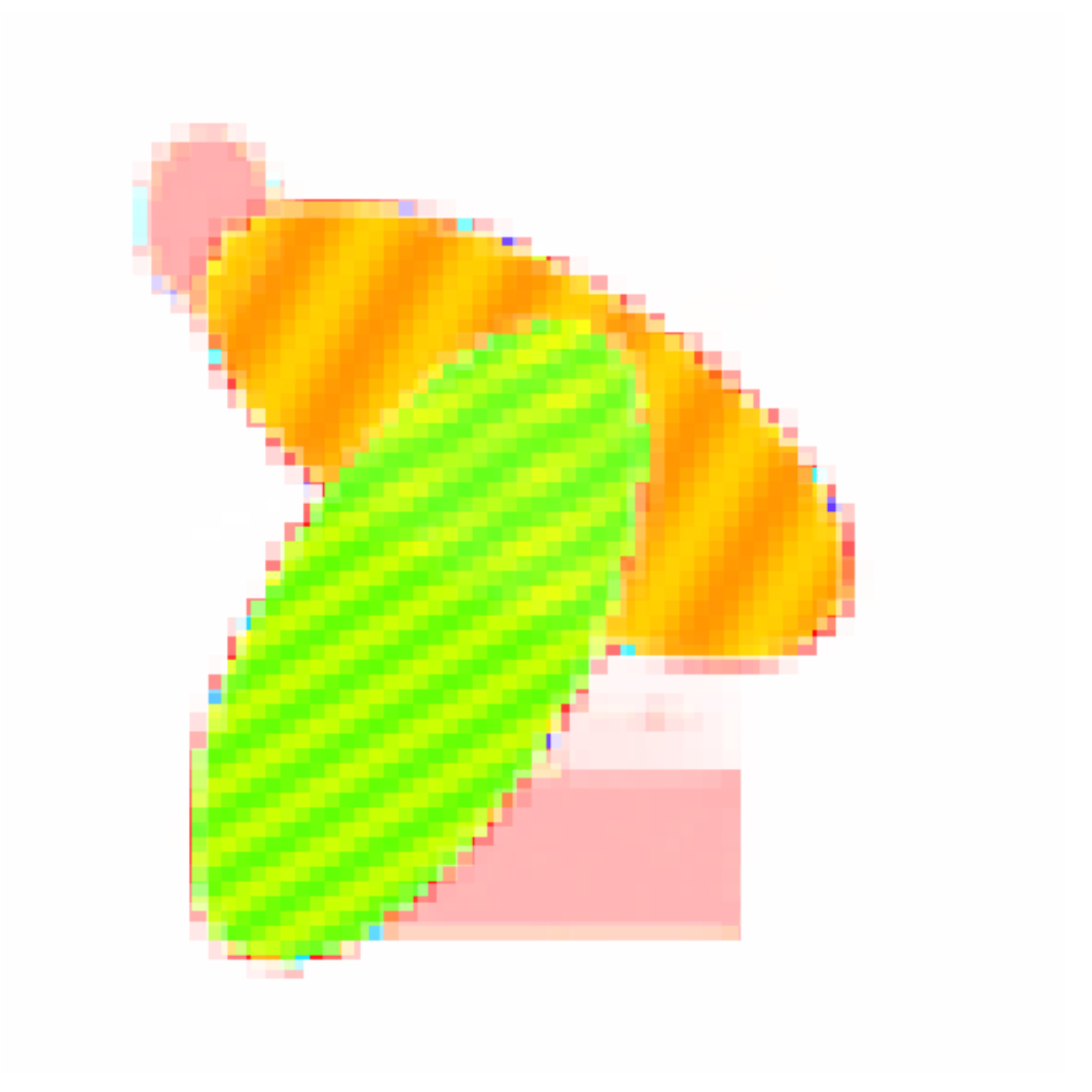} &
\includegraphics[width=0.17in]{figures/colorbar_freq-eps-converted-to.pdf} &
\includegraphics[width=1.1in]{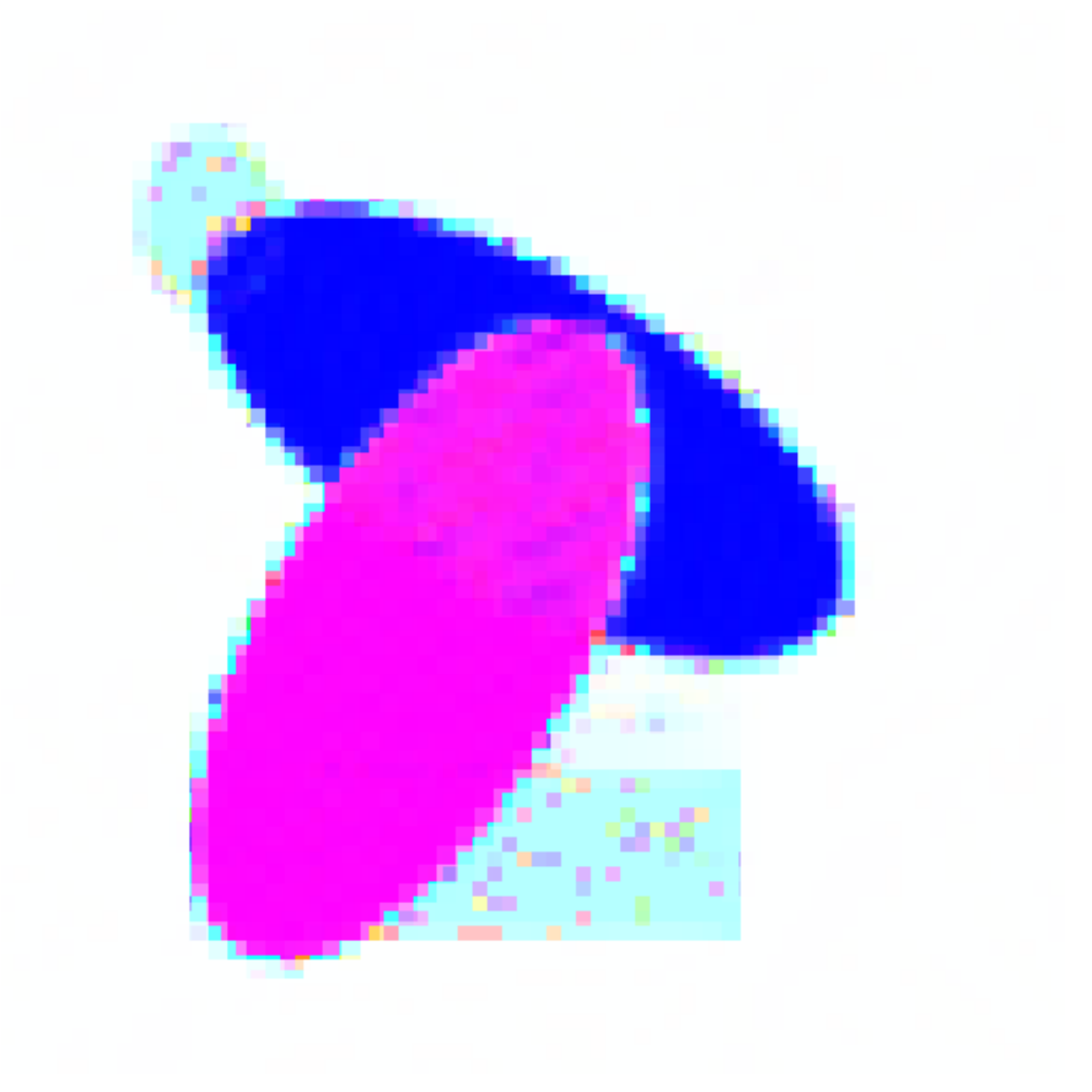} &
\includegraphics[width=0.2in]{figures/colorbar_orientation-eps-converted-to.pdf}
\\

\end{tabular}
\normalsize
\caption{Spectral analysis on the 1st mode using different methods. 1st row: Riesz-Laplace~\cite{Unser_M_2009_j-ieee-tip_multiresolution_msa}. 2nd row:  G2D--HHT (G2D--EMD + monogenic analysis). 3rd row: G2D--PHT (G2D--EMD + annihilation based spectral analysis). 4th row:  P2D--HHT (P2D--EMD + monogenic analysis)~\cite{Schmitt_J_2013_p-icassp_2D_HHT}. 5th row: P2D--PHT (P2D--EMD + annihilation based spectral analysis). From left to right: mode $\mathbf{d}^{(1)}$, amplitude $\boldsymbol{\alpha}^{(1)}$, frequency $\boldsymbol{\eta}^{(1)}$ and orientation $\boldsymbol{\theta}^{(1)}$.
}
\label{fig:simspectralimf1}

\end{center}
\end{figure*}

\begin{figure*}
\begin{center}
\footnotesize
\begin{tabular}{p{0.2cm}ccccccc}


&
2nd mode&
 Amplitude &
 &
 Frequency &
 &
 Orientation \\

\rotatebox{90}{Riesz-Laplace~\cite{Unser_M_2009_j-ieee-tip_multiresolution_msa}} &
 \includegraphics[width=1.1in]{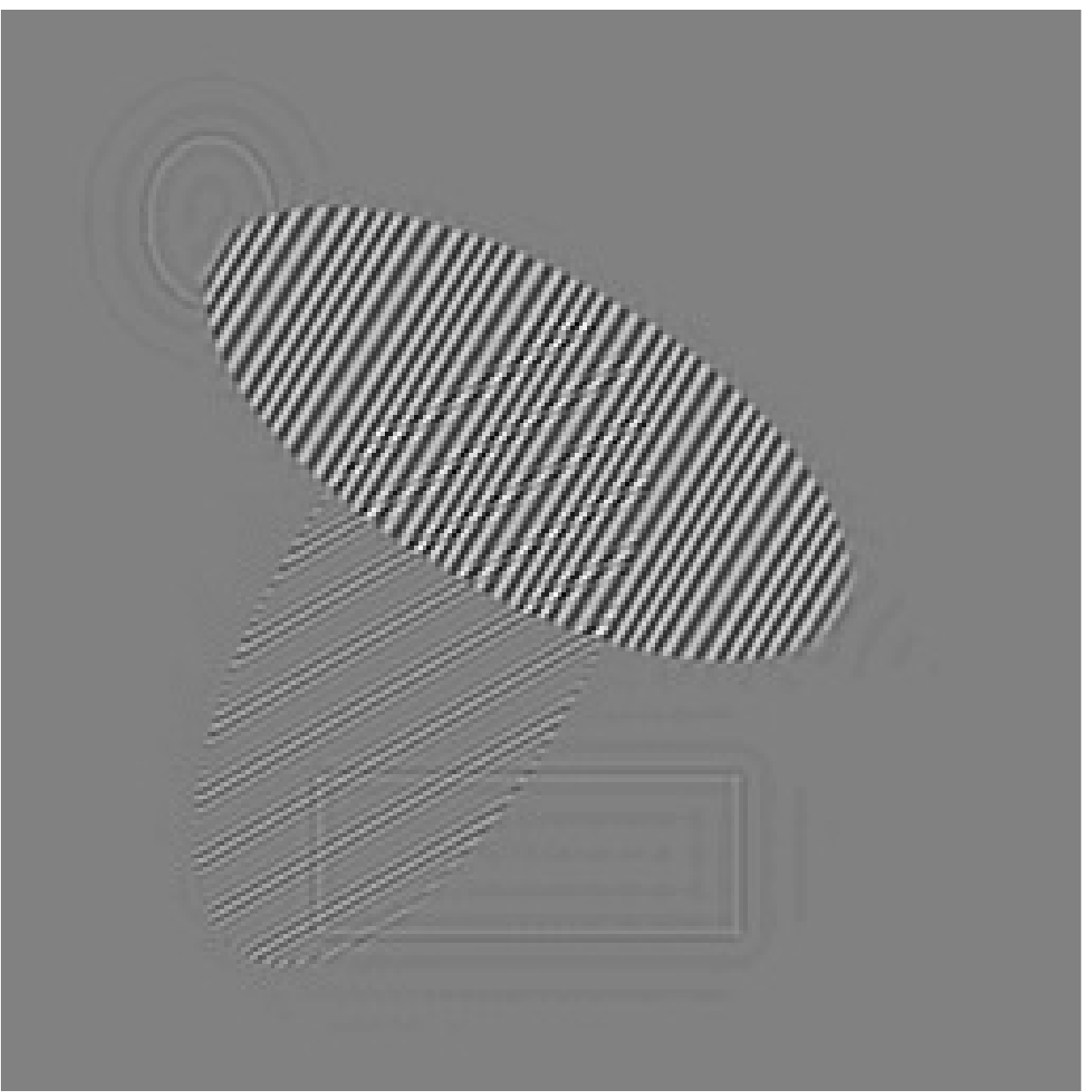} &
\includegraphics[width=1.1in]{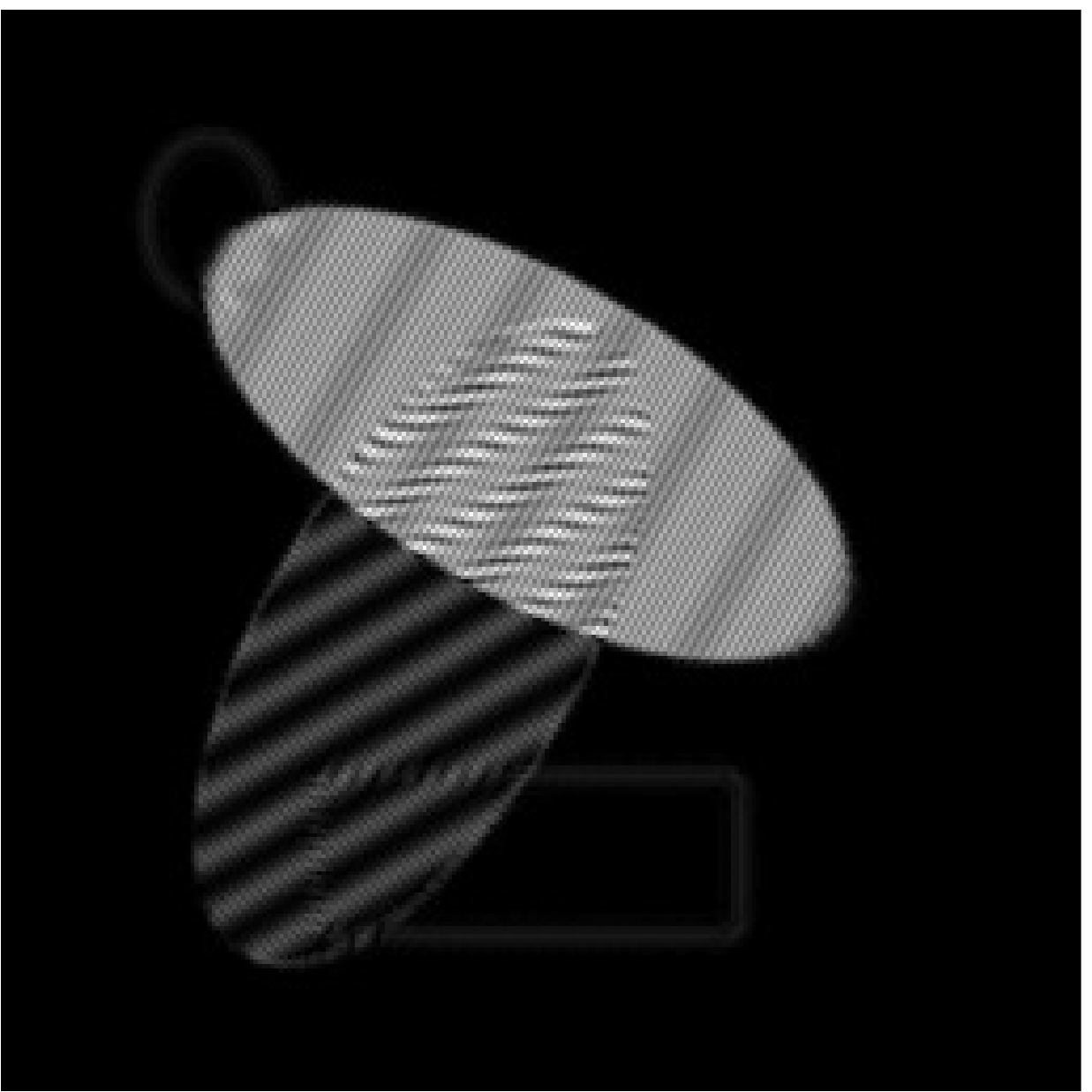} &
\includegraphics[width=0.16in]{figures/colorbar_amplitude_simu-eps-converted-to.pdf} &
\includegraphics[width=1.1in]{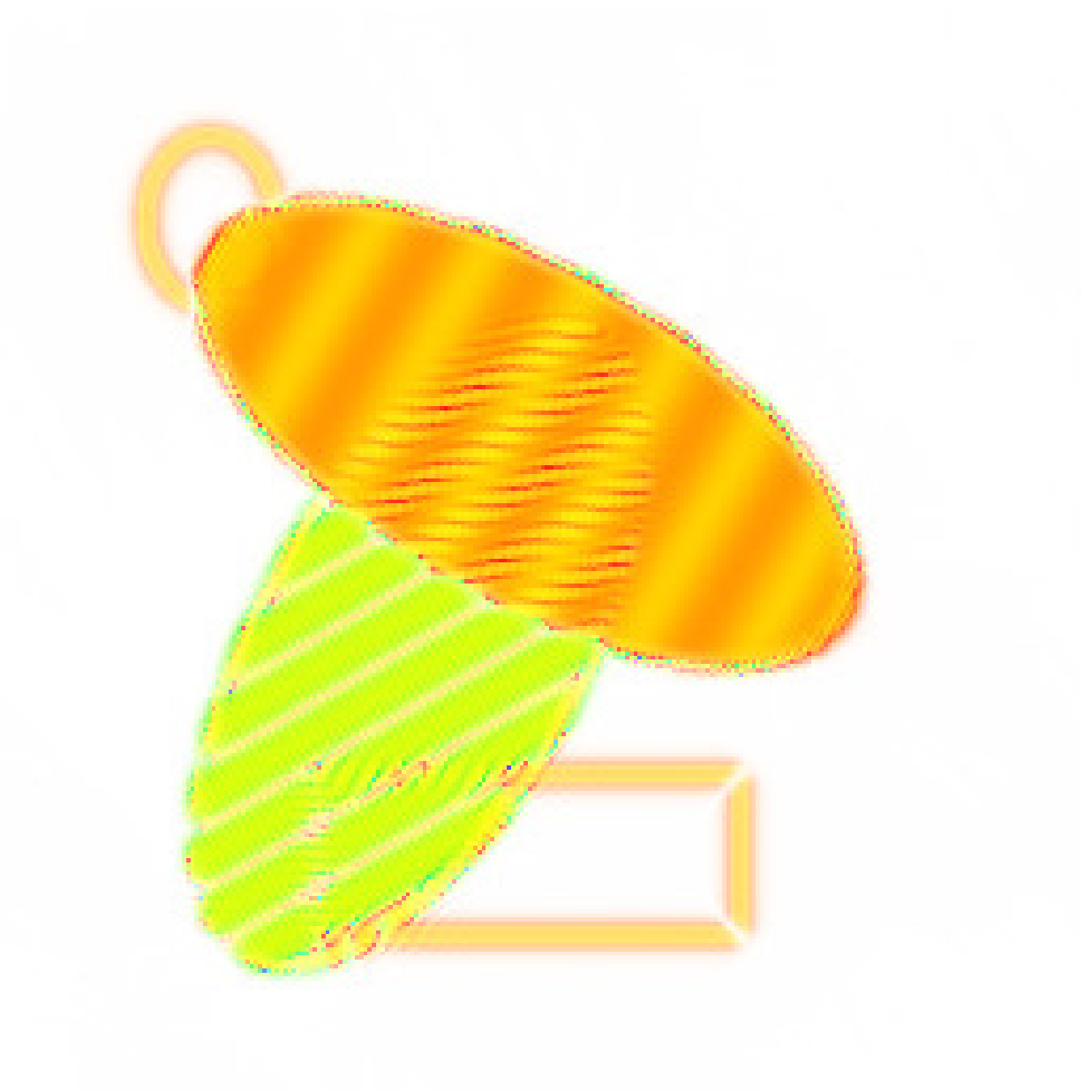} &
\includegraphics[width=0.17in]{figures/colorbar_freq-eps-converted-to.pdf} &
\includegraphics[width=1.1in]{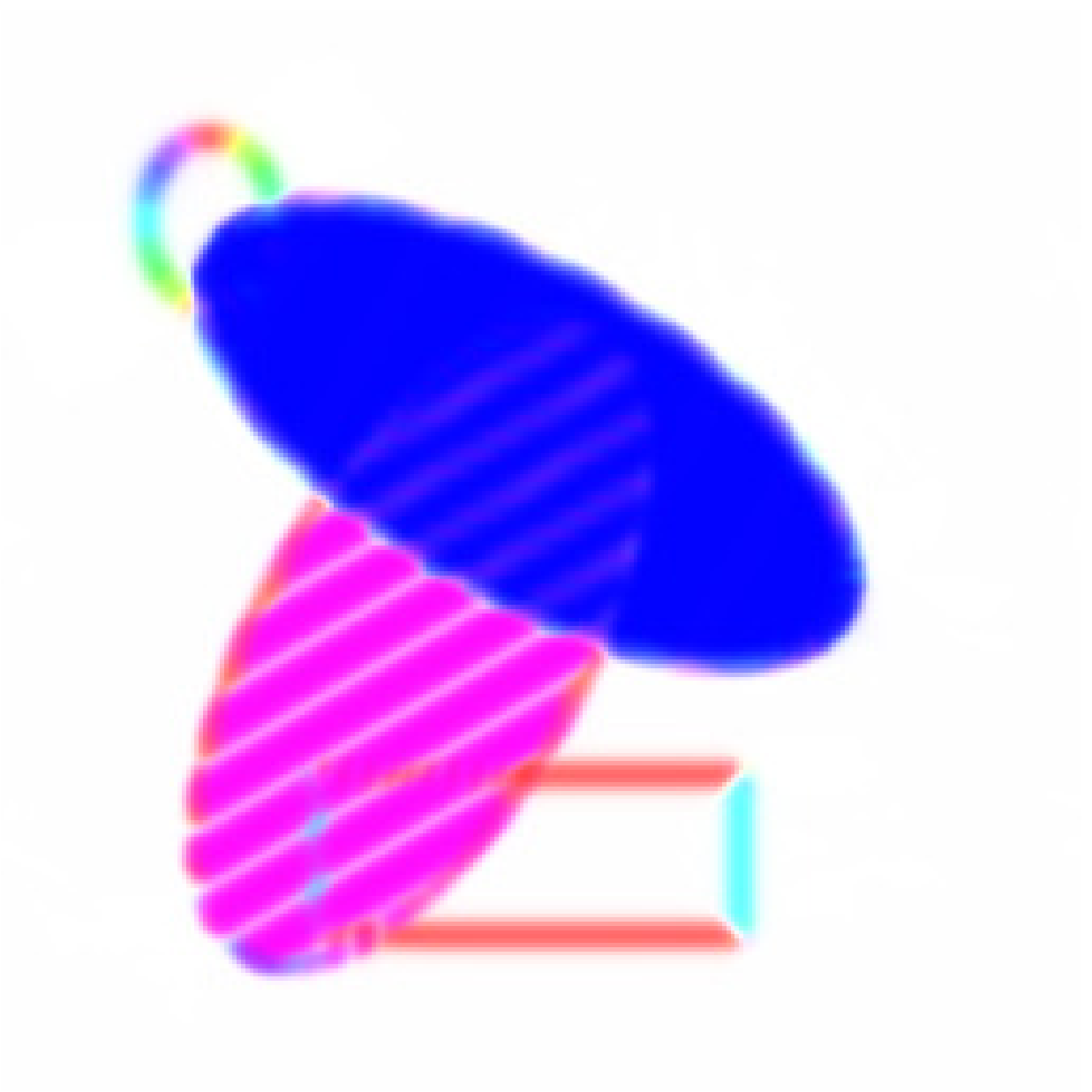} &
\includegraphics[width=0.2in]{figures/colorbar_orientation-eps-converted-to.pdf}
 \\

\rotatebox{90}{\hspace{0.5cm} G2D--HHT }&
\includegraphics[width=1.1in]{figures/imf2_g2d_0p02_1000_0p05_1-eps-converted-to.pdf} &
\includegraphics[width=1.1in]{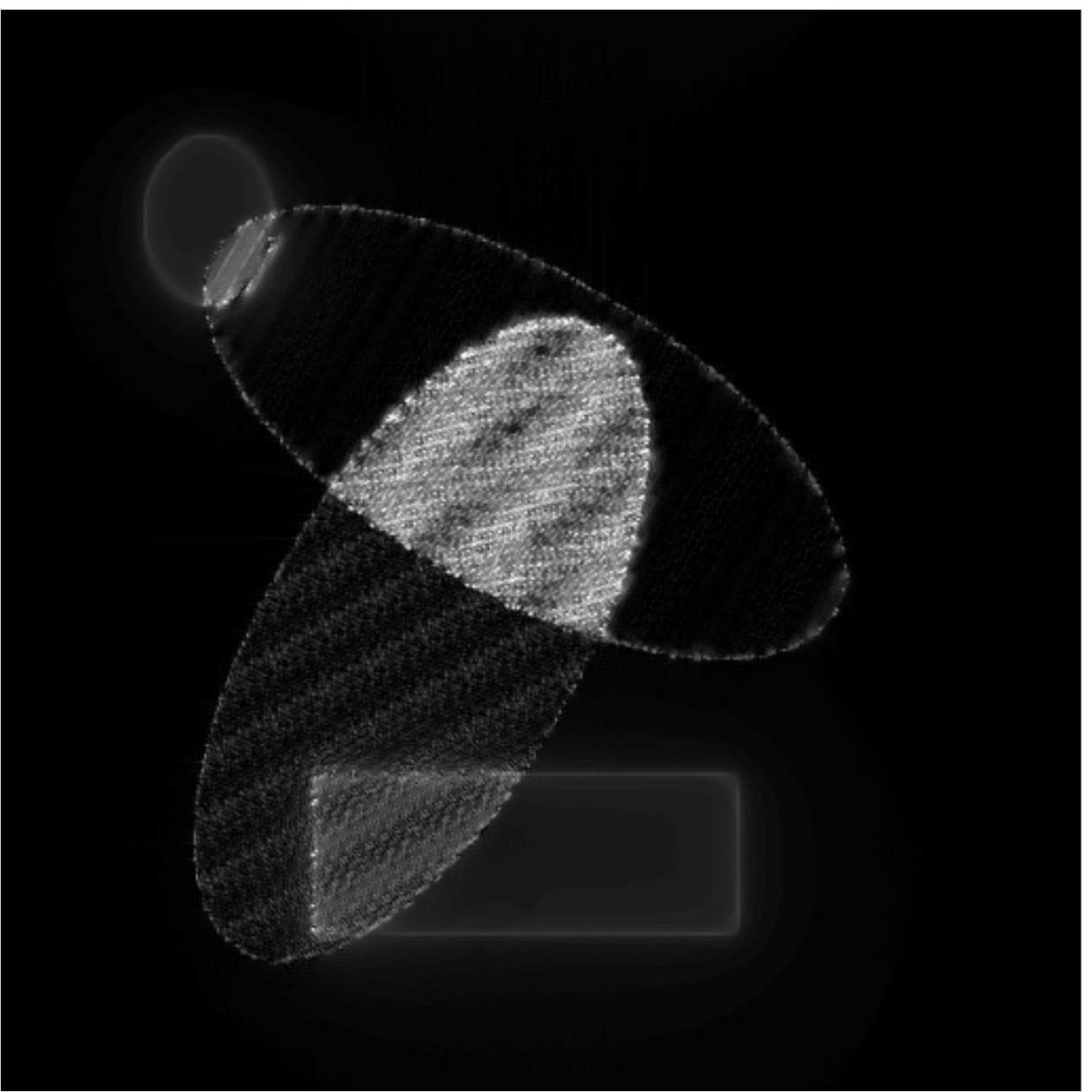} &
\includegraphics[width=0.16in]{figures/colorbar_amplitude_simu-eps-converted-to.pdf} &
\includegraphics[width=1.1in]{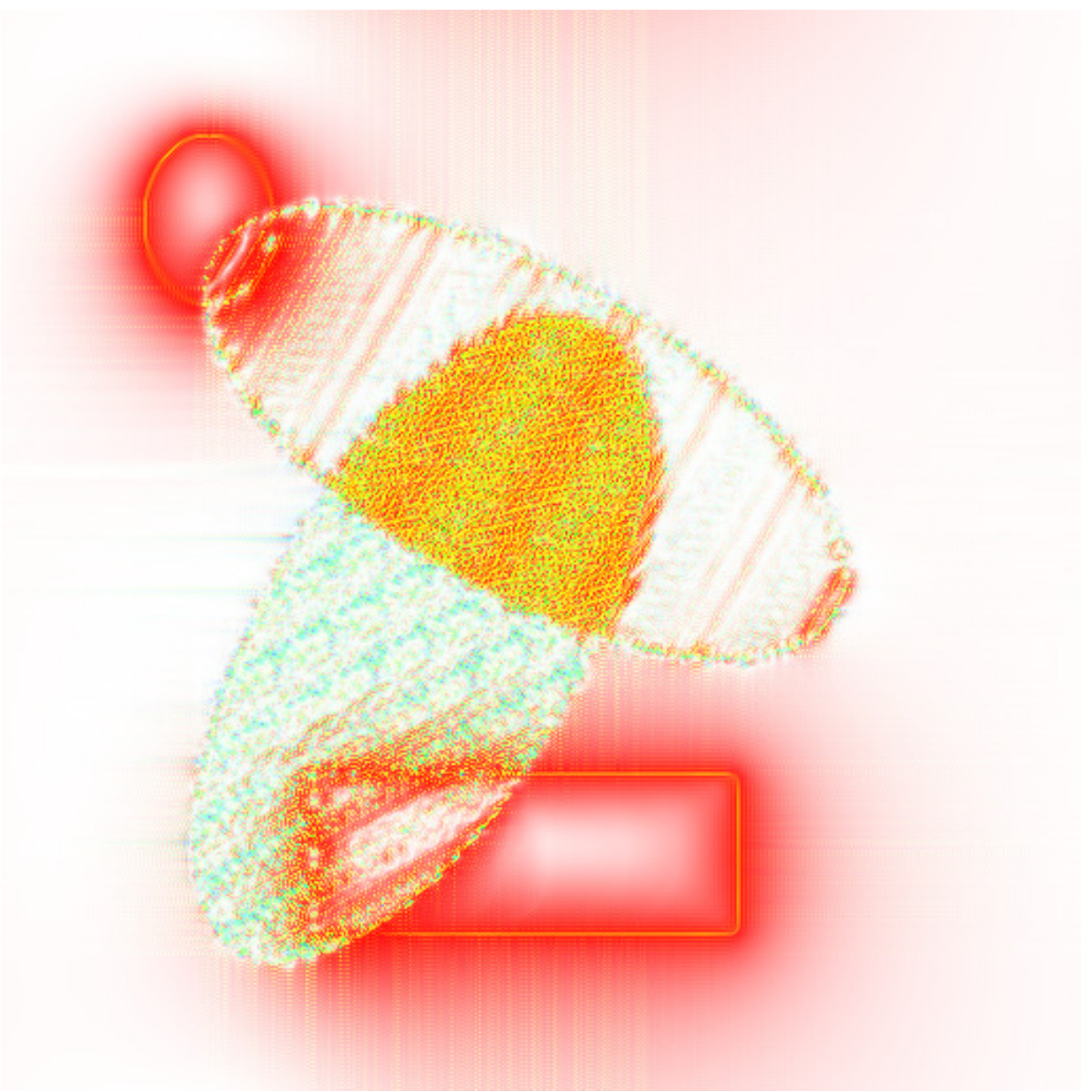} &
\includegraphics[width=0.17in]{figures/colorbar_freq-eps-converted-to.pdf} &
\includegraphics[width=1.1in]{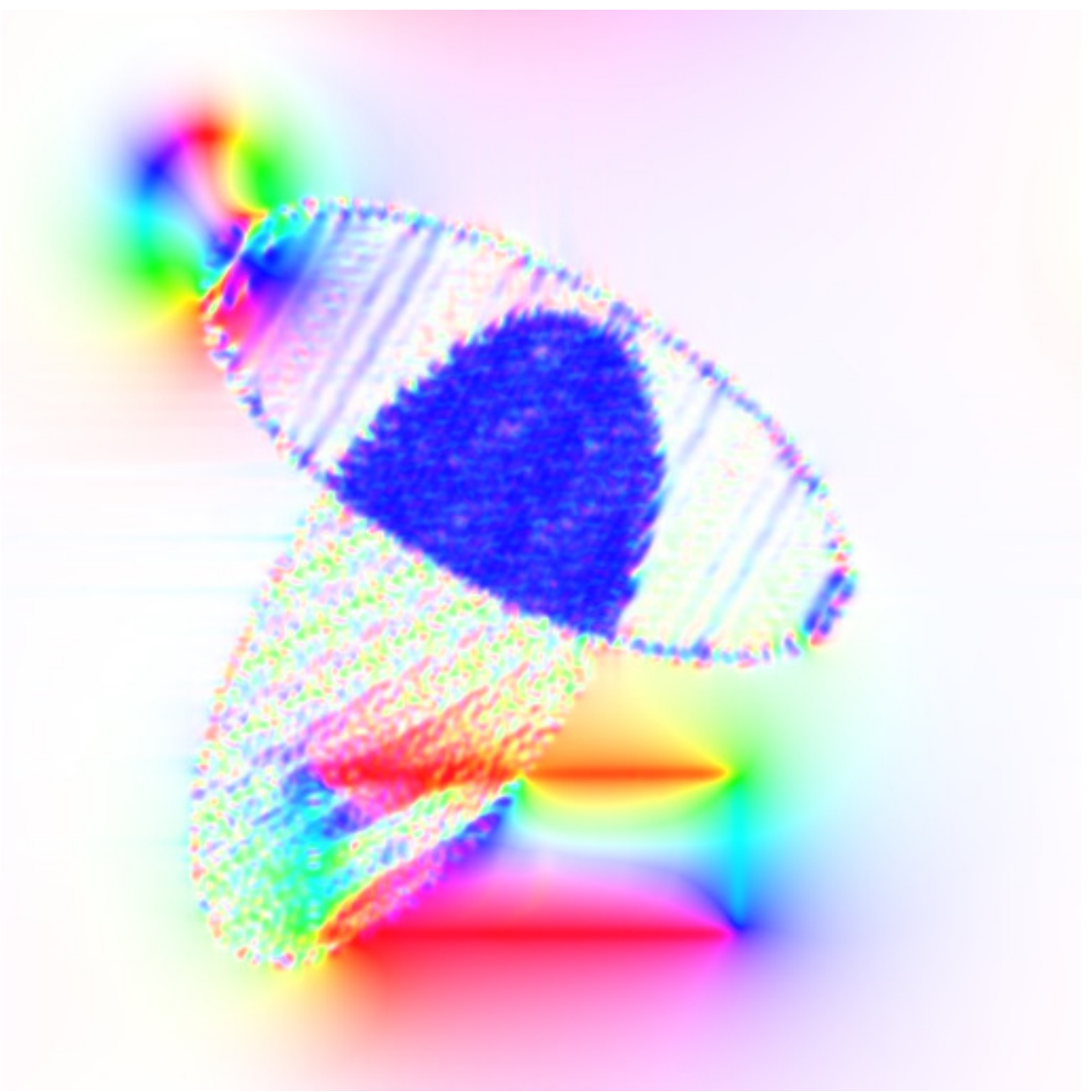} &
\includegraphics[width=0.2in]{figures/colorbar_orientation-eps-converted-to.pdf}
 \\

\rotatebox{90}{Proposed G2D--PHT} &
\includegraphics[width=1.1in]{figures/imf2_g2d_0p02_1000_0p05_1-eps-converted-to.pdf} &
\includegraphics[width=1.1in]{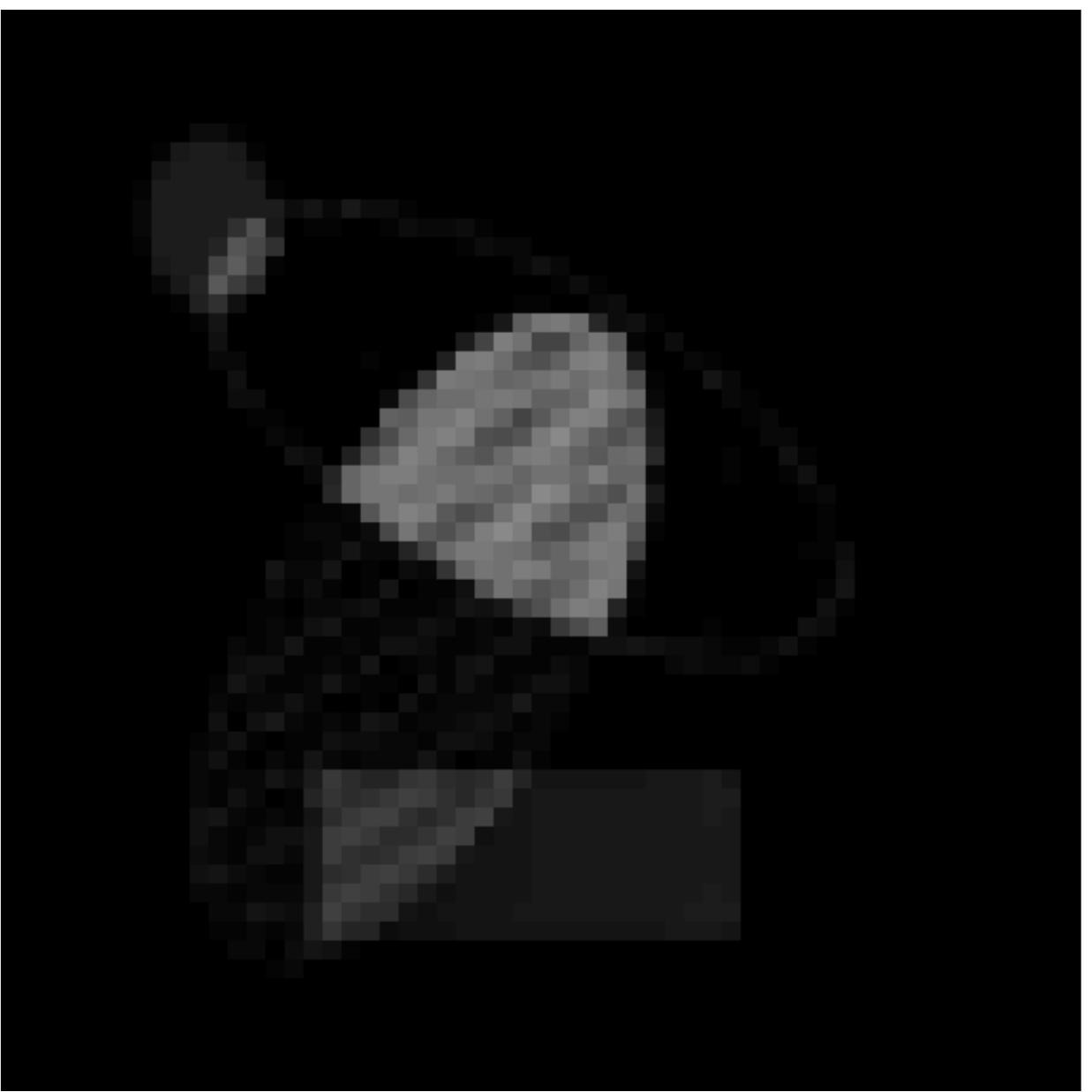} &
\includegraphics[width=0.16in]{figures/colorbar_amplitude_simu-eps-converted-to.pdf} &
\includegraphics[width=1.1in]{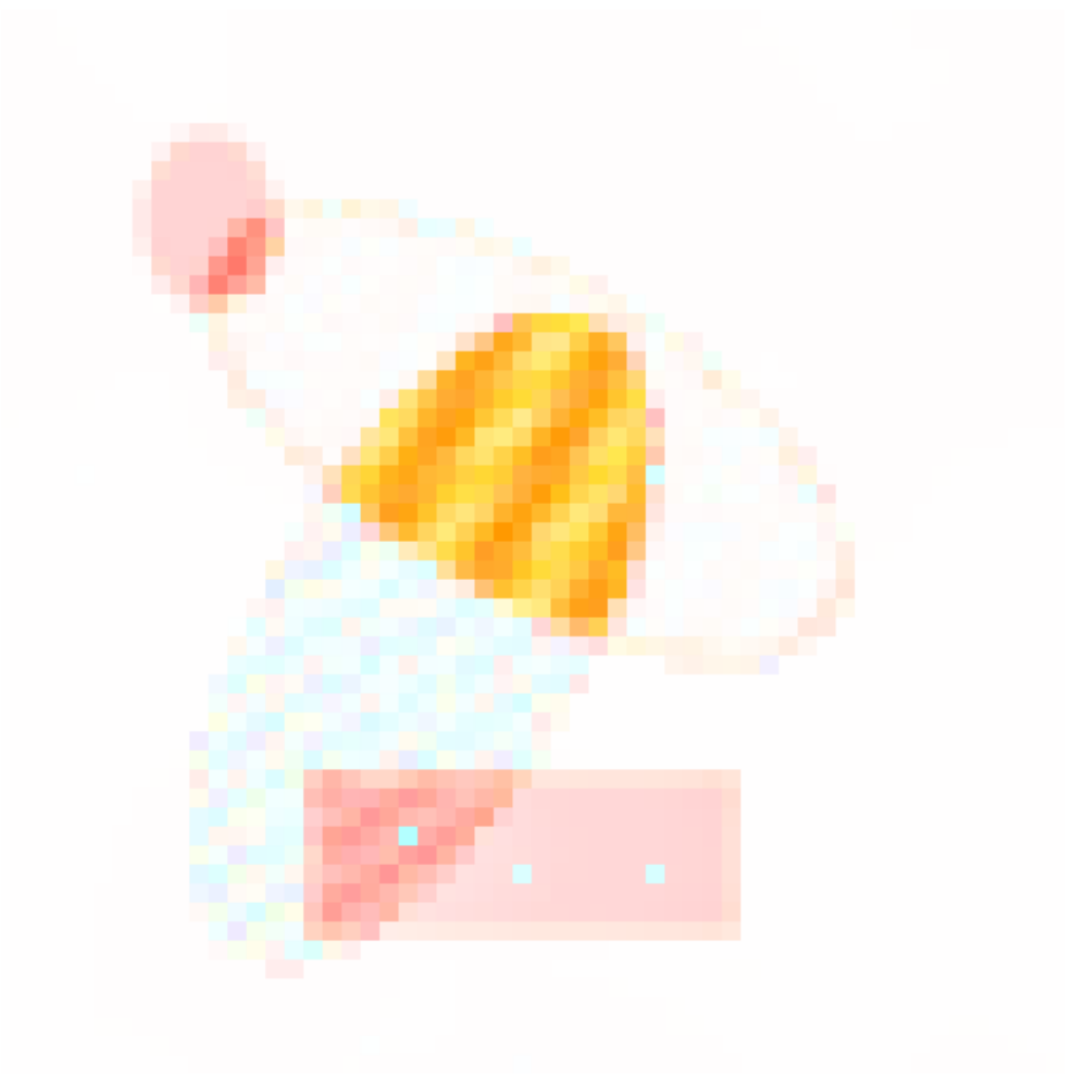} &
\includegraphics[width=0.17in]{figures/colorbar_freq-eps-converted-to.pdf} &
\includegraphics[width=1.1in]{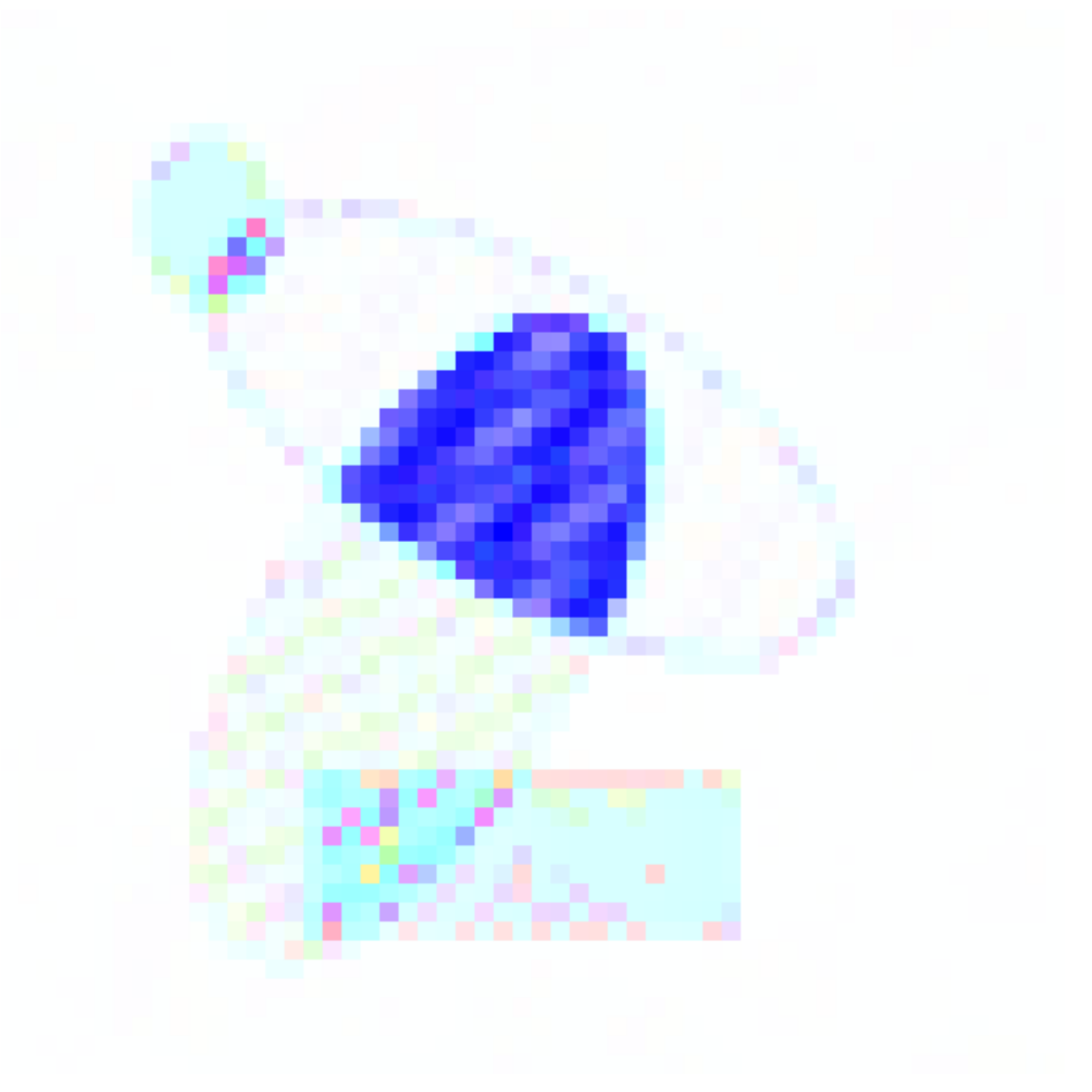} &
\includegraphics[width=0.2in]{figures/colorbar_orientation-eps-converted-to.pdf}
\\

\rotatebox{90}{\hspace{0.2cm} P2D--HHT~\cite{Schmitt_J_2013_p-icassp_2D_HHT}} &
\includegraphics[width=1.1in]{figures/exp3_vemd_imf2_0p3_0p3_1_0p1-eps-converted-to.pdf} &
\includegraphics[width=1.1in]{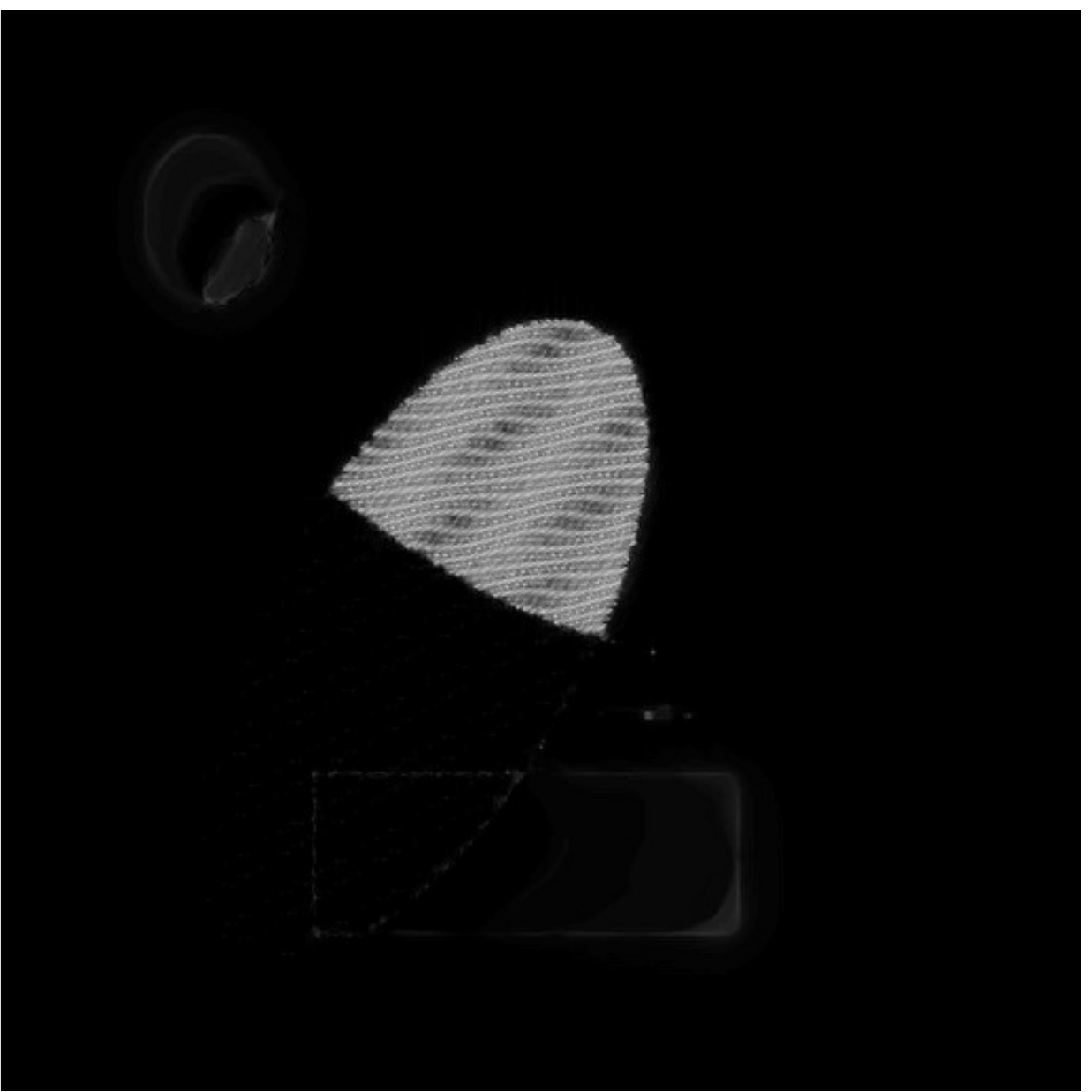} &
\includegraphics[width=0.16in]{figures/colorbar_amplitude_simu-eps-converted-to.pdf} &
\includegraphics[width=1.1in]{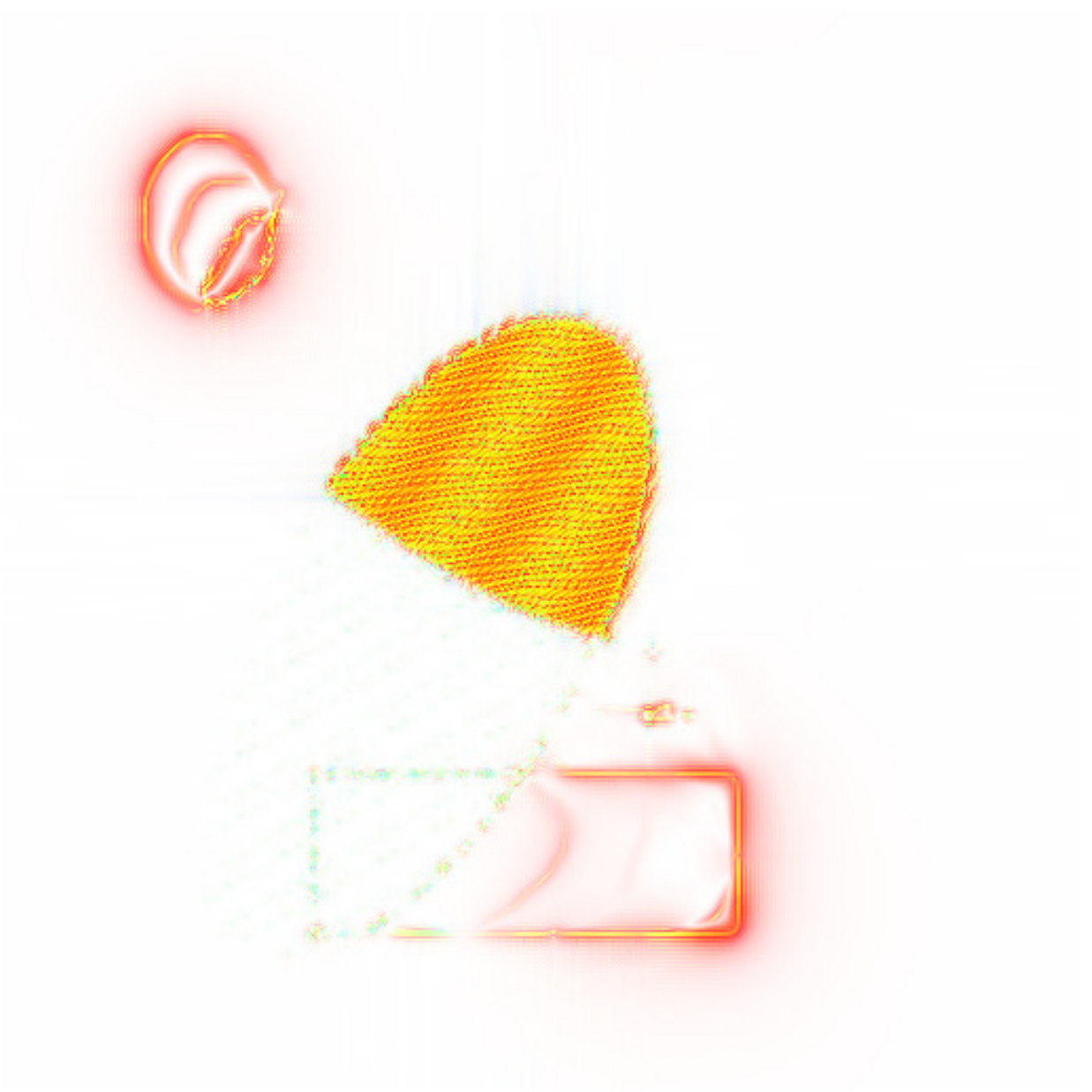} &
\includegraphics[width=0.17in]{figures/colorbar_freq-eps-converted-to.pdf} &
\includegraphics[width=1.1in]{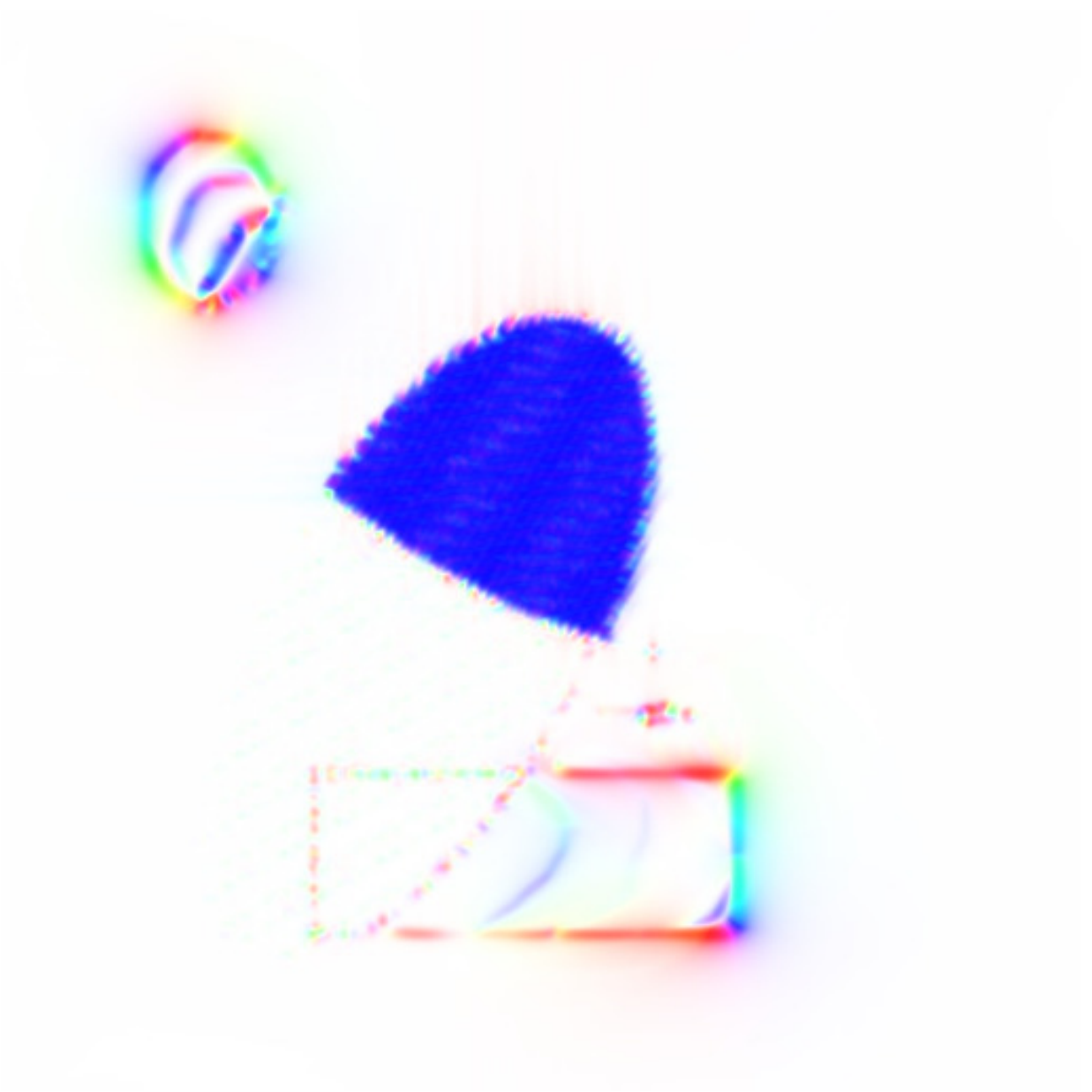} &
\includegraphics[width=0.2in]{figures/colorbar_orientation-eps-converted-to.pdf}
 \\

\rotatebox{90}{Proposed P2D--PHT} &
\includegraphics[width=1.1in]{figures/exp3_vemd_imf2_0p3_0p3_1_0p1-eps-converted-to.pdf} &
\includegraphics[width=1.1in]{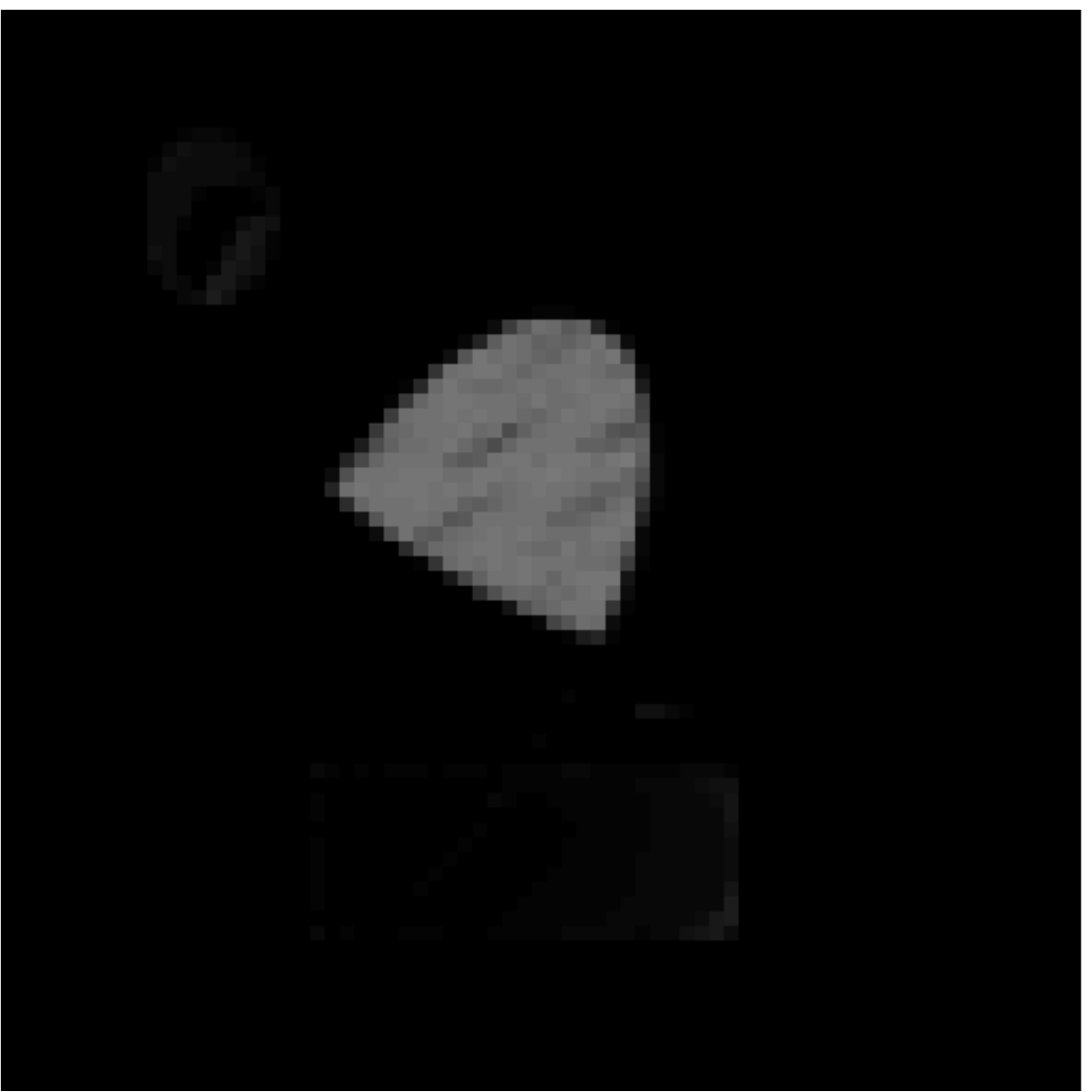} &
\includegraphics[width=0.16in]{figures/colorbar_amplitude_simu-eps-converted-to.pdf} &
\includegraphics[width=1.1in]{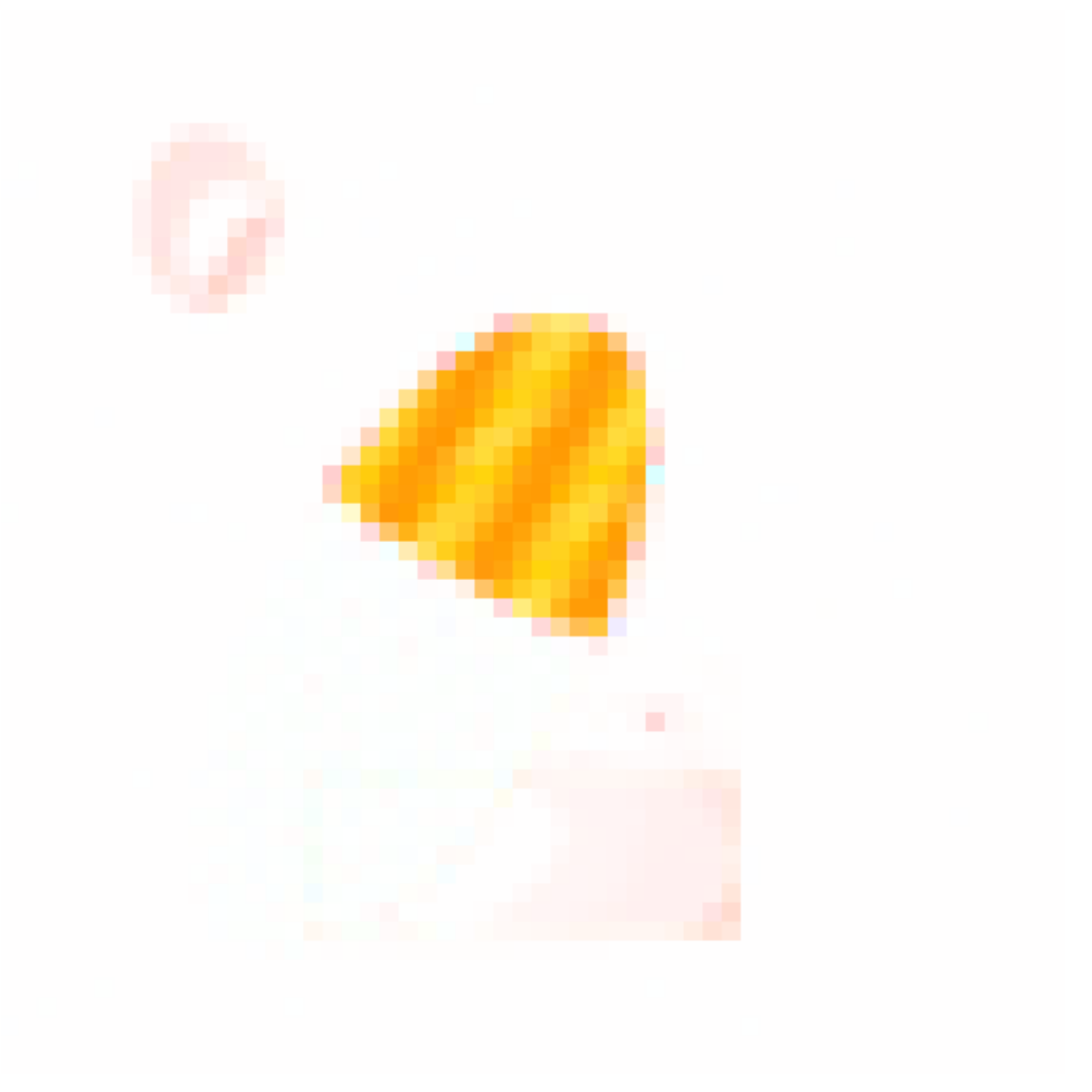} &
\includegraphics[width=0.17in]{figures/colorbar_freq-eps-converted-to.pdf} &
\includegraphics[width=1.1in]{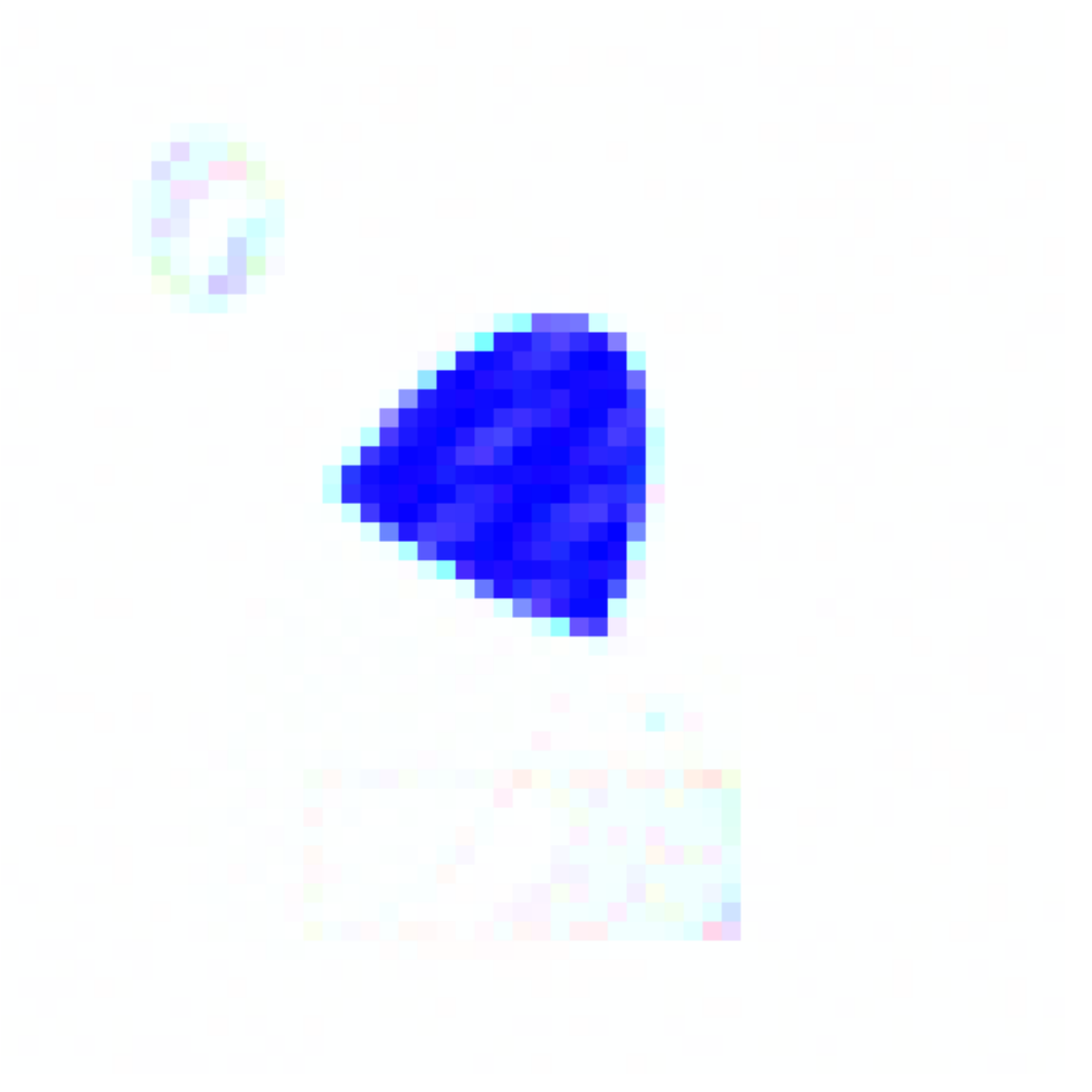} &
\includegraphics[width=0.2in]{figures/colorbar_orientation-eps-converted-to.pdf}
 \\

\end{tabular}
\normalsize
\caption{Spectral analysis on 2nd mode using different methods. 1st row: Riesz-Laplace~\cite{Unser_M_2009_j-ieee-tip_multiresolution_msa}. 2nd row:  G2D--HHT (G2D--EMD + monogenic analysis). 3rd row: G2D--PHT (G2D--EMD + annihilation based spectral analysis). 4th row:  P2D--HHT (P2D--EMD + monogenic analysis)~\cite{Schmitt_J_2013_p-icassp_2D_HHT}. 5th row: P2D--PHT (P2D--EMD + annihilation based spectral analysis). From left to right: mode $\mathbf{d}^{(2)}$, amplitude $\boldsymbol{\alpha}^{(2)}$, frequency $\boldsymbol{\eta}^{(2)}$ and orientation $\boldsymbol{\theta}^{(2)}$.
}
\label{fig:simspectralimf2}

\end{center}
\end{figure*}

We perform two types of spectral analysis on the IMFs obtained by G2D--EMD and P2D--EMD. The first approach, that we proposed in \cite{Schmitt_J_2013_p-icassp_2D_HHT}, is based on a monogenic analysis. We refer to this approach as G2D--HHT and P2D--HHT. On the other hand, the proposed method based on Prony's annihilation property  is referred as G2D--PHT and P2D--PHT. Results on simulations are shown on Figures~\ref{fig:simspectralimf1} and~\ref{fig:simspectralimf2}. For the annihilation based method, we have chosen $\overline{N}^{(k)} = 7$ for both IMFs. A comparison with the Riesz-Laplace transform analysis proposed in \cite{Unser_M_2009_j-ieee-tip_multiresolution_msa} is also performed. For the three methods, frequency and orientation maps are composed by the coherency index, in order to have better visual results.

The EMD based procedure appears to be more adaptive and better suited for nonstationary data than Riesz-Laplace transform. Moreover, one should notice that the PHT based spectral estimation gives smoother results and performs better for higher order IMFs, especially for the frequency estimation. Indeed, the denoising step of the Prony based estimation achieves a better robustness with respect to errors linked to the EMD decomposition step. The drawbacks of the proposed annihilation based method are the loss of resolution and the computational time.



%
%
%
%
%
%

\subsection{Real data}

\begin{figure*}
\begin{center}
\footnotesize
\begin{tabular}{ccccc}

Data: $\mathbf{x}$ &
1st IMF: $\mathbf{d}^{(1)}$ &
2nd IMF: $\mathbf{d}^{(2)}$ &
3rd IMF: $\mathbf{d}^{(3)}$ &
Residual: $\mathbf{a}^{(3)}$ \\

\includegraphics[width=1.1in]{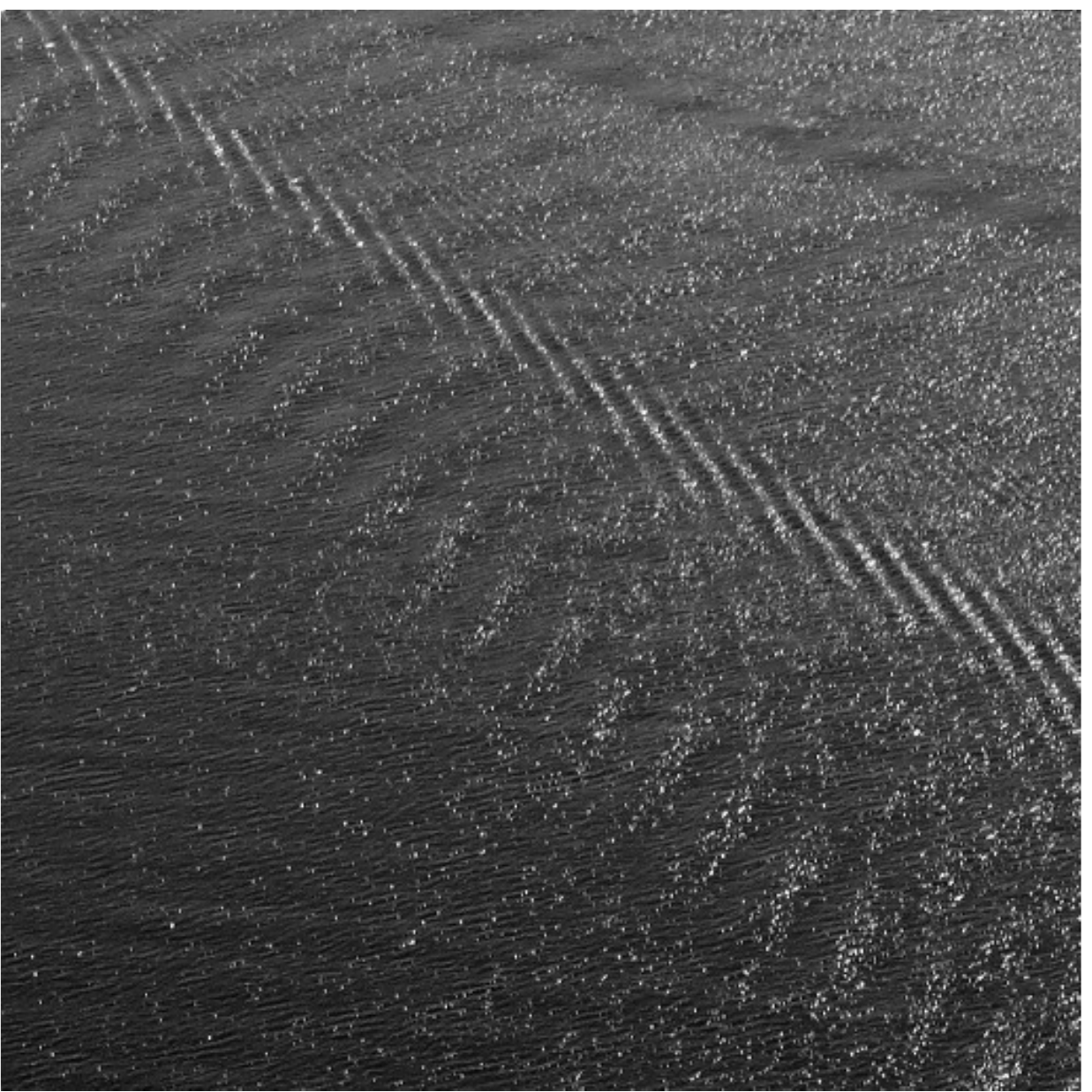} &
\includegraphics[width=1.1in]{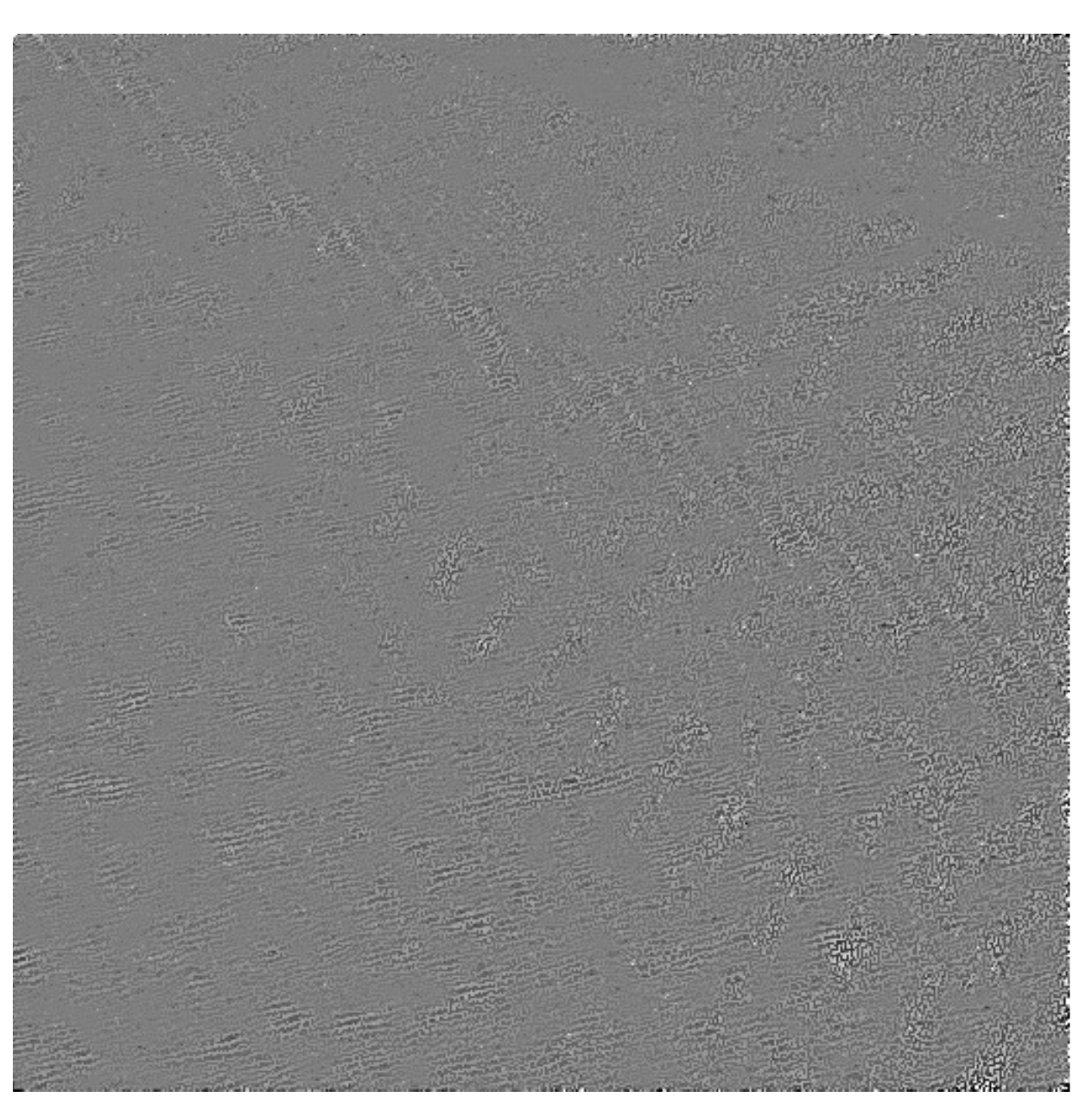} &
 \includegraphics[width=1.1in]{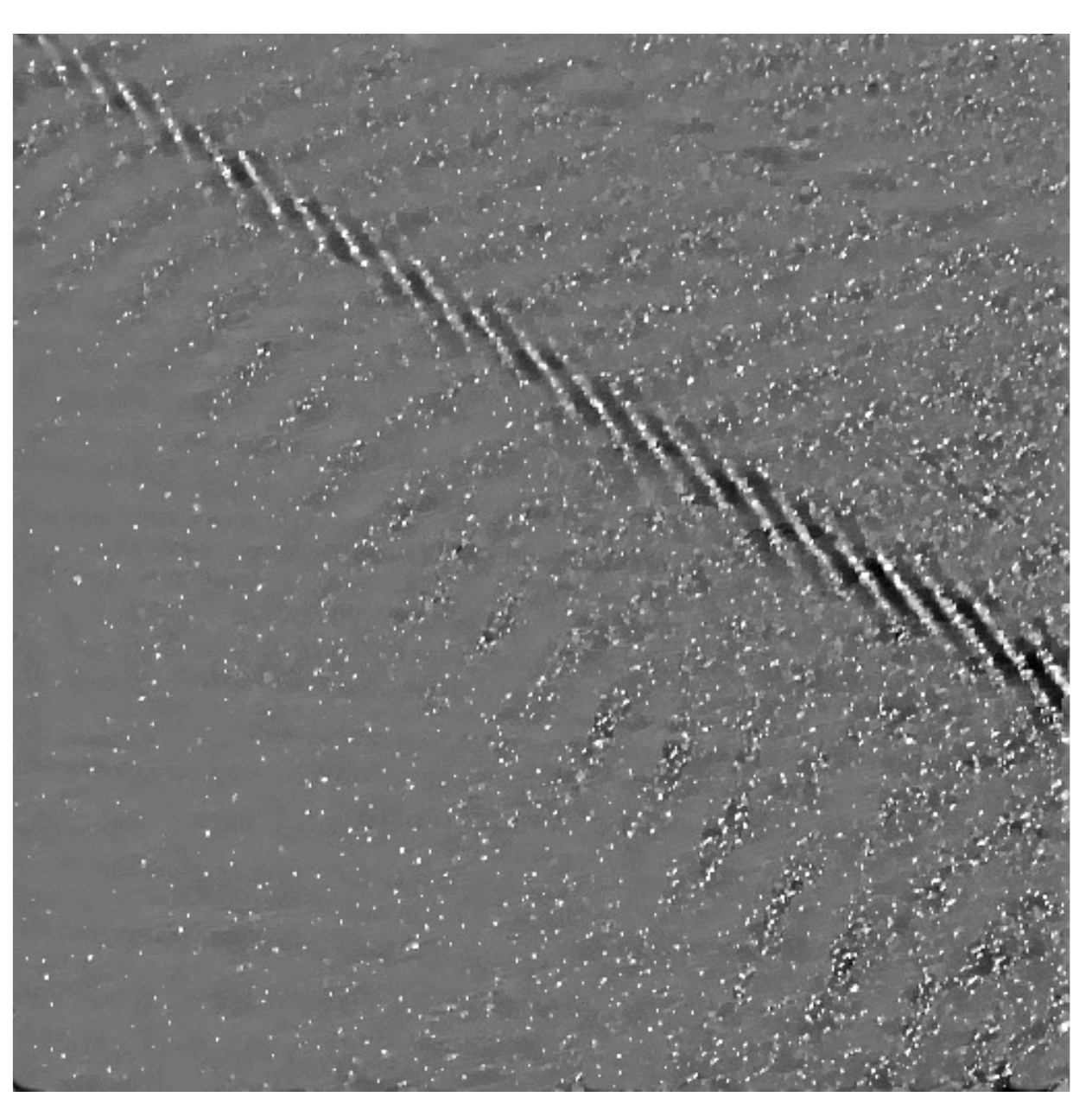} &
  \includegraphics[width=1.1in]{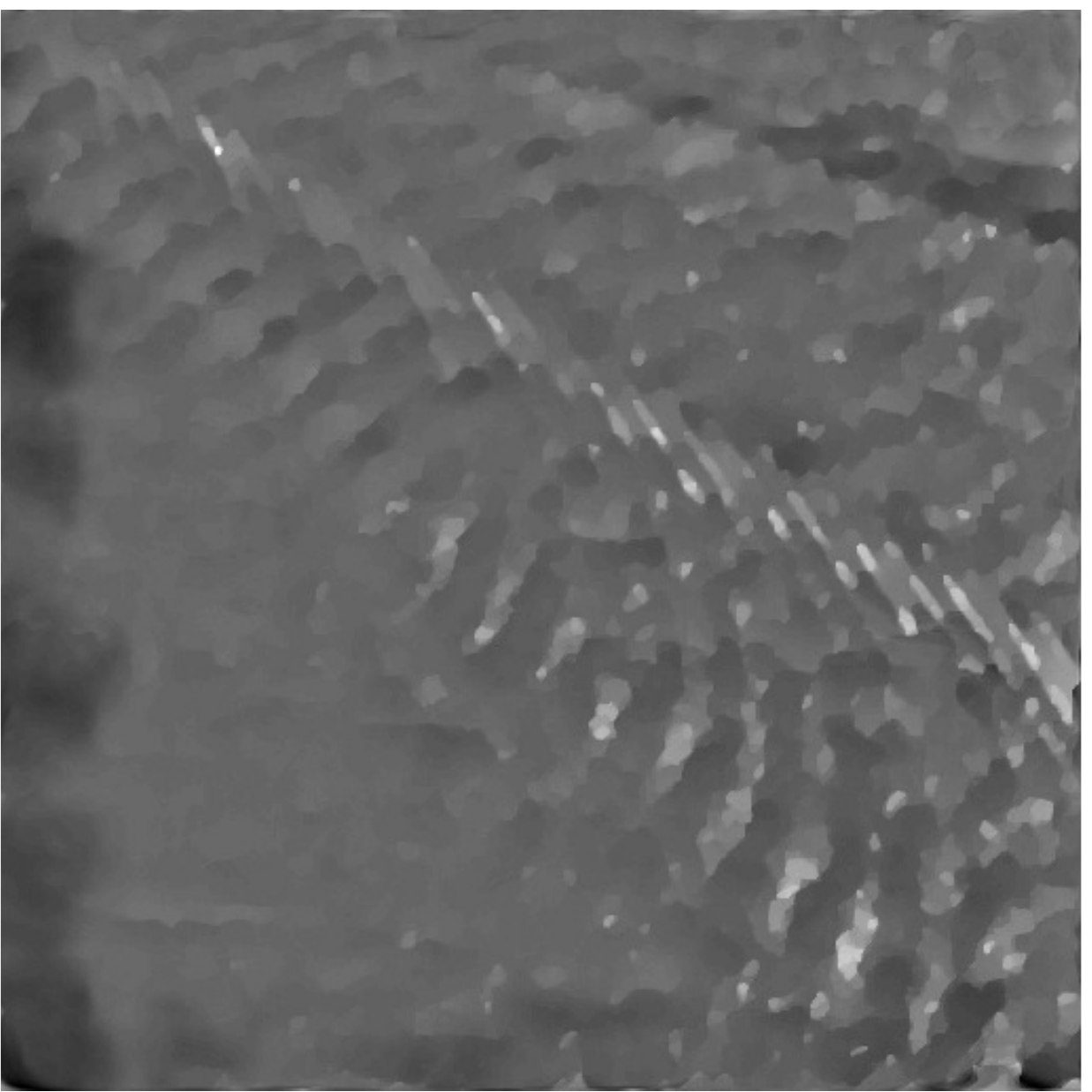} &
\includegraphics[width=1.1in]{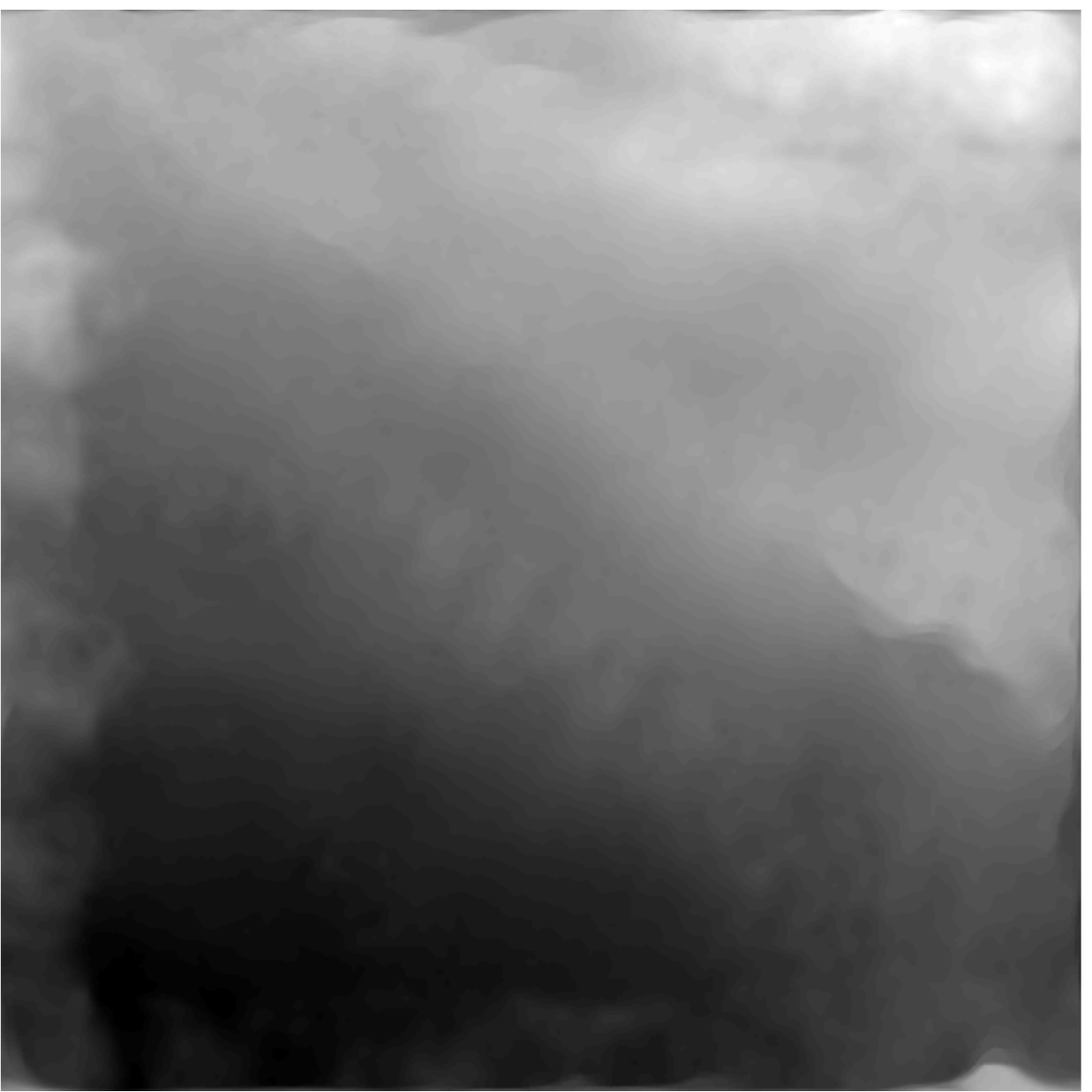} \\
\end{tabular}
\normalsize

\caption{P2D--EMD decomposition of wake image. 1st IMF: $\rho^{(1)} = 50$, $\nu^{(1)} = 50$. 2nd IMF: $\rho^{(2)} = 20$, $\nu^{(2)} = 5$. 3rd IMF: $\rho^{(3)} = 20$, $\nu^{(3)} = 1$.}
\label{fig:psamtikemd}
\end{center}
\end{figure*}

\begin{figure*}
\begin{center}
\footnotesize
\begin{tabular}{p{0.2cm}p{0.2cm}ccccccc}


&
&
&
Amplitude&
&
Frequency&
&
Orientation&
\\

\rotatebox{90}{\hspace{0.4cm}Riesz-Laplace} &
\rotatebox{90}{\hspace{0.9cm}Scale 1} &
\includegraphics[width=1.1in]{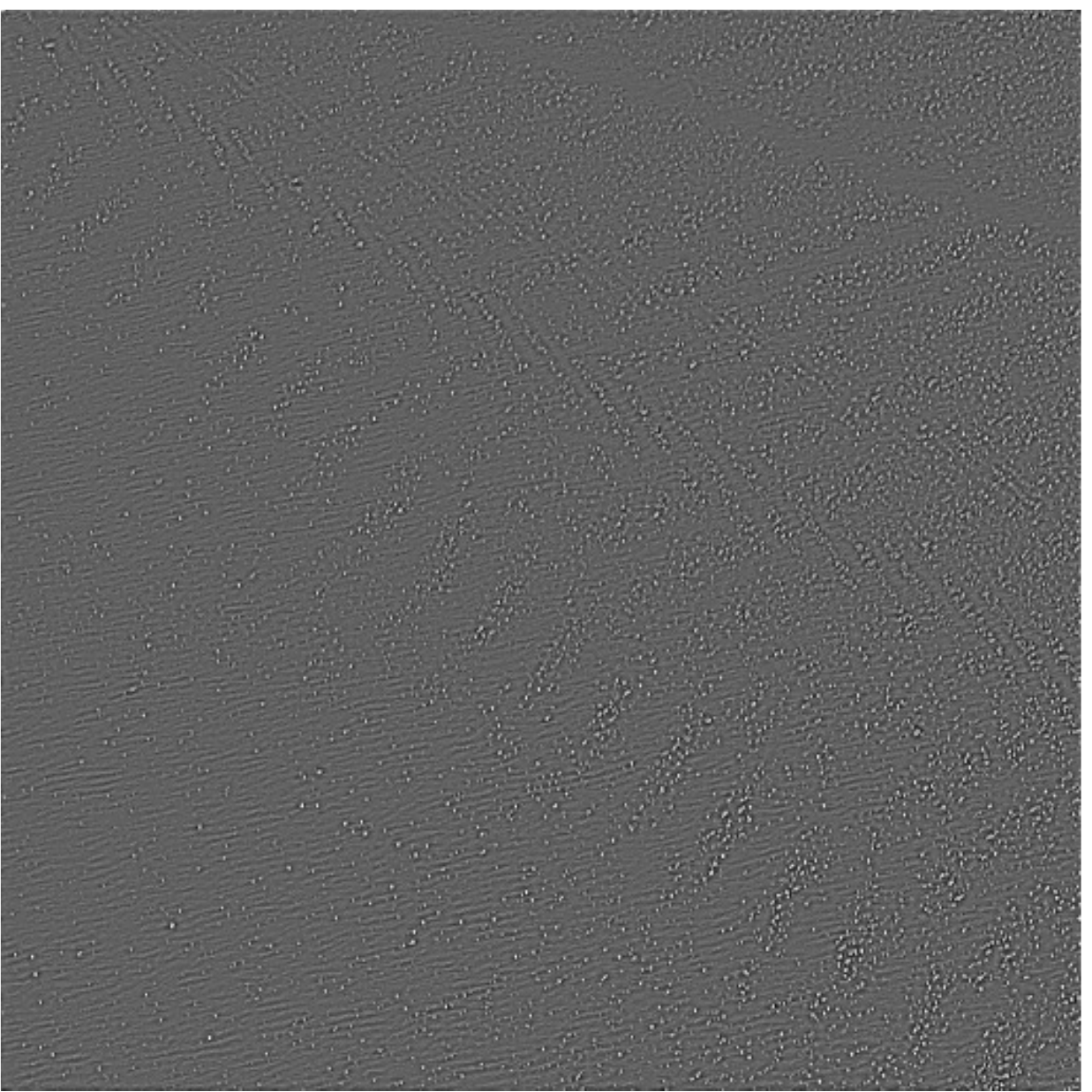} &
\includegraphics[width=1.1in]{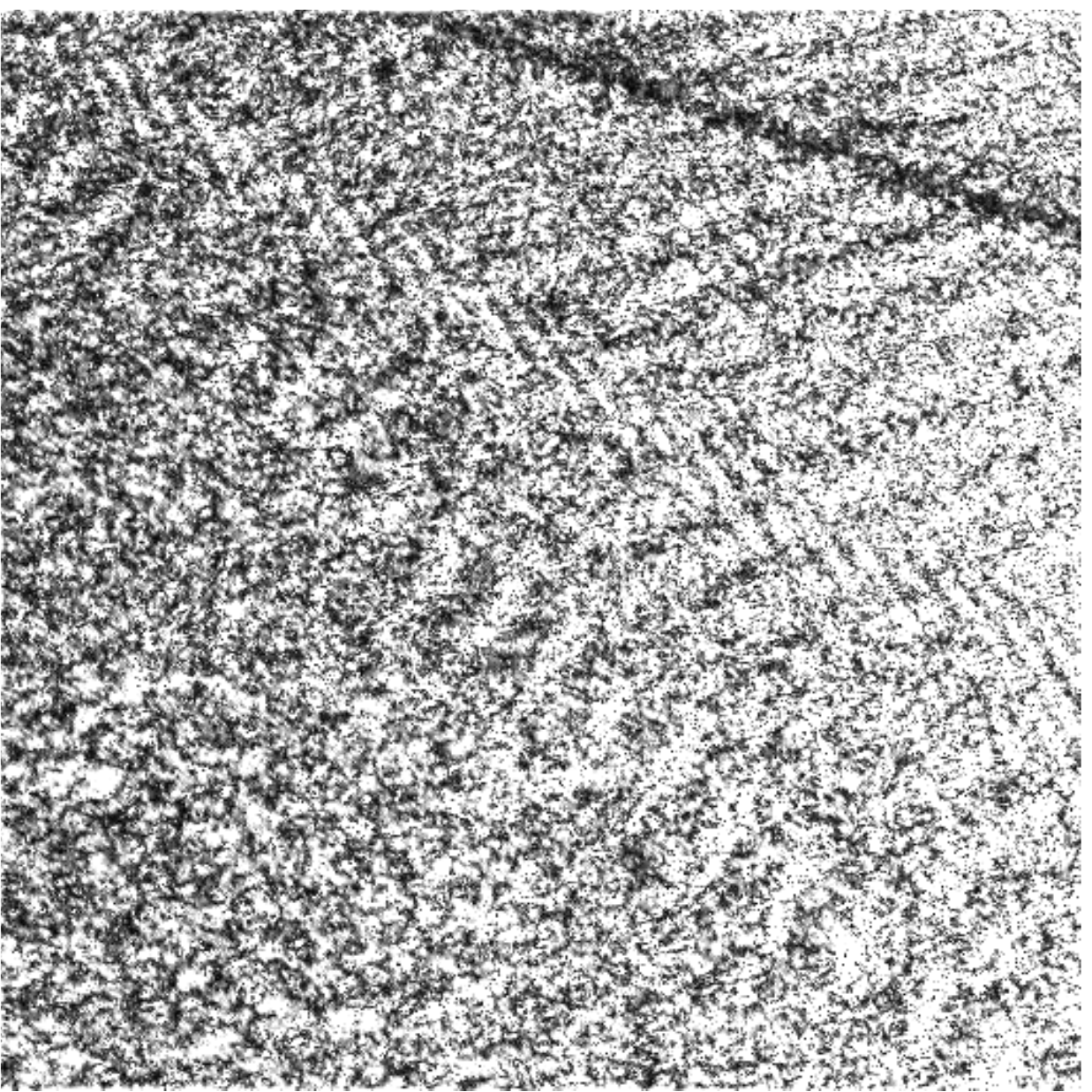} &
\includegraphics[width=0.14in]{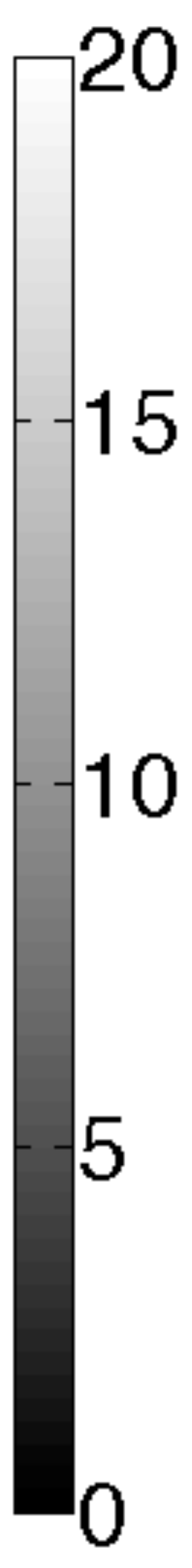} &
\includegraphics[width=1.1in]{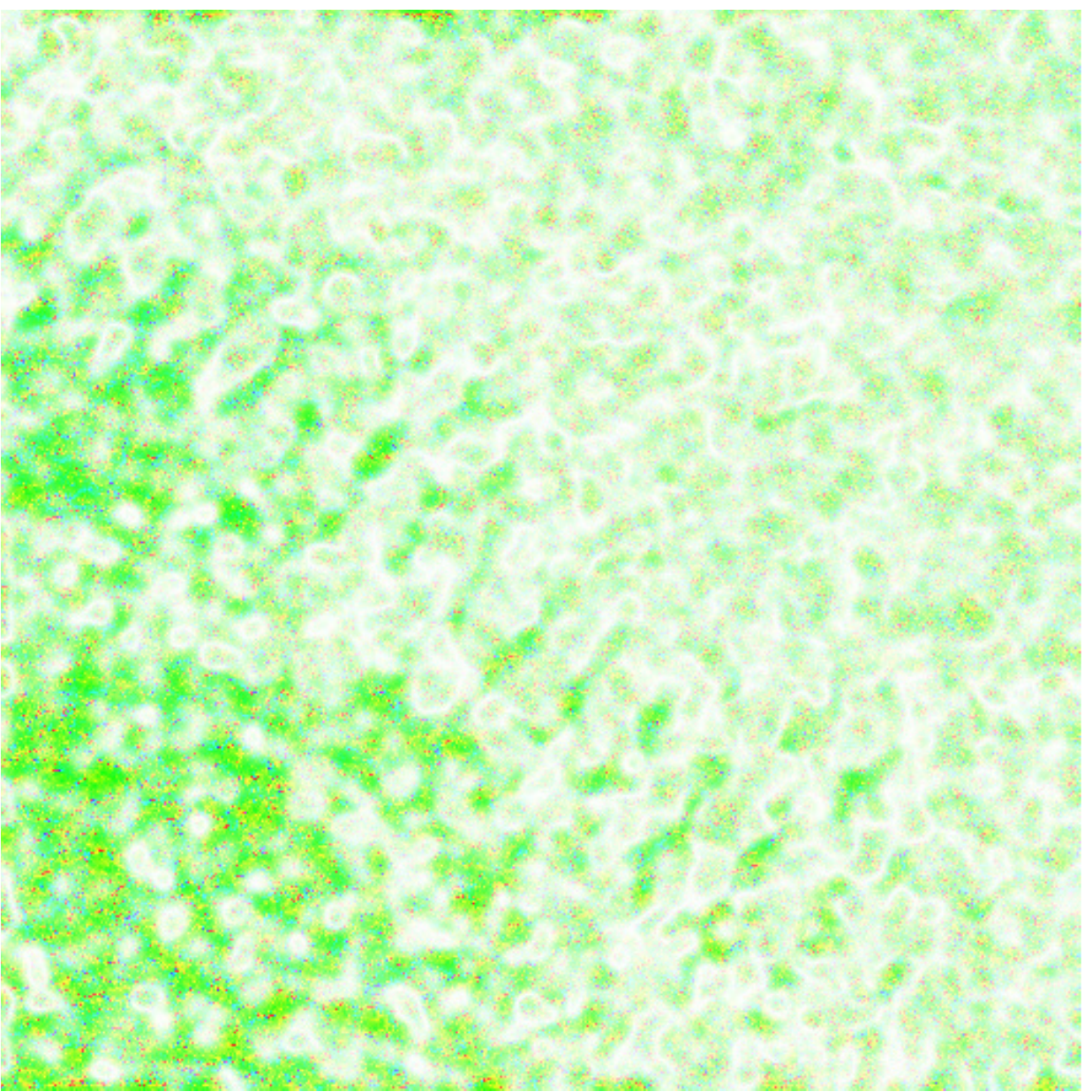} &
\includegraphics[width=0.17in]{figures/colorbar_freq-eps-converted-to.pdf} &
\includegraphics[width=1.1in]{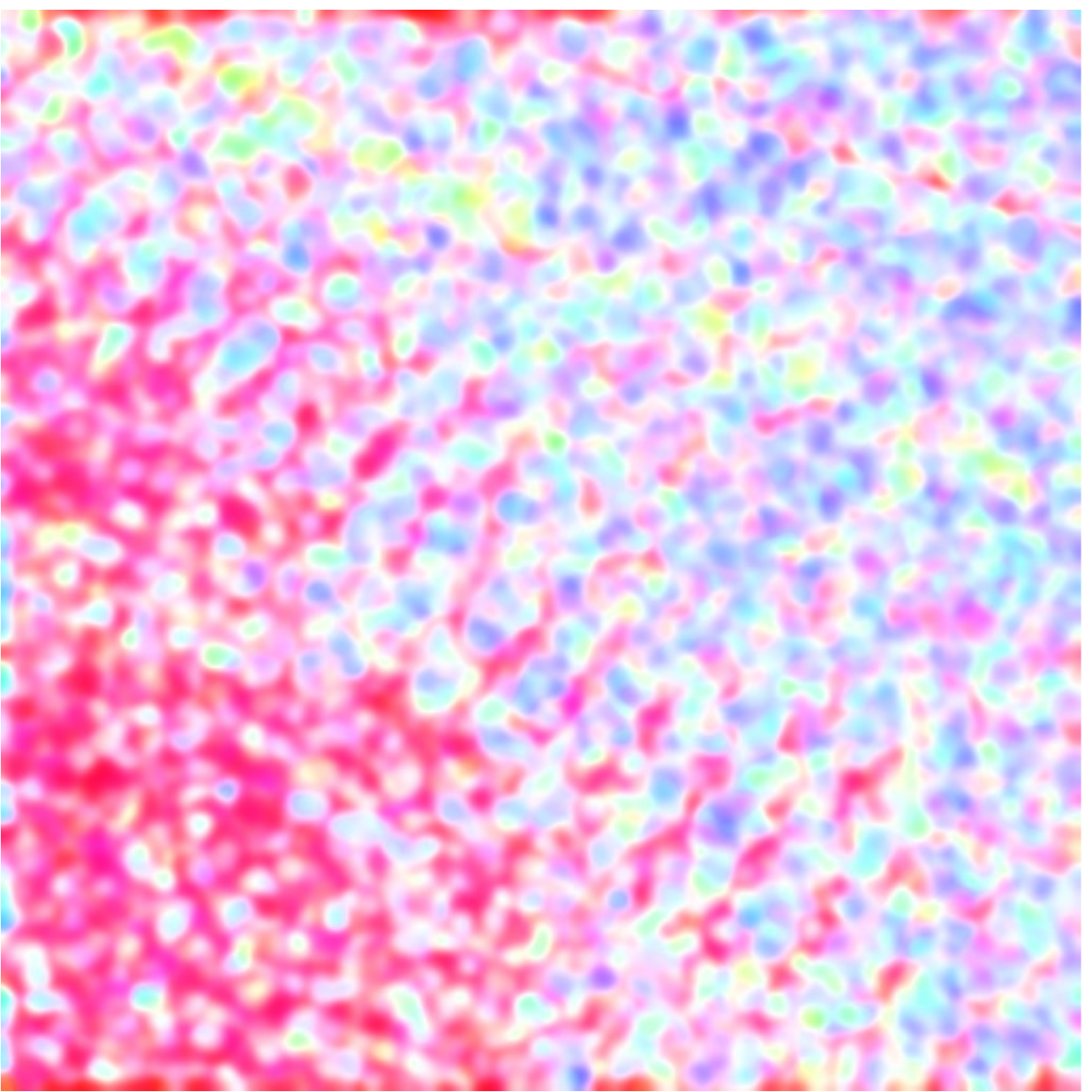} &
\includegraphics[width=0.2in]{figures/colorbar_orientation-eps-converted-to.pdf}
 \\

\rotatebox{90}{\hspace{0.6cm}P2D--HHT} &
\rotatebox{90}{\hspace{1cm}IMF 1} &
\includegraphics[width=1.1in]{figures/psamtik_imf1_50_50_thresh40-eps-converted-to.pdf} &
\includegraphics[width=1.1in]{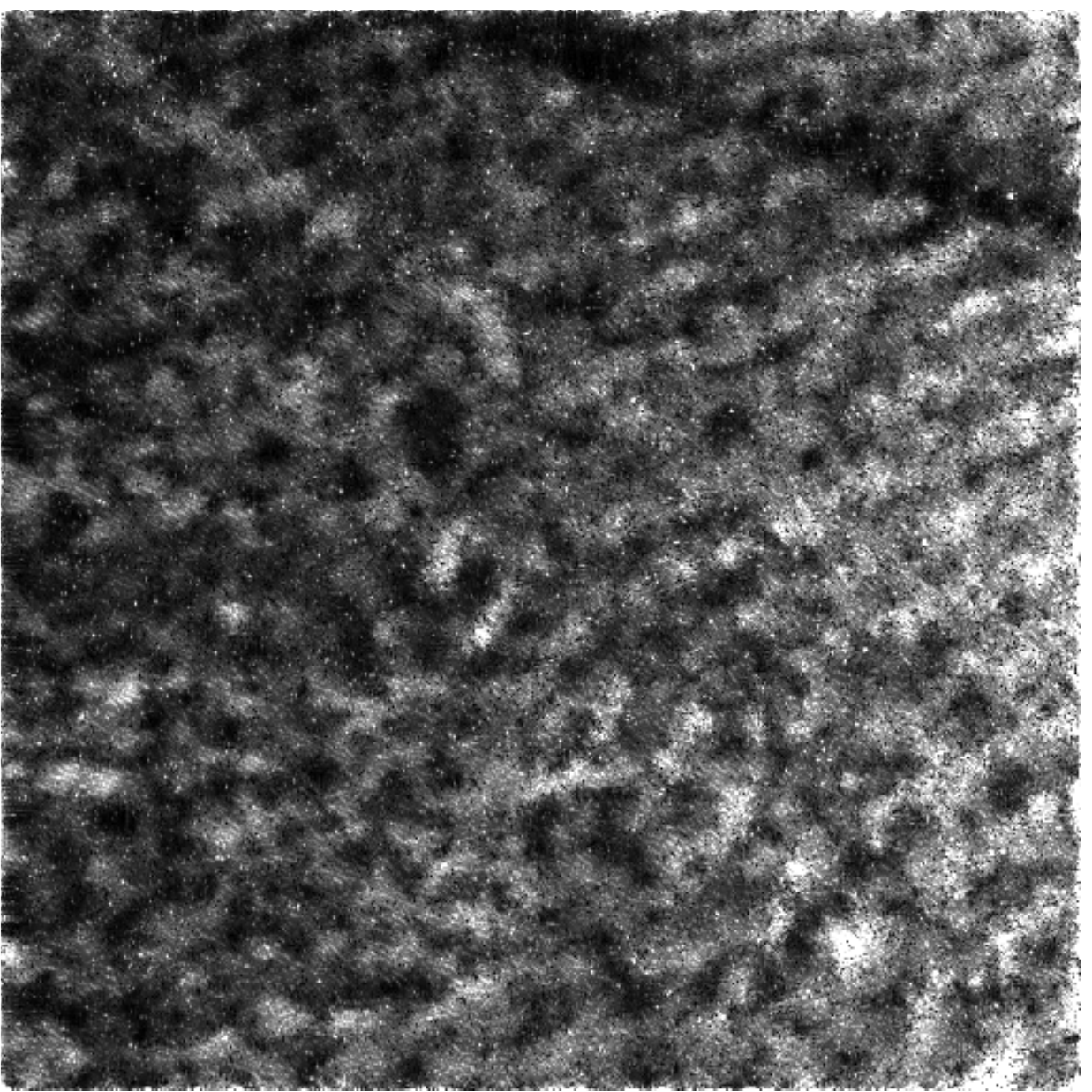} &
\includegraphics[width=0.14in]{figures/colorbar_amplitude_psamtik_20-eps-converted-to.pdf} &
\includegraphics[width=1.1in]{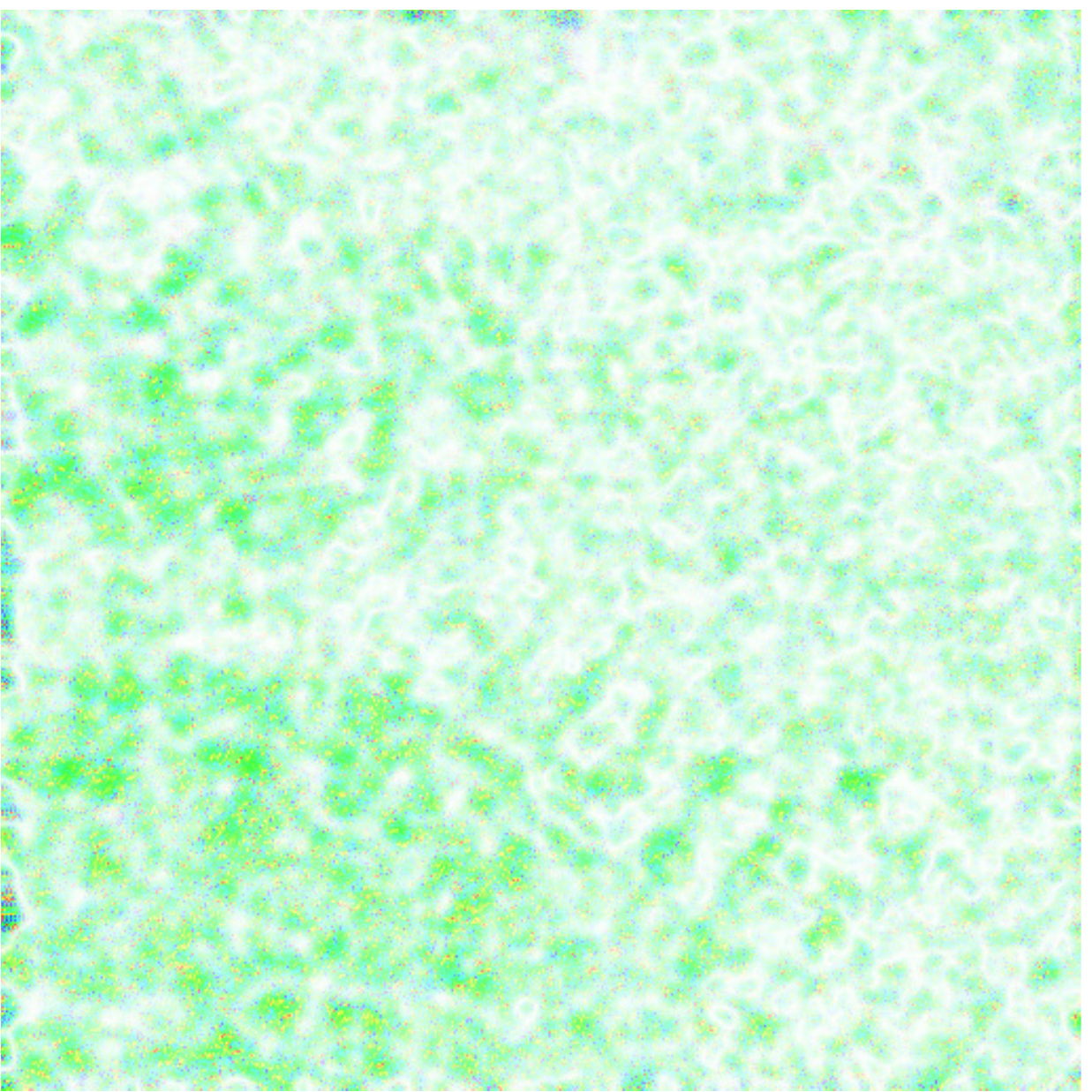} &
\includegraphics[width=0.17in]{figures/colorbar_freq-eps-converted-to.pdf} &
\includegraphics[width=1.1in]{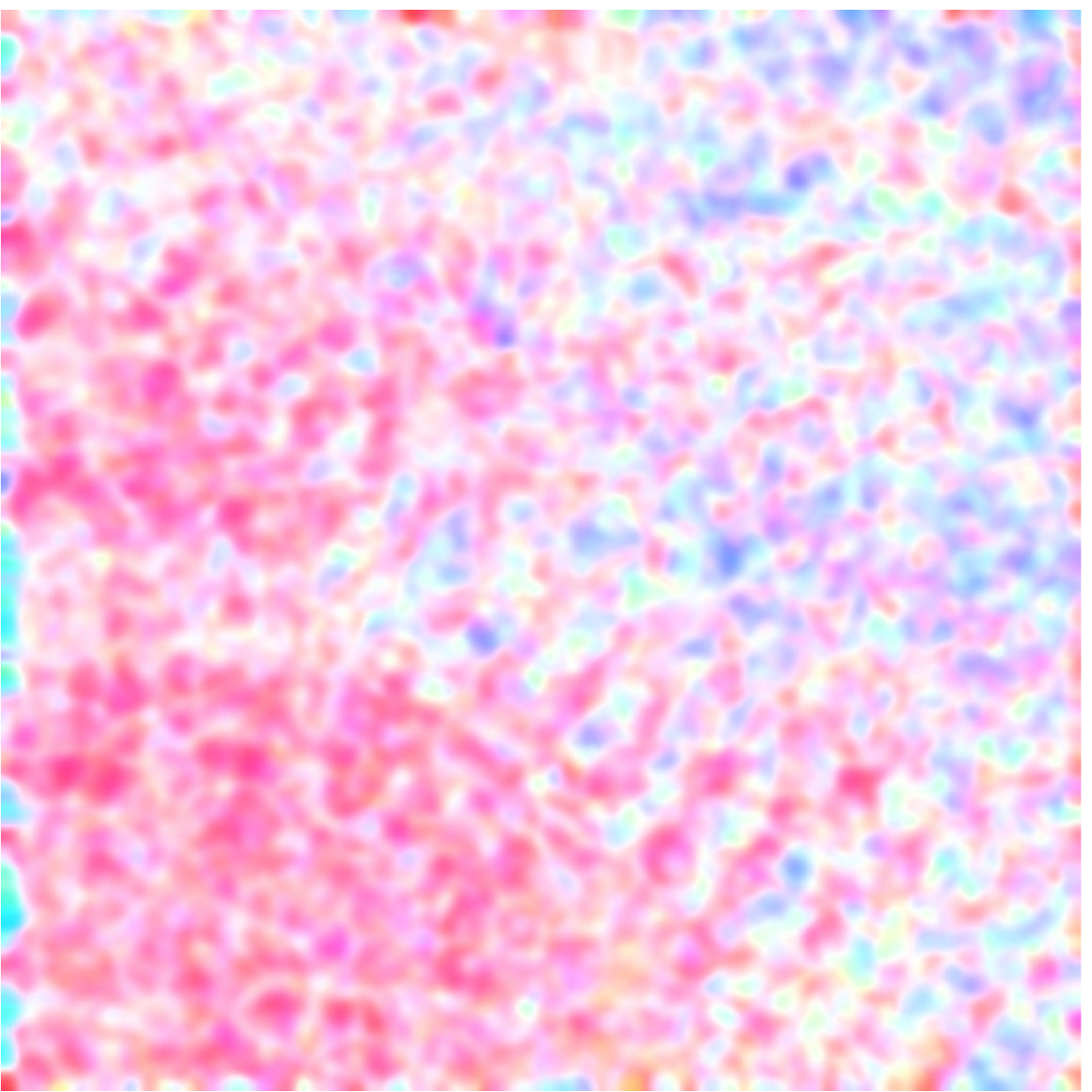} &
\includegraphics[width=0.2in]{figures/colorbar_orientation-eps-converted-to.pdf}

 \\


\rotatebox{90}{\hspace{0.6cm}P2D--PHT} &
\rotatebox{90}{\hspace{0.3cm}Denoised IMF 1} &
\includegraphics[width=1.1in]{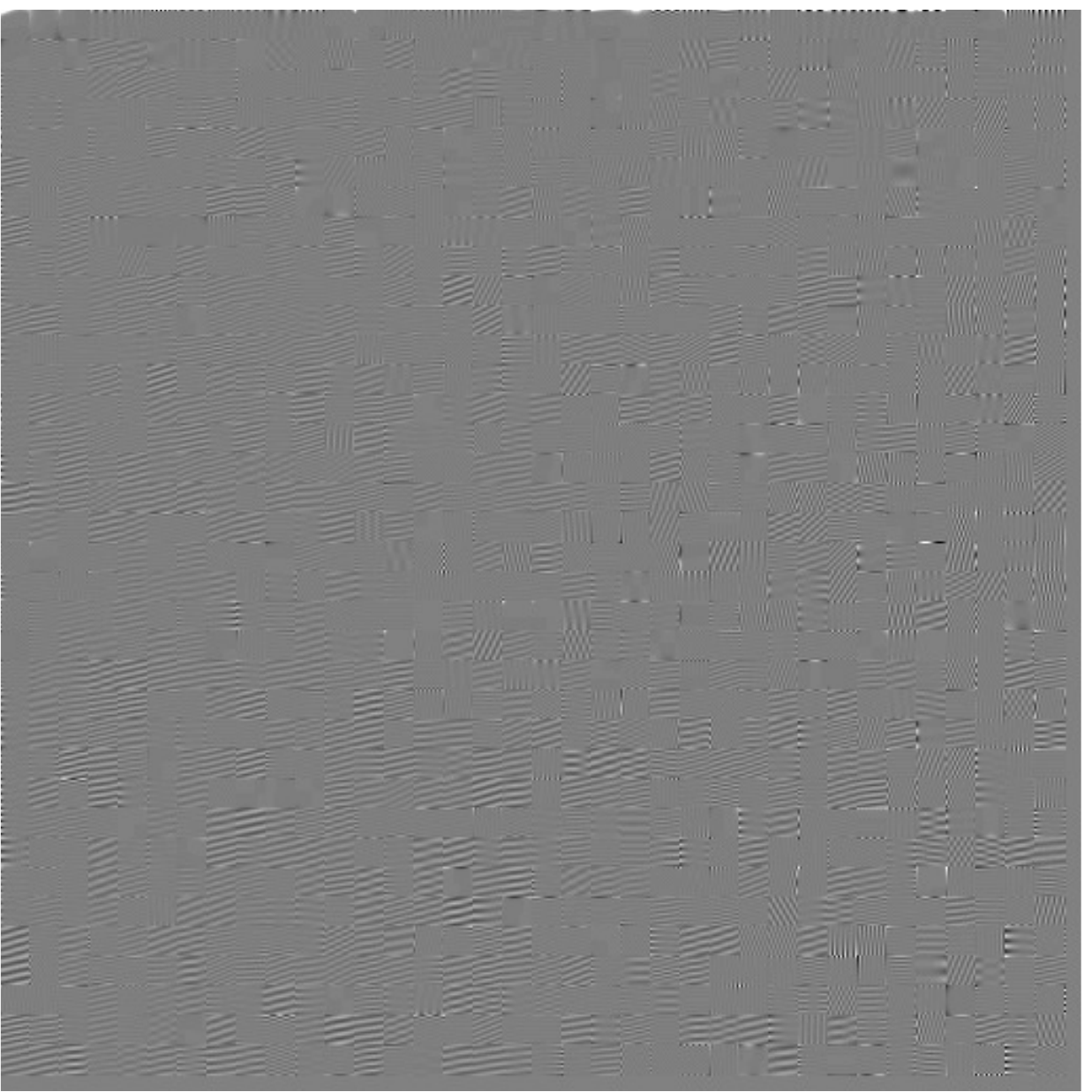}&
\includegraphics[width=1.1in]{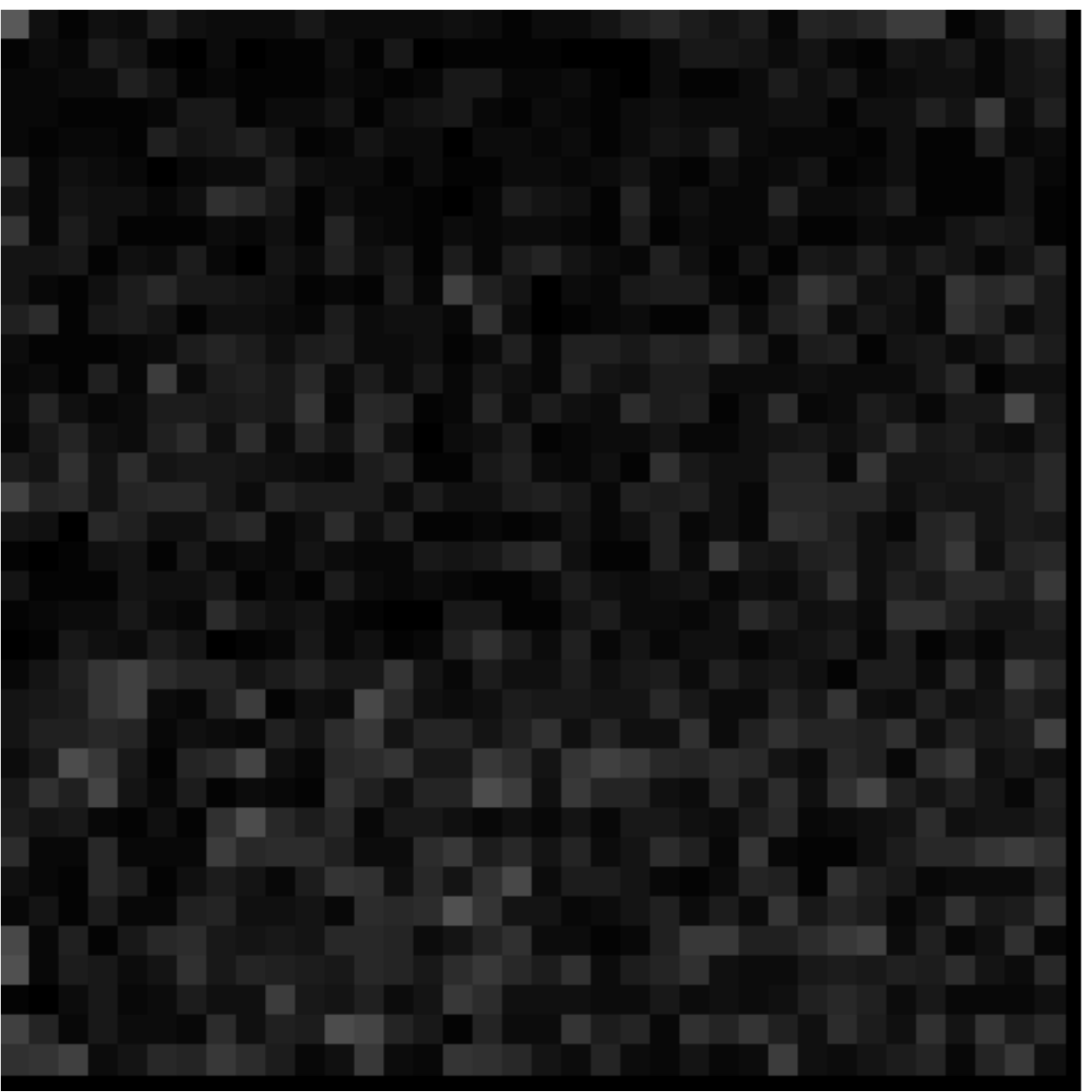} &
\includegraphics[width=0.14in]{figures/colorbar_amplitude_psamtik_20-eps-converted-to.pdf} &
\includegraphics[width=1.1in]{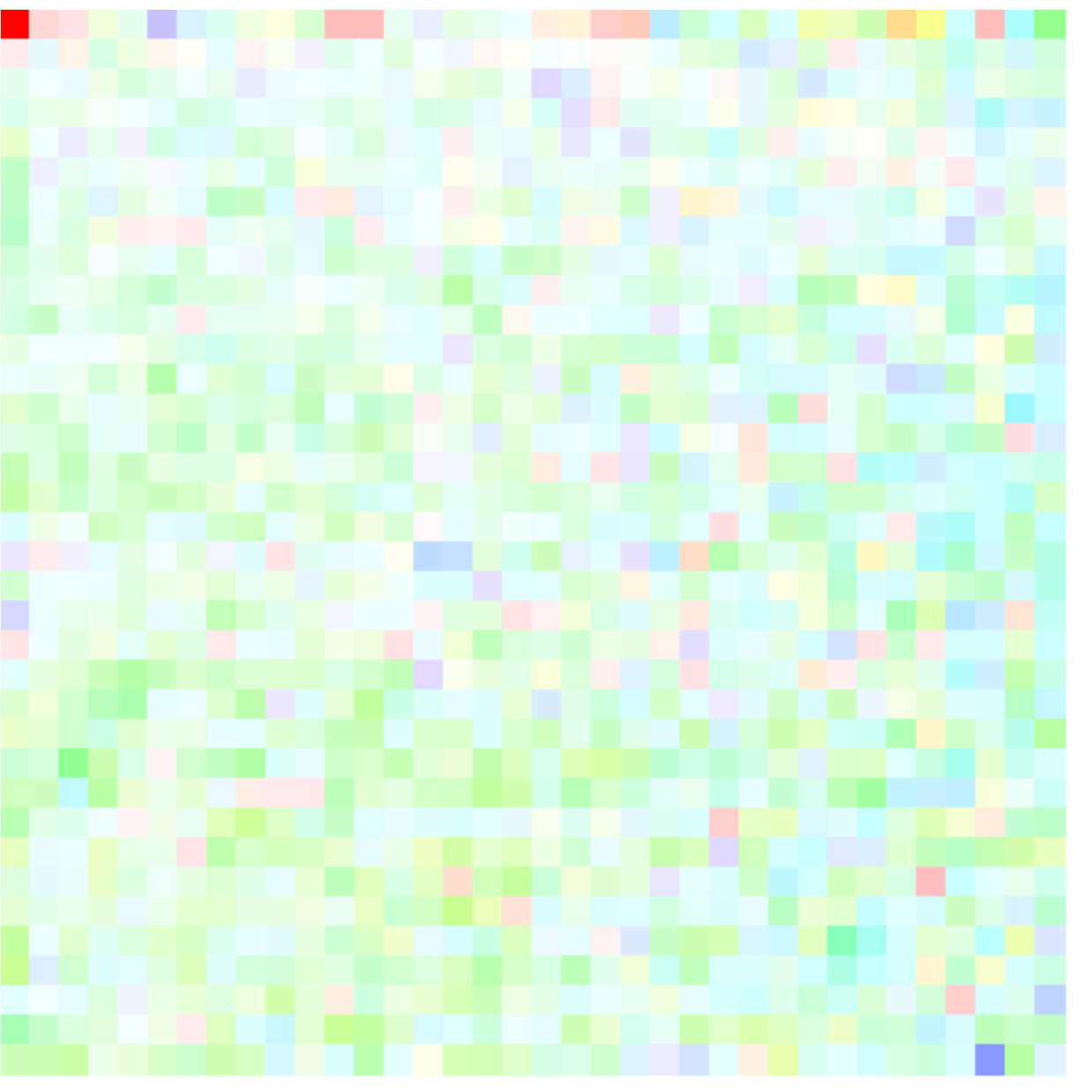} &
\includegraphics[width=0.17in]{figures/colorbar_freq-eps-converted-to.pdf} &
\includegraphics[width=1.1in]{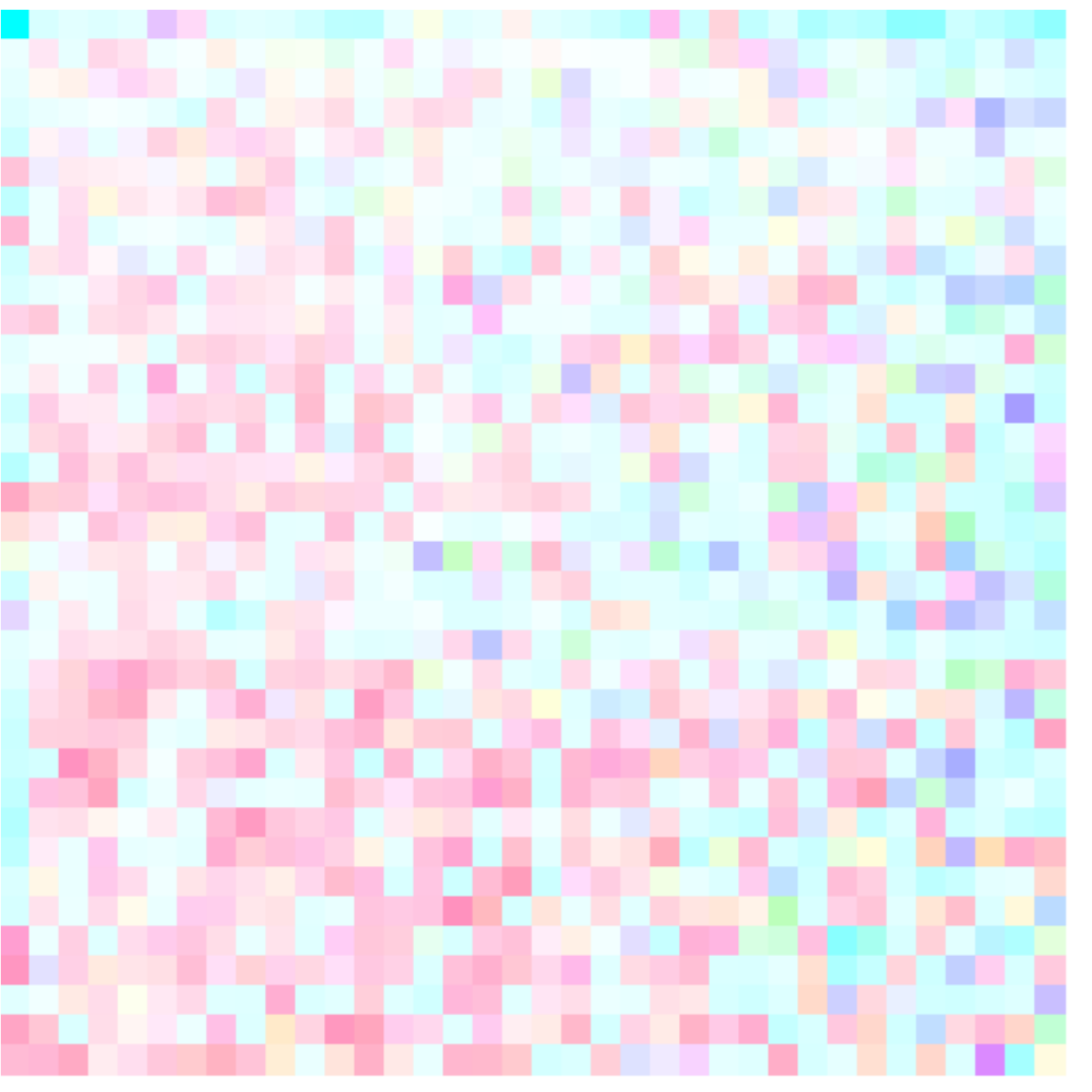} &
\includegraphics[width=0.2in]{figures/colorbar_orientation-eps-converted-to.pdf}
 \\

 \end{tabular}
 \normalsize
\caption{Spectral analysis on 1st mode of wake image. 1st row: 1st scale of Riesz-Laplace wavelet transform. From left to right: mode $\mathbf{d}^{(1)}$, amplitude $\boldsymbol{\alpha}^{(1)}$, frequency $\boldsymbol{\eta}^{(1)}$ and orientation $\boldsymbol{\theta}^{(1)}$. 2nd row: 1st IMF P2D--HHT. From left to right: mode $\mathbf{d}^{(1)}$, amplitude $\boldsymbol{\alpha}^{(1)}$, frequency $\boldsymbol{\eta}^{(1)}$ and orientation $\boldsymbol{\theta}^{(1)}$. 3rd row: 1st IMF P2D--PHT ($\overline{N}^{(1)} = 14$). From left to right: denoised mode, amplitude $\boldsymbol{\alpha}^{(1)}$, frequency $\boldsymbol{\eta}^{(1)}$ and orientation $\boldsymbol{\theta}^{(1)}$.}
\label{fig:psamtikimf1}
\end{center}
\end{figure*}

\begin{figure*}
\begin{center}
\footnotesize
\begin{tabular}{p{0.2cm}p{0.2cm}ccccccc}


&
&
&
 Amplitude&
&
 Frequency&
&
 Orientation&
\\

\rotatebox{90}{\hspace{0.4cm}Riesz-Laplace} &
\rotatebox{90}{\hspace{0.9cm}Scale 2} &
\includegraphics[width=1.1in]{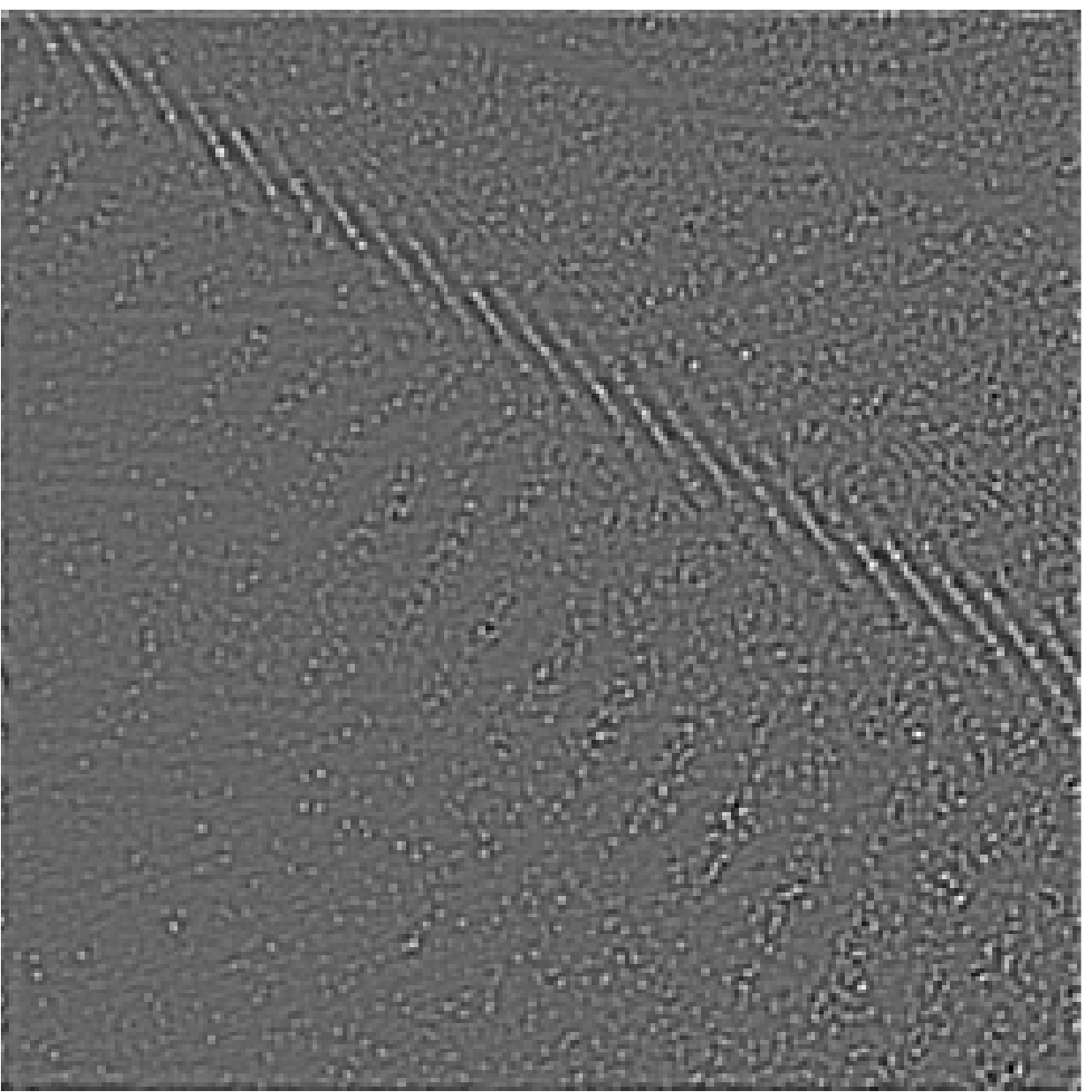} &
\includegraphics[width=1.1in]{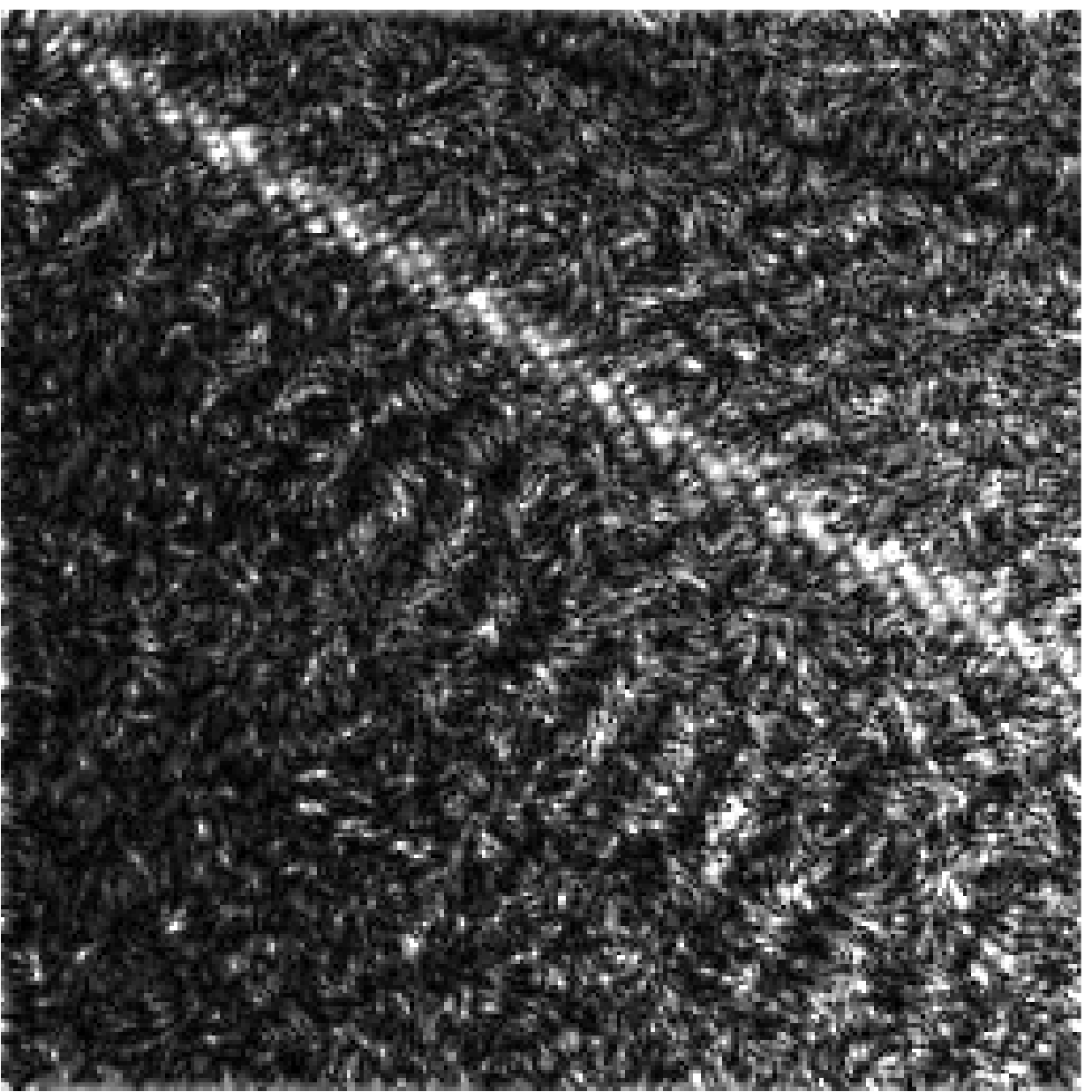} &
\includegraphics[width=0.16in]{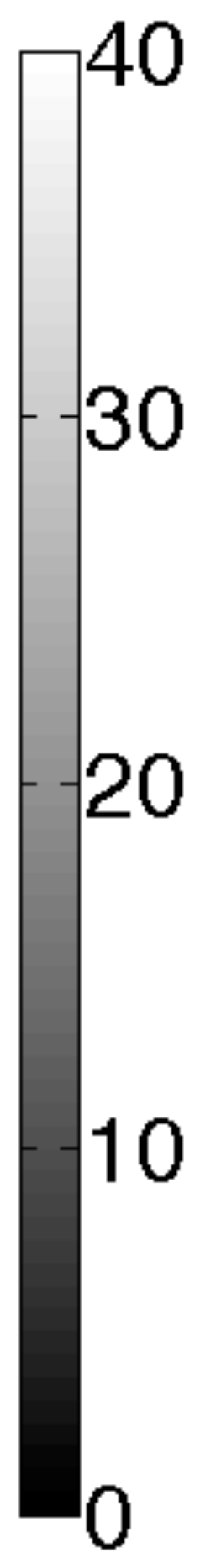} &
\includegraphics[width=1.1in]{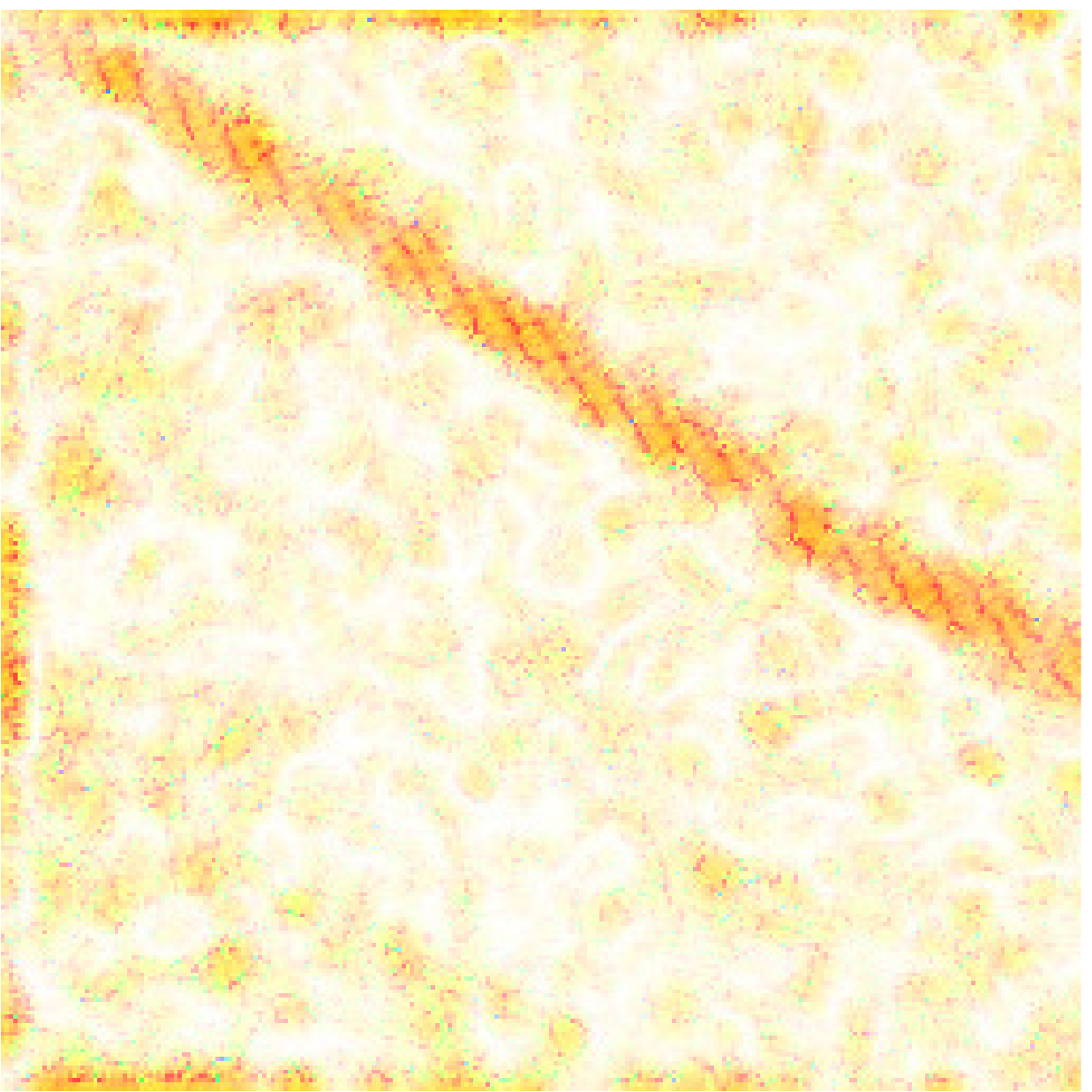} &
\includegraphics[width=0.17in]{figures/colorbar_freq-eps-converted-to.pdf} &
\includegraphics[width=1.1in]{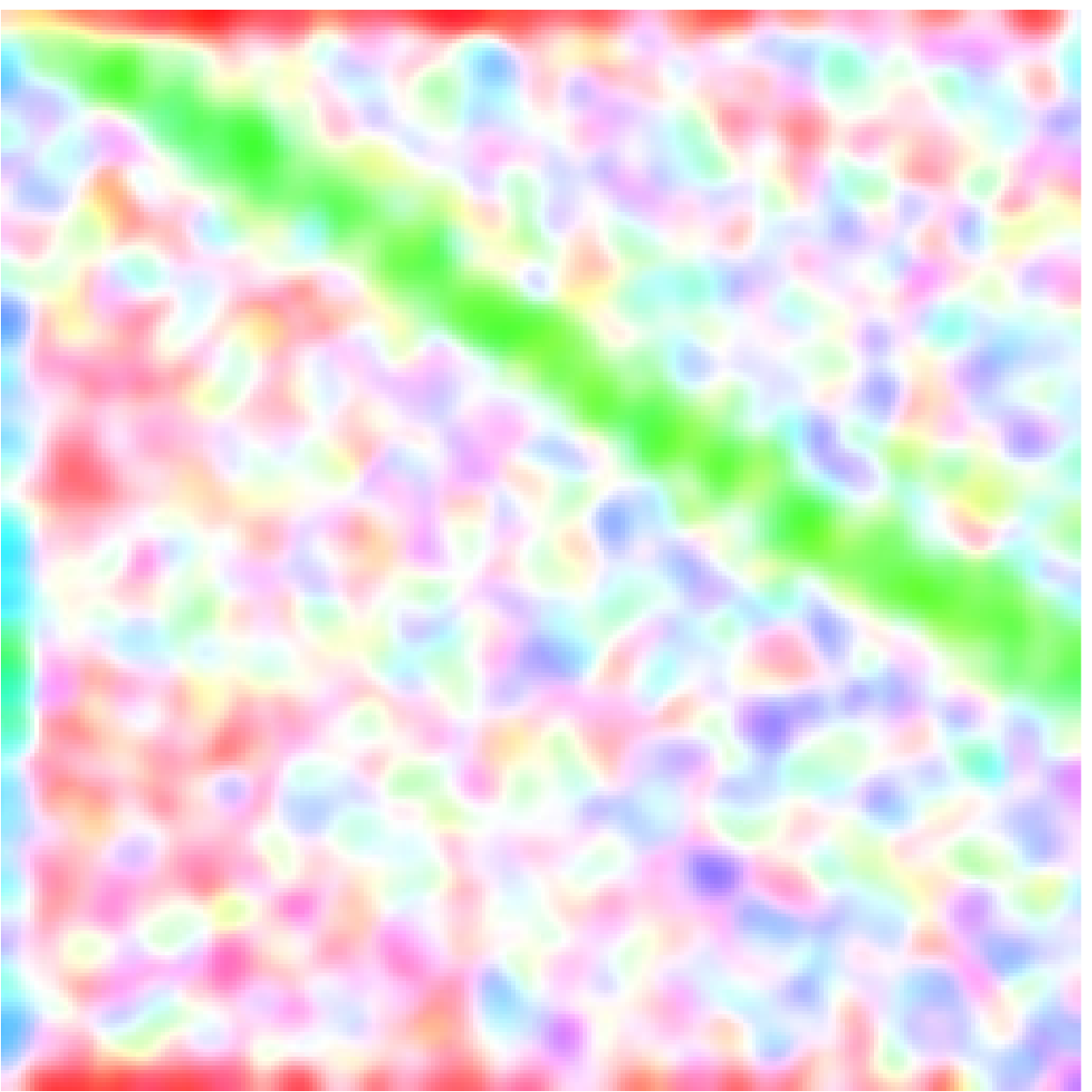} &
\includegraphics[width=0.2in]{figures/colorbar_orientation-eps-converted-to.pdf}
 \\

\rotatebox{90}{\hspace{0.4cm}Riesz-Laplace} &
\rotatebox{90}{\hspace{0.9cm}Scale 3} &
\includegraphics[width=1.1in]{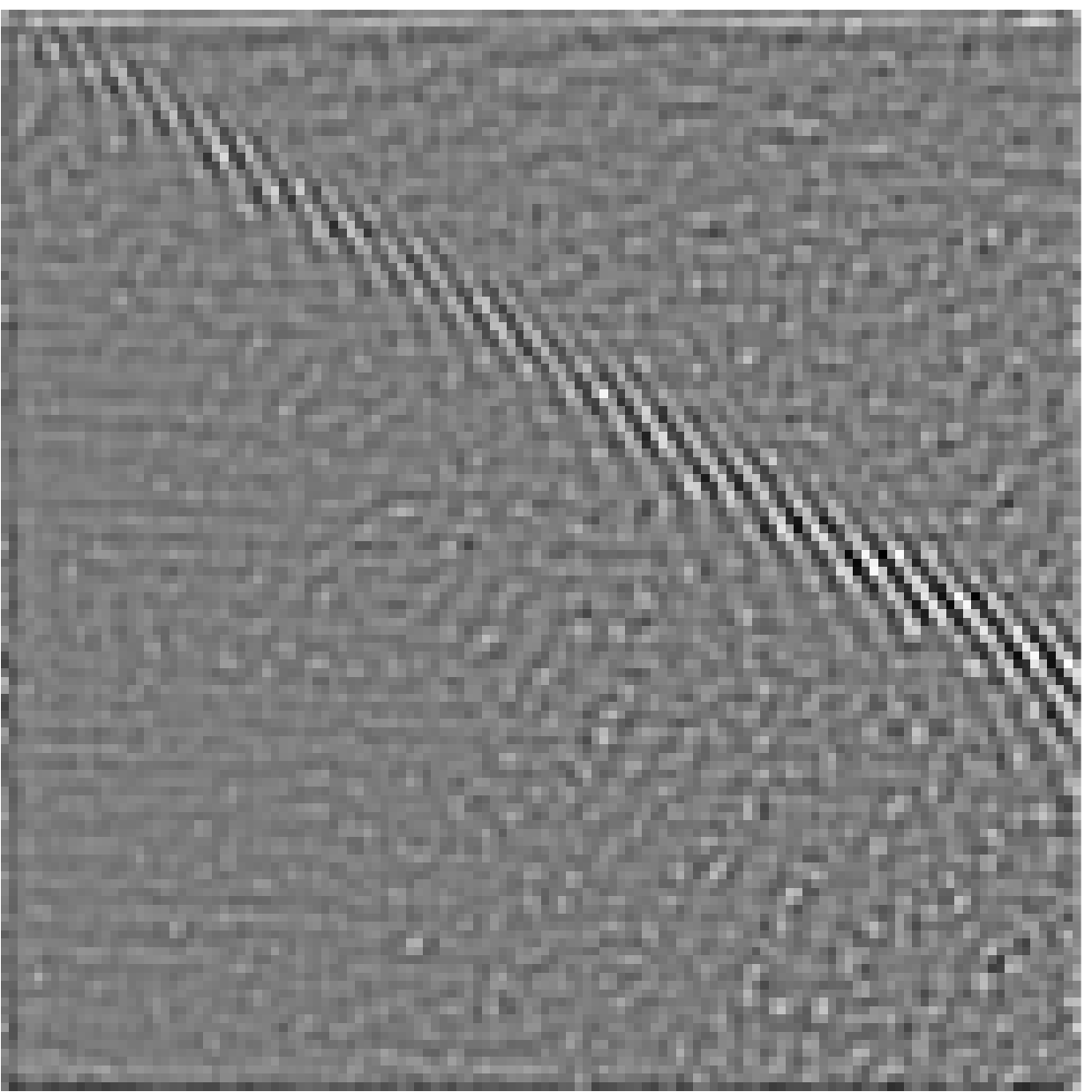} &
\includegraphics[width=1.1in]{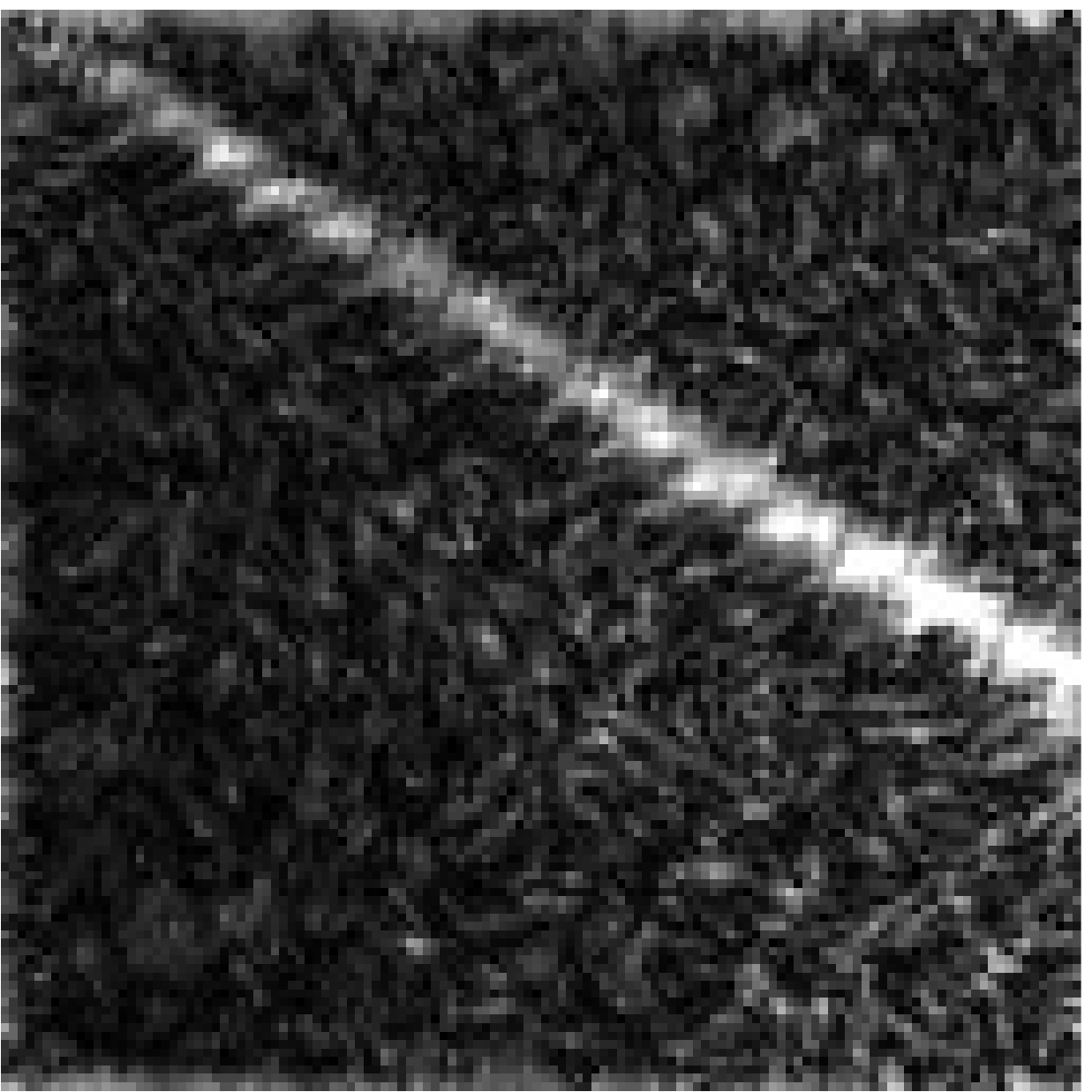} &
\includegraphics[width=0.16in]{figures/colorbar_amplitude_psamtik_40-eps-converted-to.pdf} &
\includegraphics[width=1.1in]{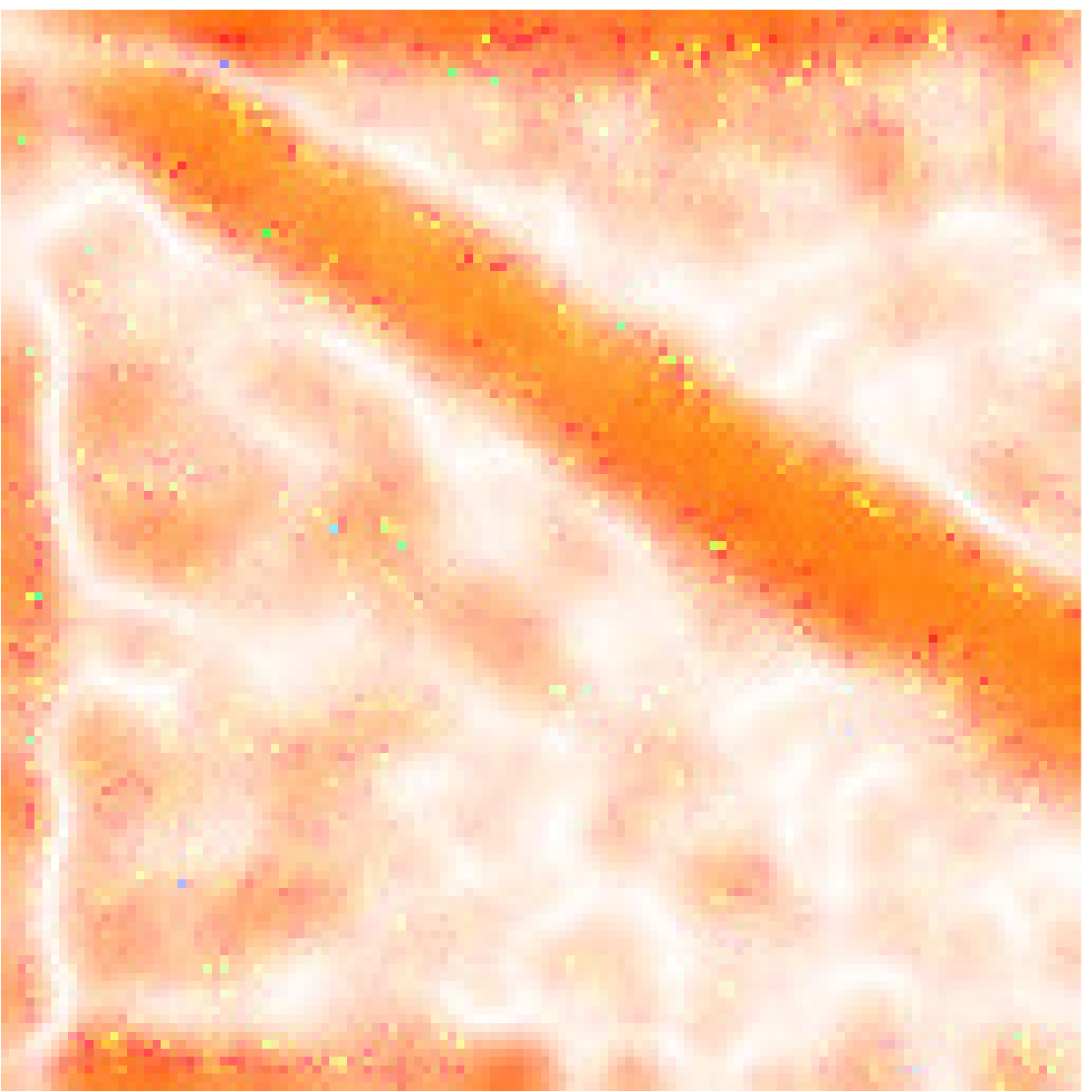} &
\includegraphics[width=0.17in]{figures/colorbar_freq-eps-converted-to.pdf} &
\includegraphics[width=1.1in]{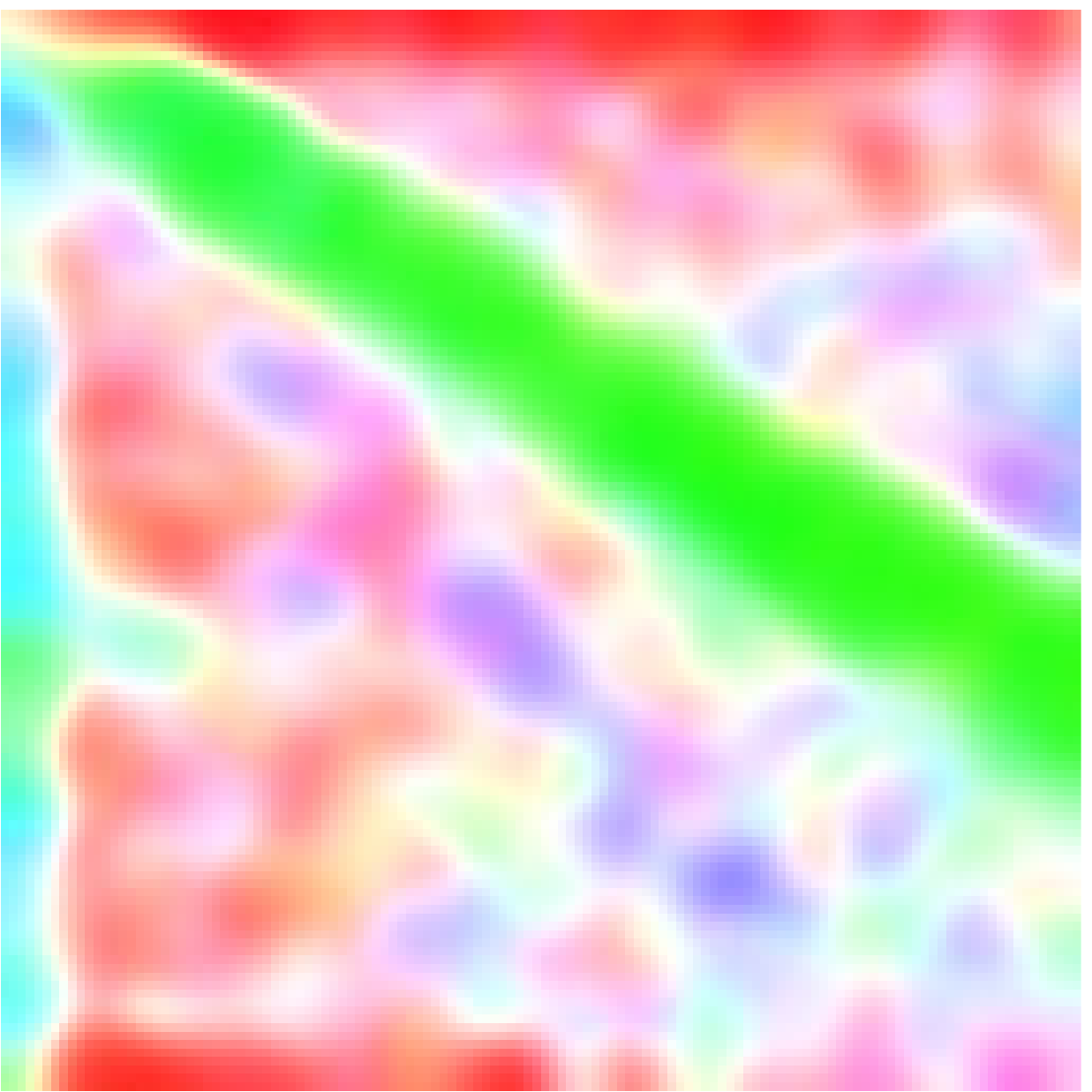} &
\includegraphics[width=0.2in]{figures/colorbar_orientation-eps-converted-to.pdf}
 \\

\rotatebox{90}{\hspace{0.6cm}P2D--HHT} &
\rotatebox{90}{\hspace{1cm}IMF 2} &
\includegraphics[width=1.1in]{figures/psamtik_imf2_50_50_20_5_thresh40-eps-converted-to.pdf} &
\includegraphics[width=1.1in]{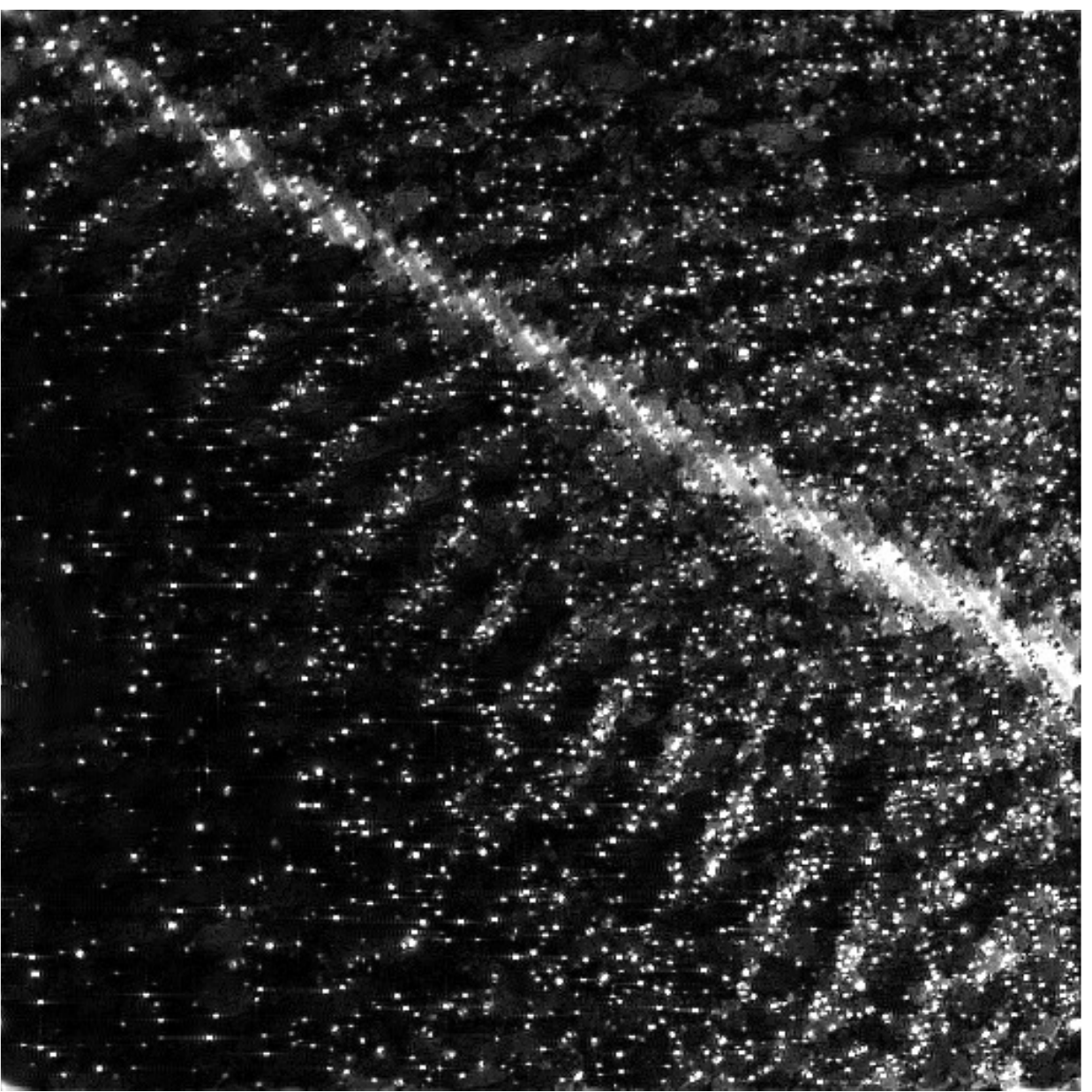} &
\includegraphics[width=0.16in]{figures/colorbar_amplitude_psamtik_40-eps-converted-to.pdf} &
\includegraphics[width=1.1in]{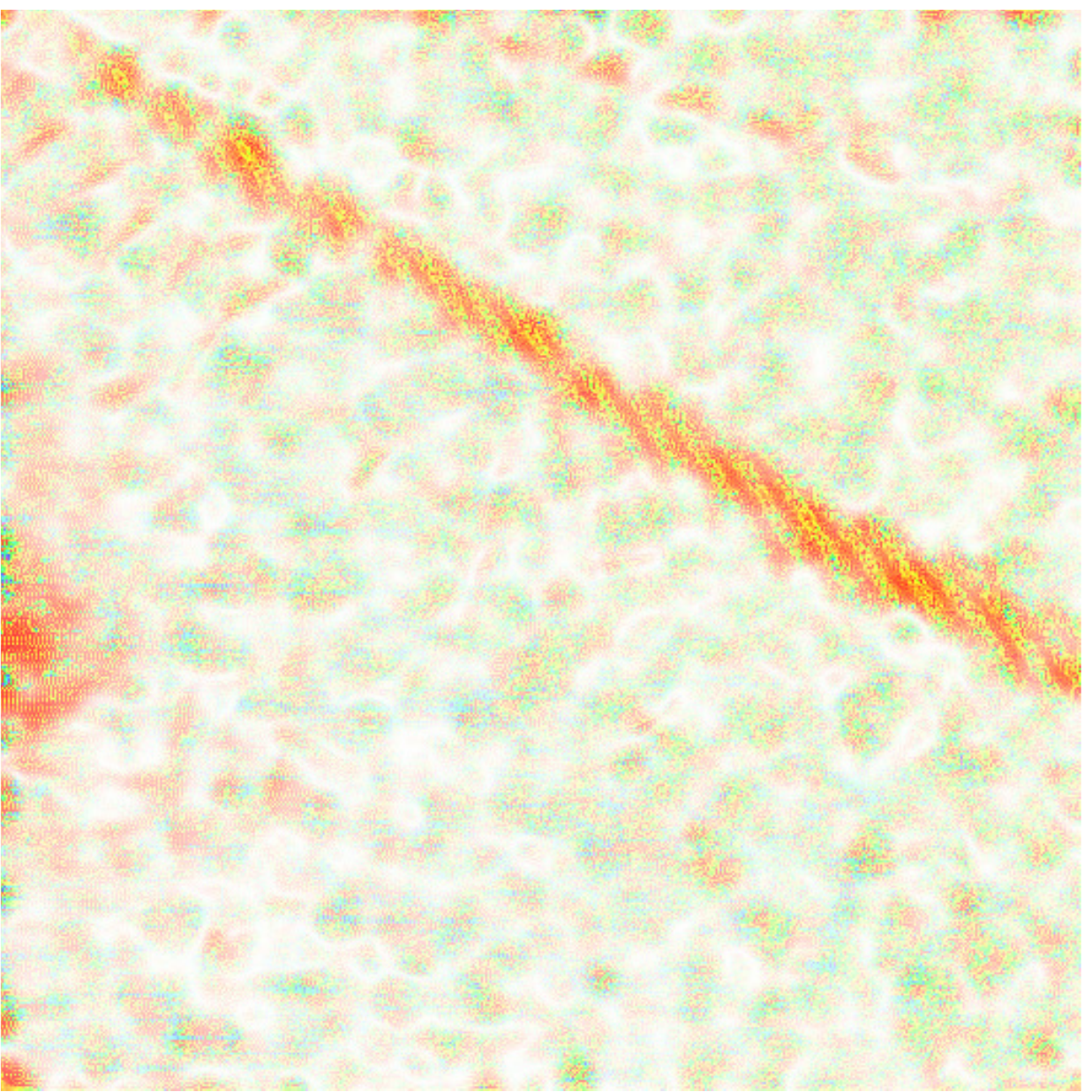} &
\includegraphics[width=0.17in]{figures/colorbar_freq-eps-converted-to.pdf} &
\includegraphics[width=1.1in]{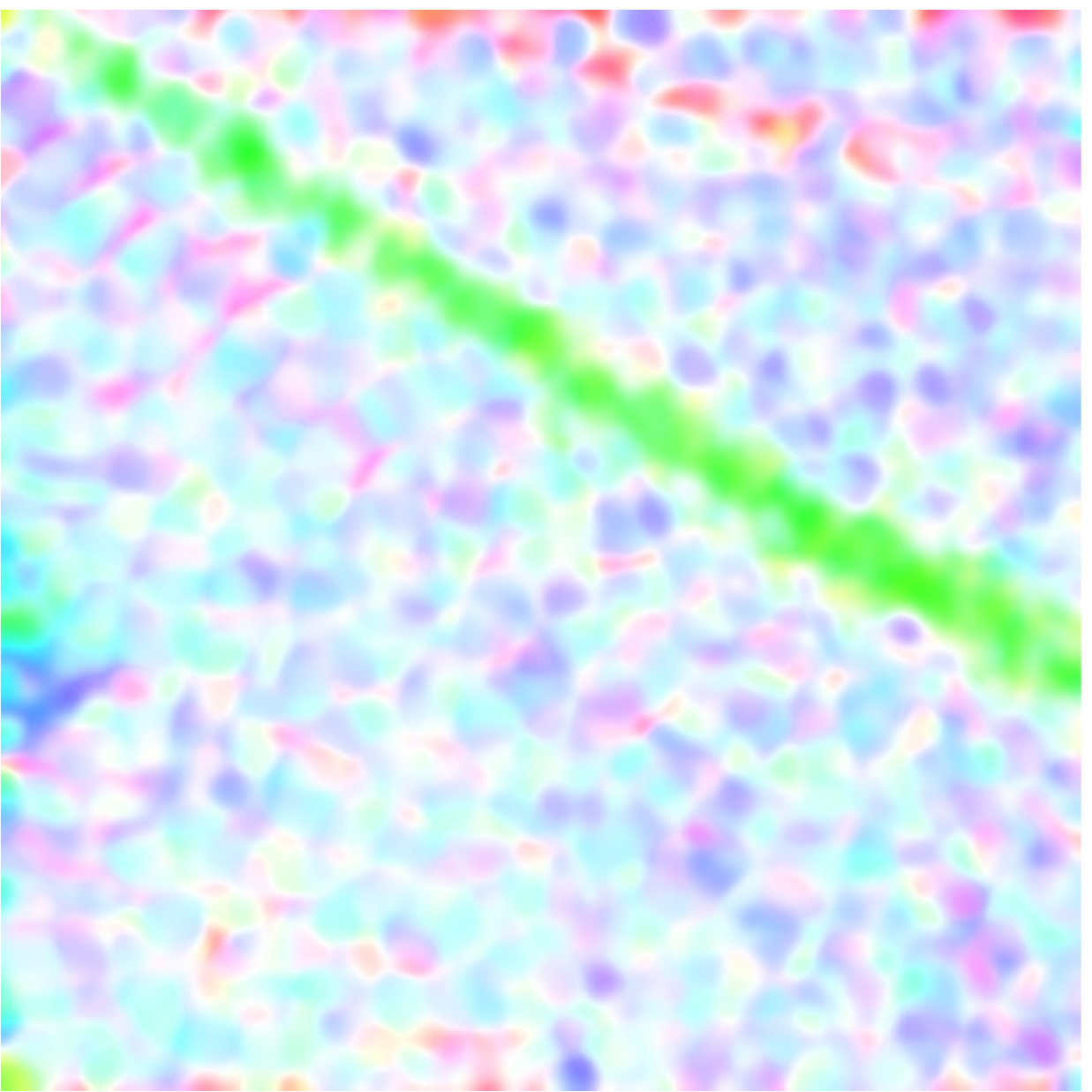} &
\includegraphics[width=0.2in]{figures/colorbar_orientation-eps-converted-to.pdf}

 \\
 

 \rotatebox{90}{\hspace{0.6cm}P2D--PHT} &
 \rotatebox{90}{\hspace{0.3cm}Denoised IMF 2} &
 \includegraphics[width=1.1in]{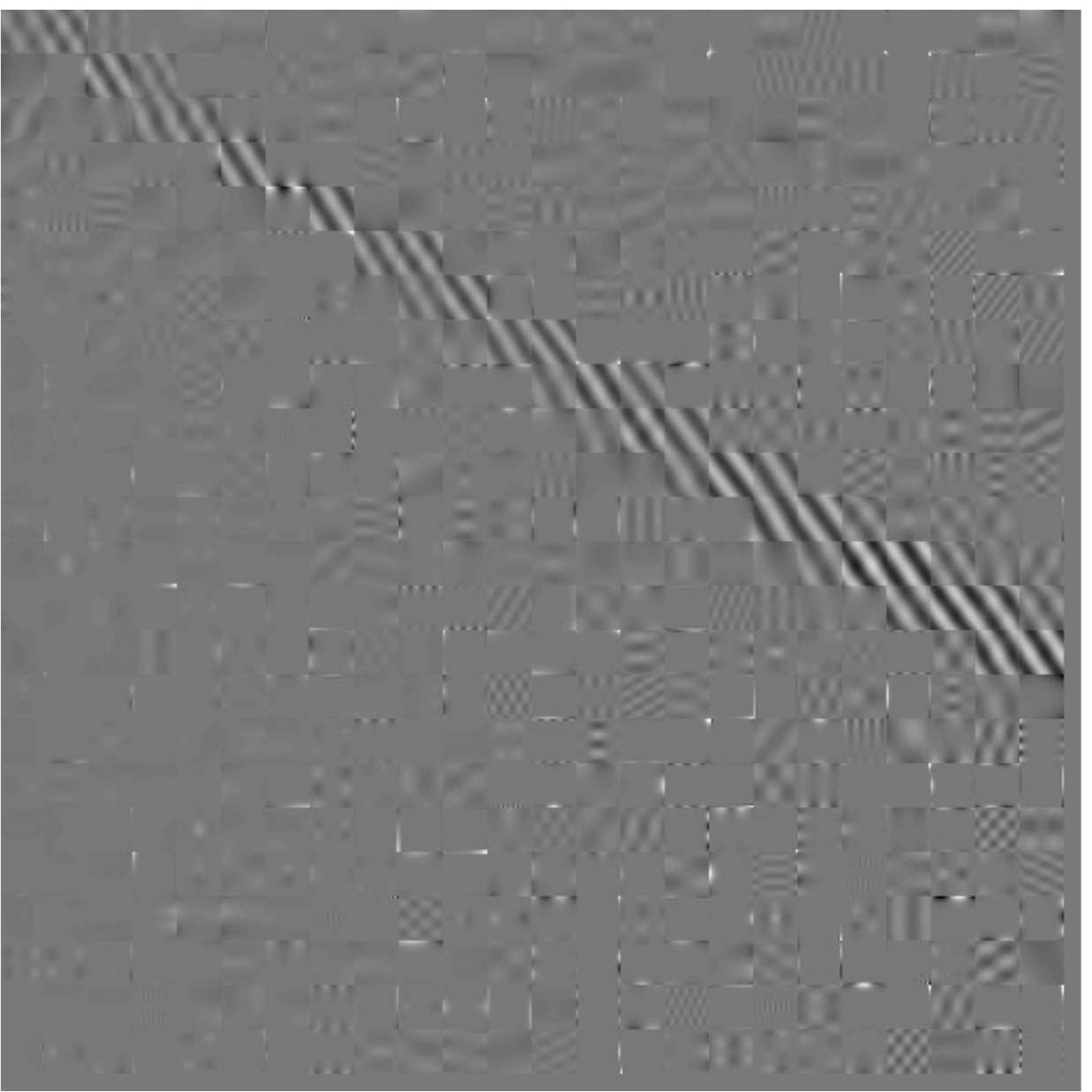} &
 \includegraphics[width=1.1in]{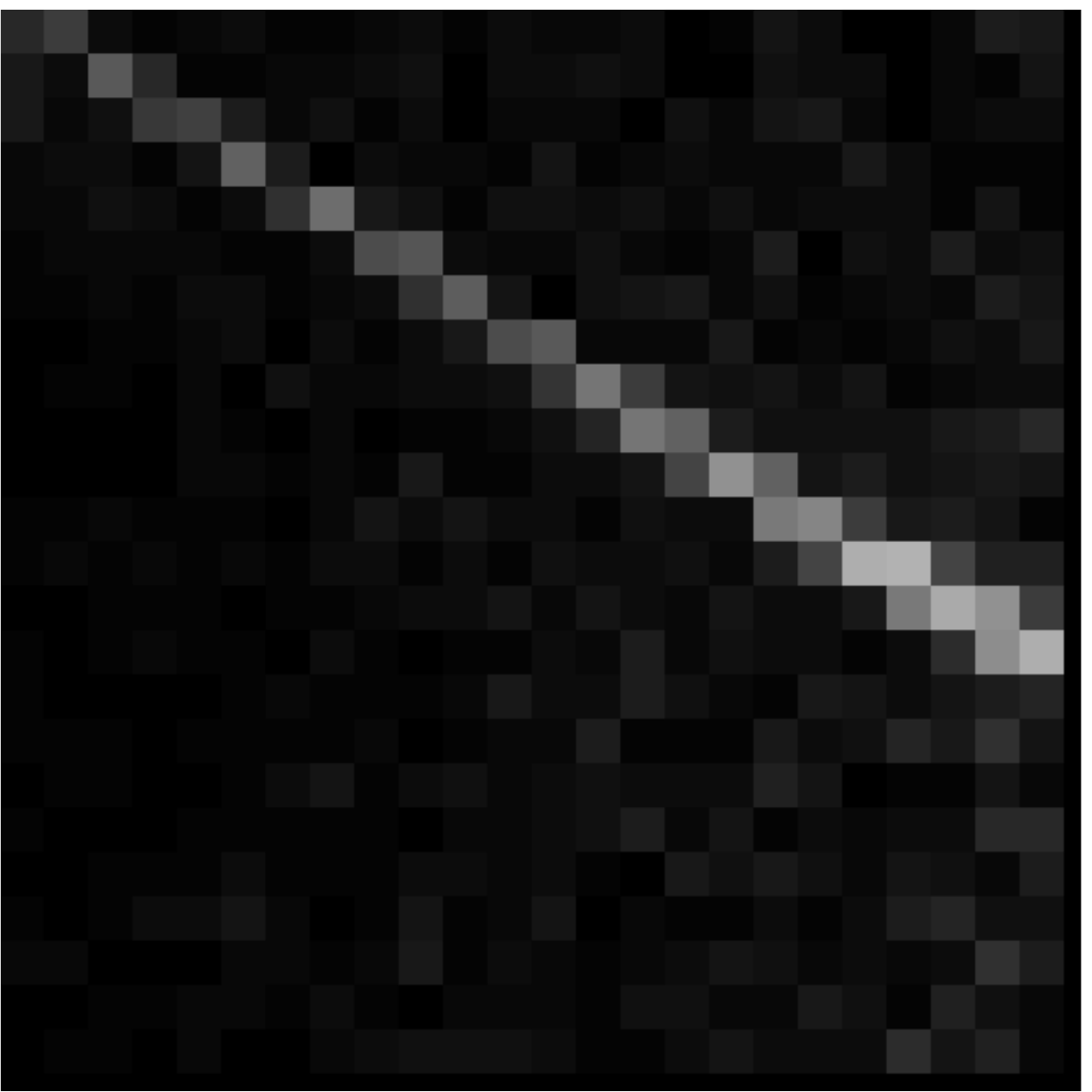} &
 \includegraphics[width=0.16in]{figures/colorbar_amplitude_psamtik_40-eps-converted-to.pdf} &
 \includegraphics[width=1.1in]{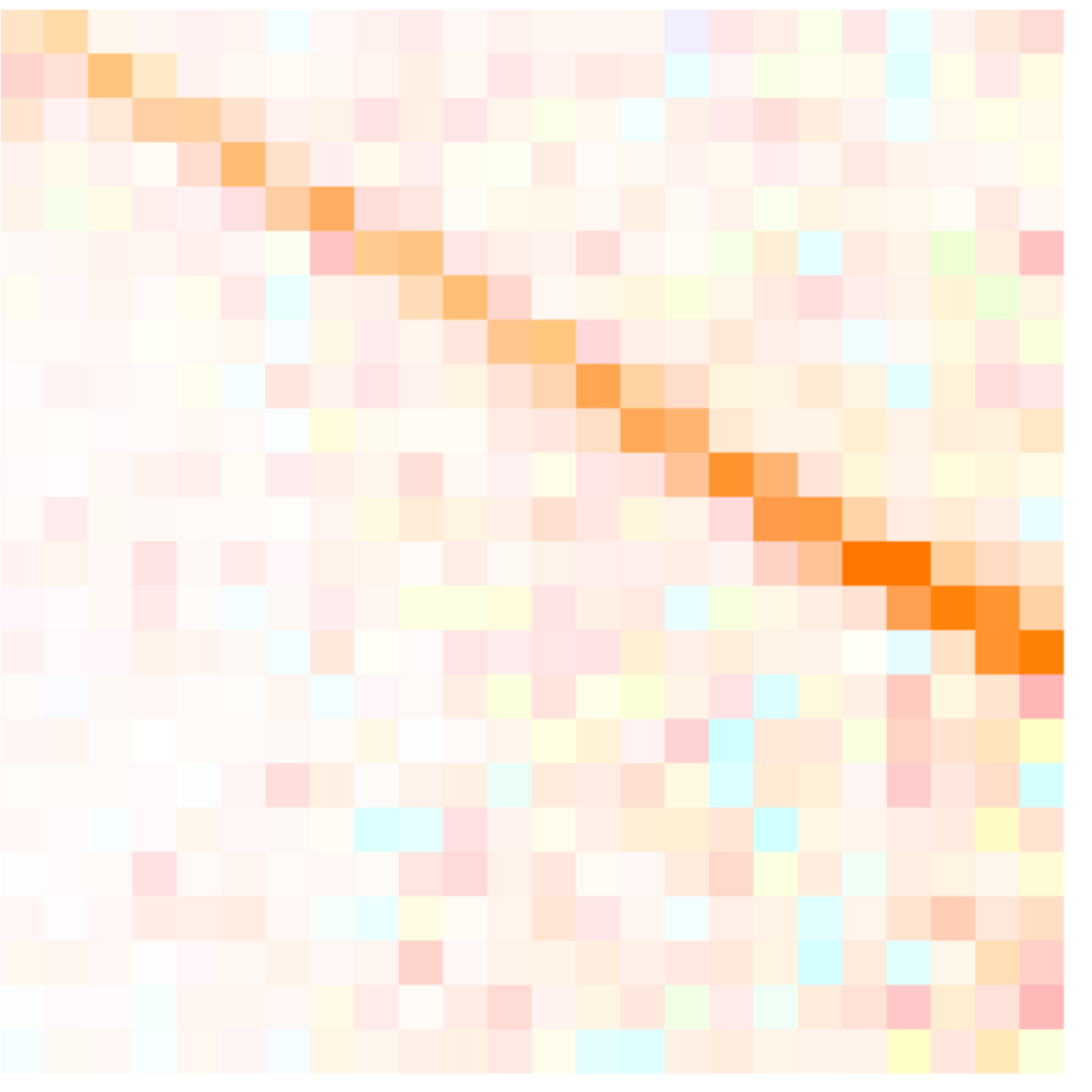} &
\includegraphics[width=0.17in]{figures/colorbar_freq-eps-converted-to.pdf} &
\includegraphics[width=1.1in]{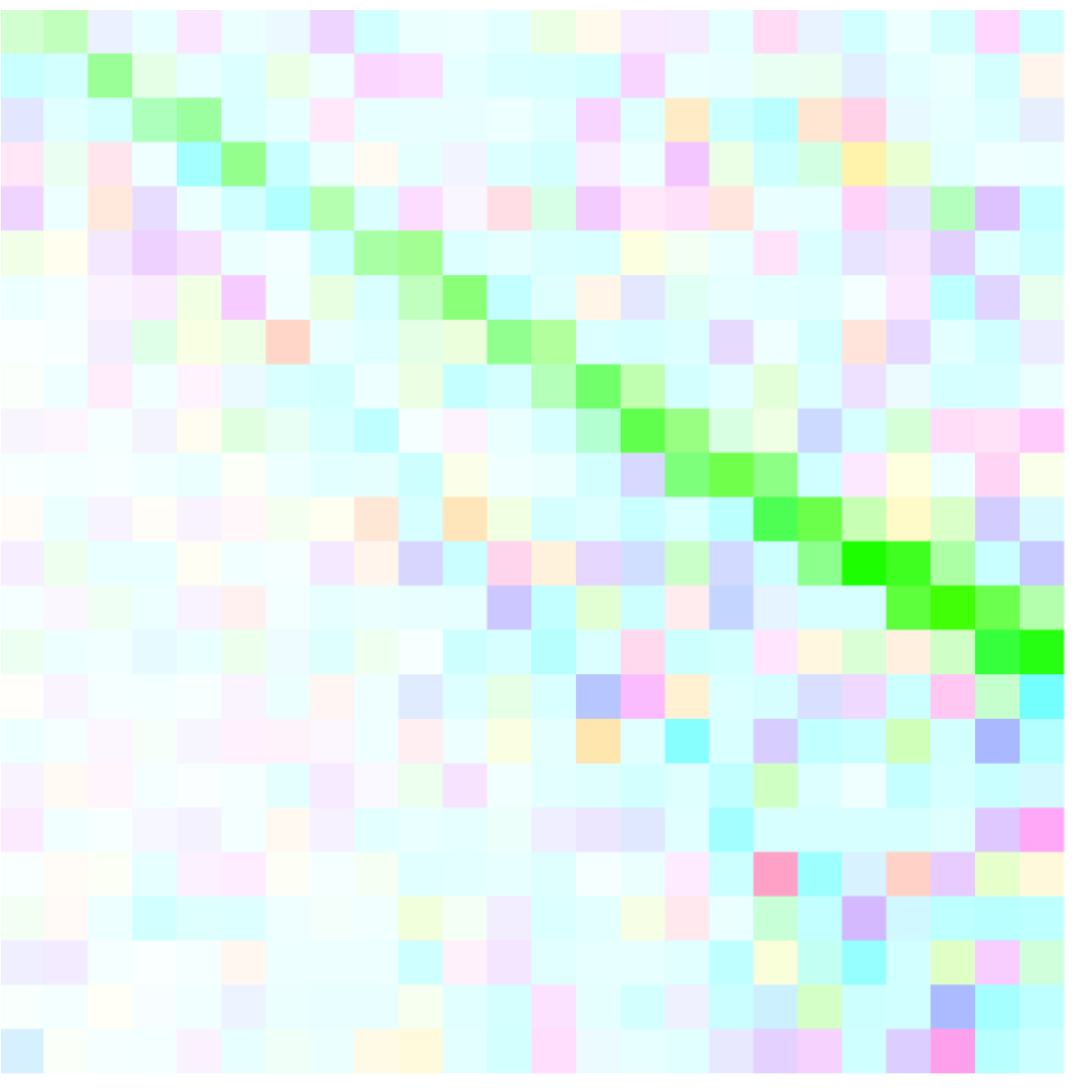} &
\includegraphics[width=0.2in]{figures/colorbar_orientation-eps-converted-to.pdf}
 \\

 \end{tabular}
 \normalsize
\caption{Spectral analysis on 2nd mode of wake image. 1st row: 2nd scale of Riesz-Laplace wavelet transform.  From left to right: mode $\mathbf{d}^{(2)}$, amplitude $\boldsymbol{\alpha}^{(2)}$, frequency $\boldsymbol{\eta}^{(2)}$ and orientation $\boldsymbol{\theta}^{(2)}$. 2nd row: 3rd scale of Riesz-Laplace wavelet transform.  From left to right: mode $\mathbf{d}^{(3)}$, amplitude $\boldsymbol{\alpha}^{(3)}$, frequency $\boldsymbol{\eta}^{(3)}$ and orientation $\boldsymbol{\theta}^{(3)}$. 3rd row: 2nd IMF P2D--HHT.  From left to right: mode $\mathbf{d}^{(2)}$, amplitude $\boldsymbol{\alpha}^{(2)}$, frequency $\boldsymbol{\eta}^{(2)}$ and orientation $\boldsymbol{\theta}^{(2)}$. 4th row: 2nd IMF P2D--PHT ($\overline{N}^{(2)} = 21$). From left to right: denoised mode, amplitude $\boldsymbol{\alpha}^{(2)}$, frequency $\boldsymbol{\eta}^{(2)}$ and orientation $\boldsymbol{\theta}^{(2)}$.}
\label{fig:psamtikimf2}
\end{center}
\end{figure*}

\begin{figure*}
\begin{center}
\footnotesize
\begin{tabular}{p{0.2cm}p{0.2cm}ccccccc}


&
&
&
 Amplitude&
&
 Frequency&
&
 Orientation&
\\

\rotatebox{90}{\hspace{0.4cm}Riesz-Laplace} &
\rotatebox{90}{\hspace{0.9cm}Scale 4} &
\includegraphics[width=1.1in]{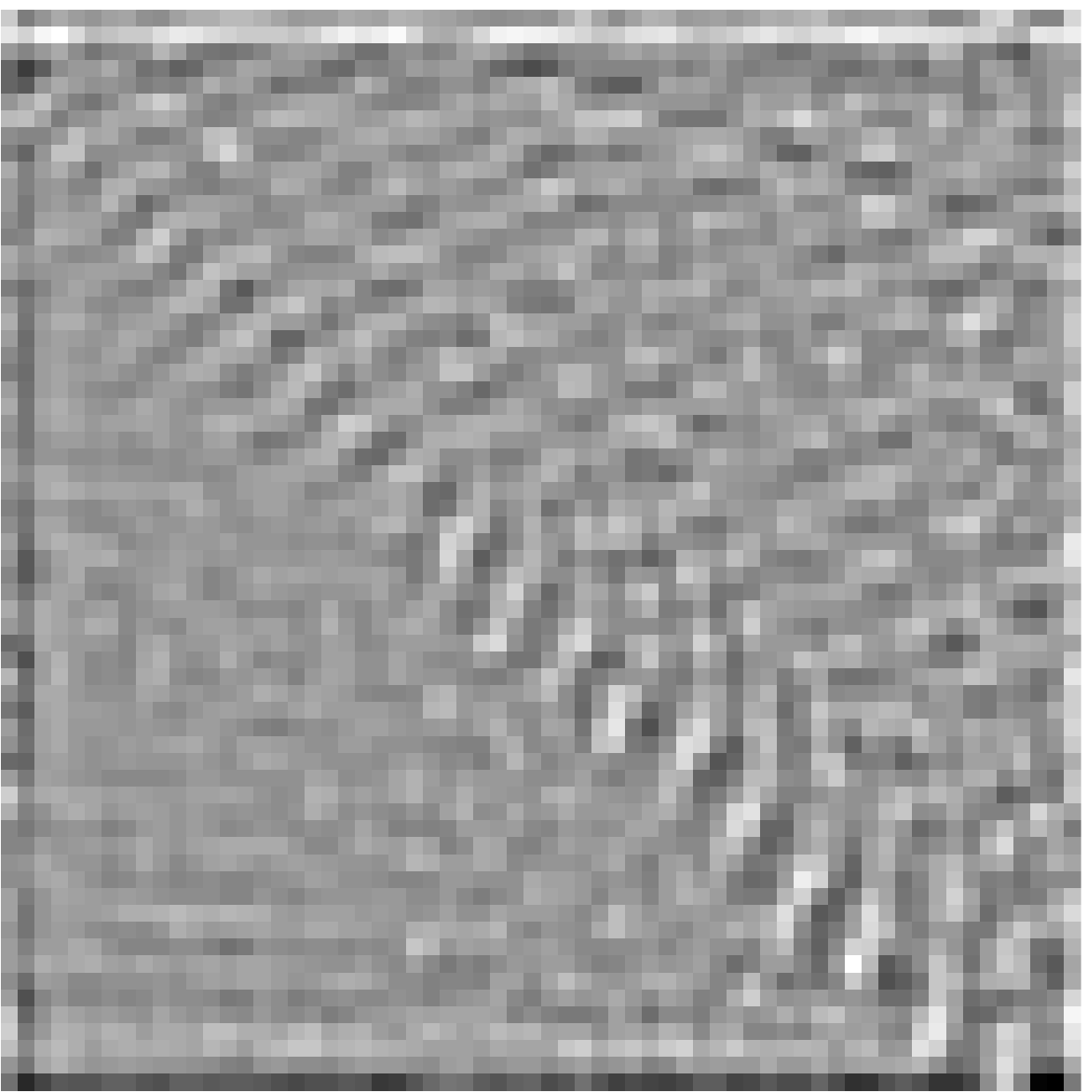} &
\includegraphics[width=1.1in]{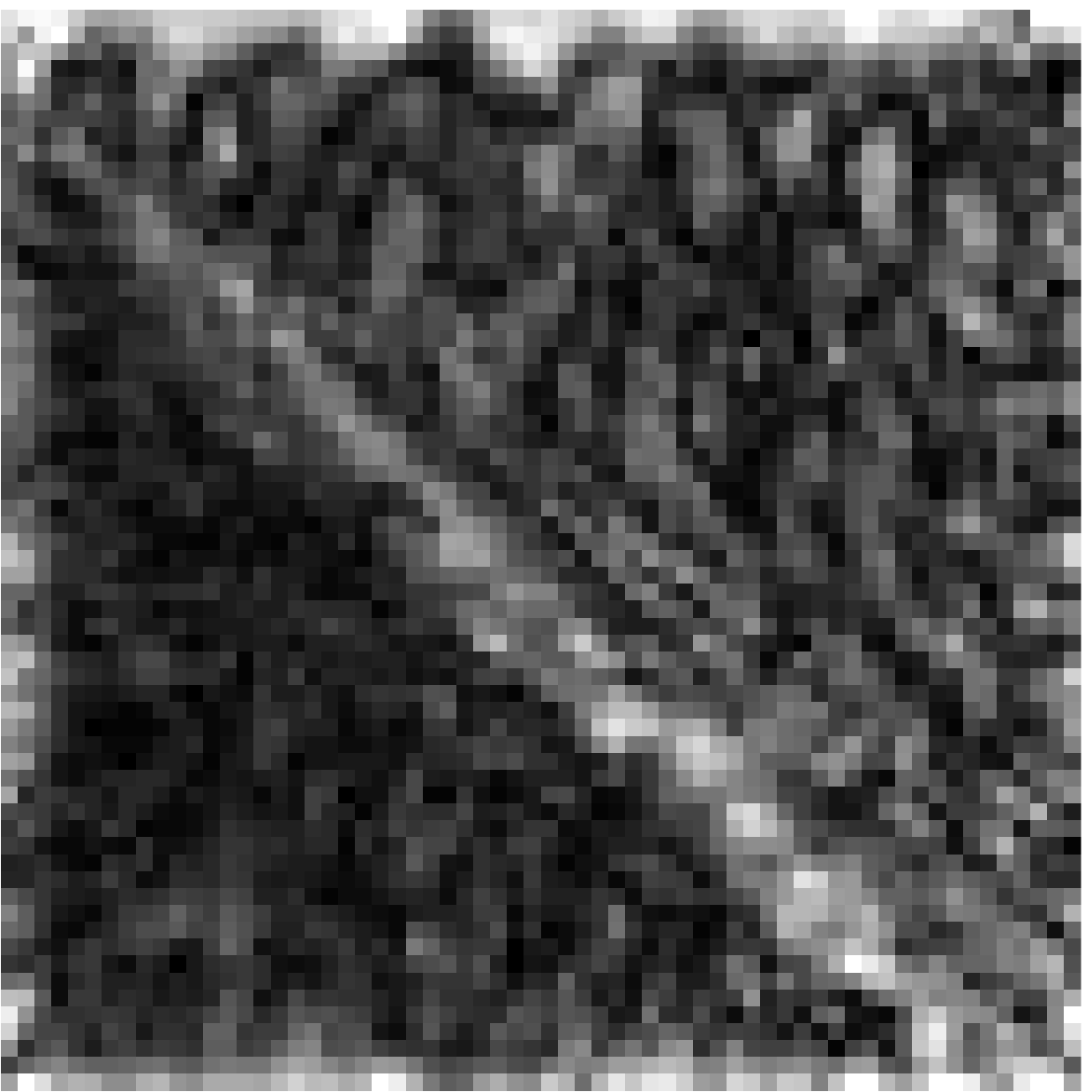} &
\includegraphics[width=0.14in]{figures/colorbar_amplitude_psamtik_20-eps-converted-to.pdf} &
\includegraphics[width=1.1in]{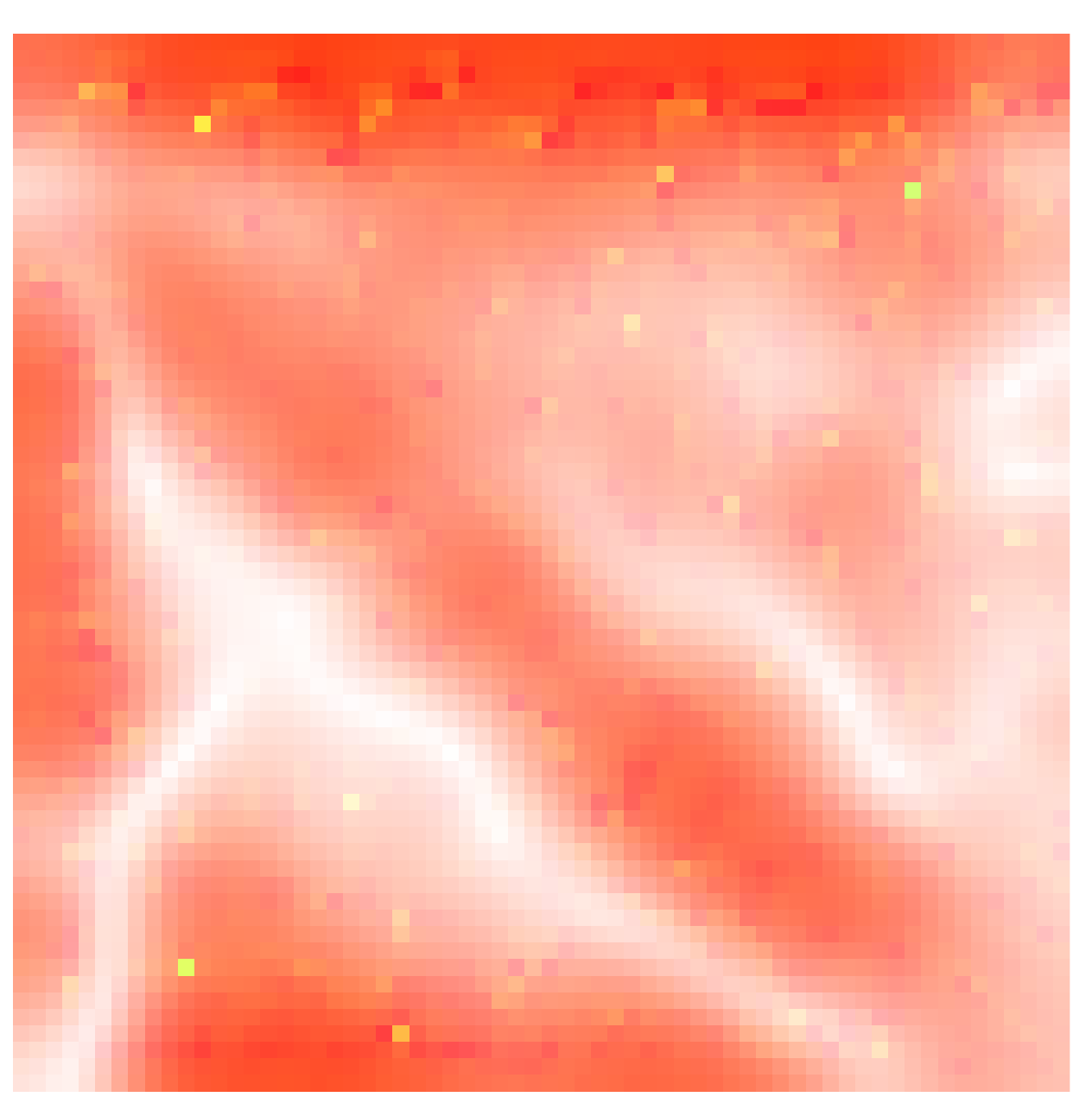} &
\includegraphics[width=0.17in]{figures/colorbar_freq-eps-converted-to.pdf} &
\includegraphics[width=1.1in]{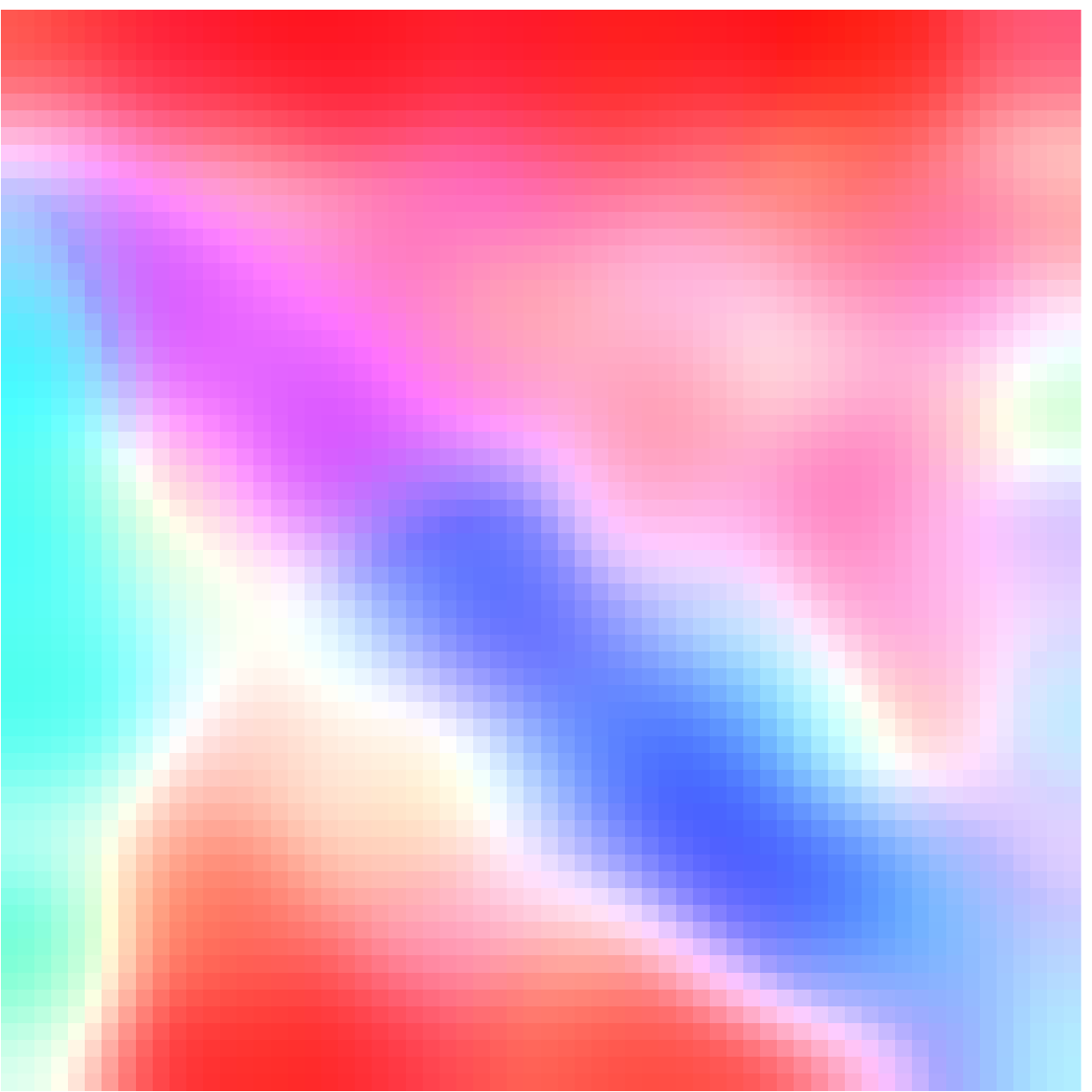} &
\includegraphics[width=0.2in]{figures/colorbar_orientation-eps-converted-to.pdf}
 \\

\rotatebox{90}{\hspace{0.6cm}P2D--HHT} &
\rotatebox{90}{\hspace{1cm}IMF 3} &
\includegraphics[width=1.1in]{figures/psamtik_imf3_50_50_20_5_20_1-eps-converted-to.pdf} &
\includegraphics[width=1.1in]{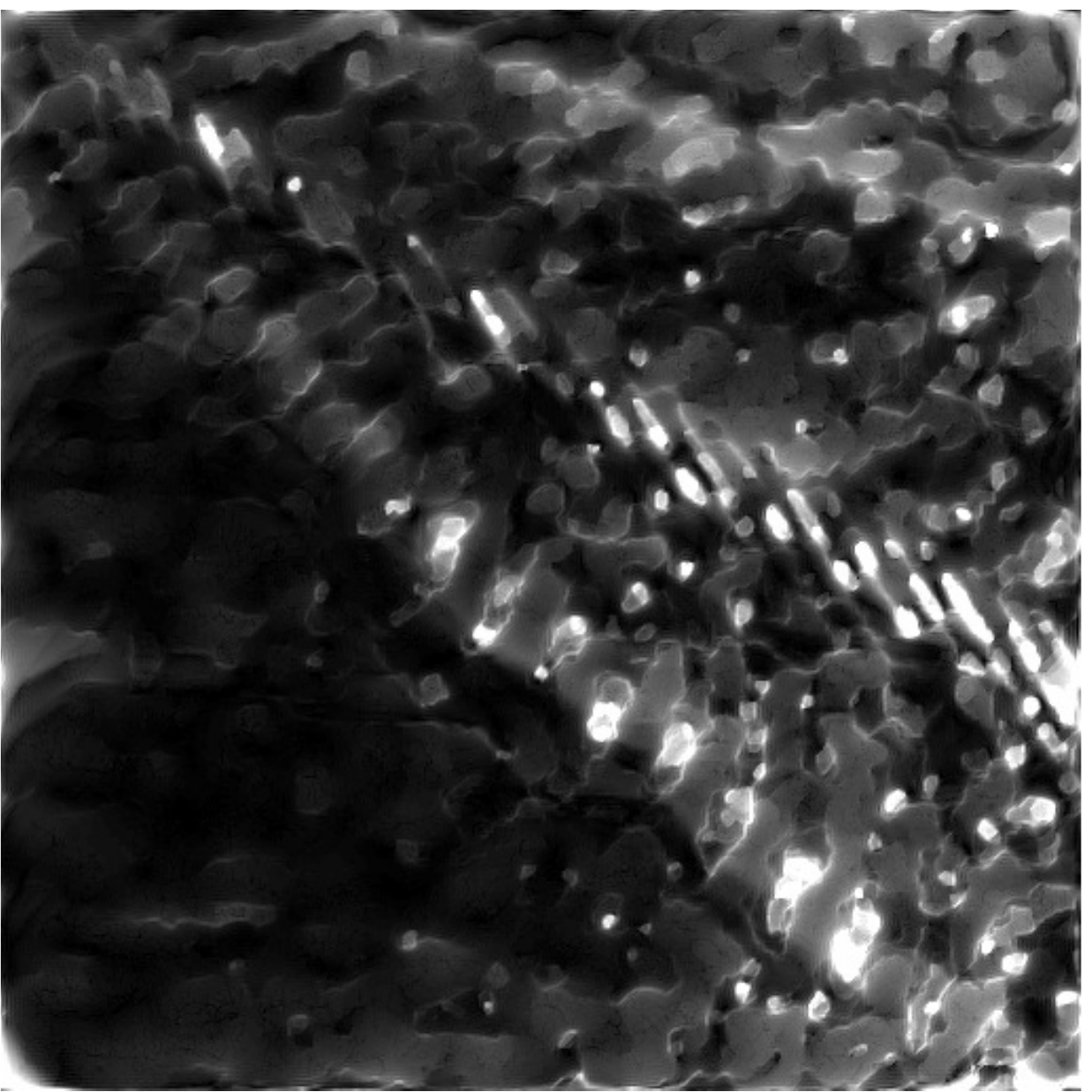} &
\includegraphics[width=0.14in]{figures/colorbar_amplitude_psamtik_20-eps-converted-to.pdf} &
\includegraphics[width=1.1in]{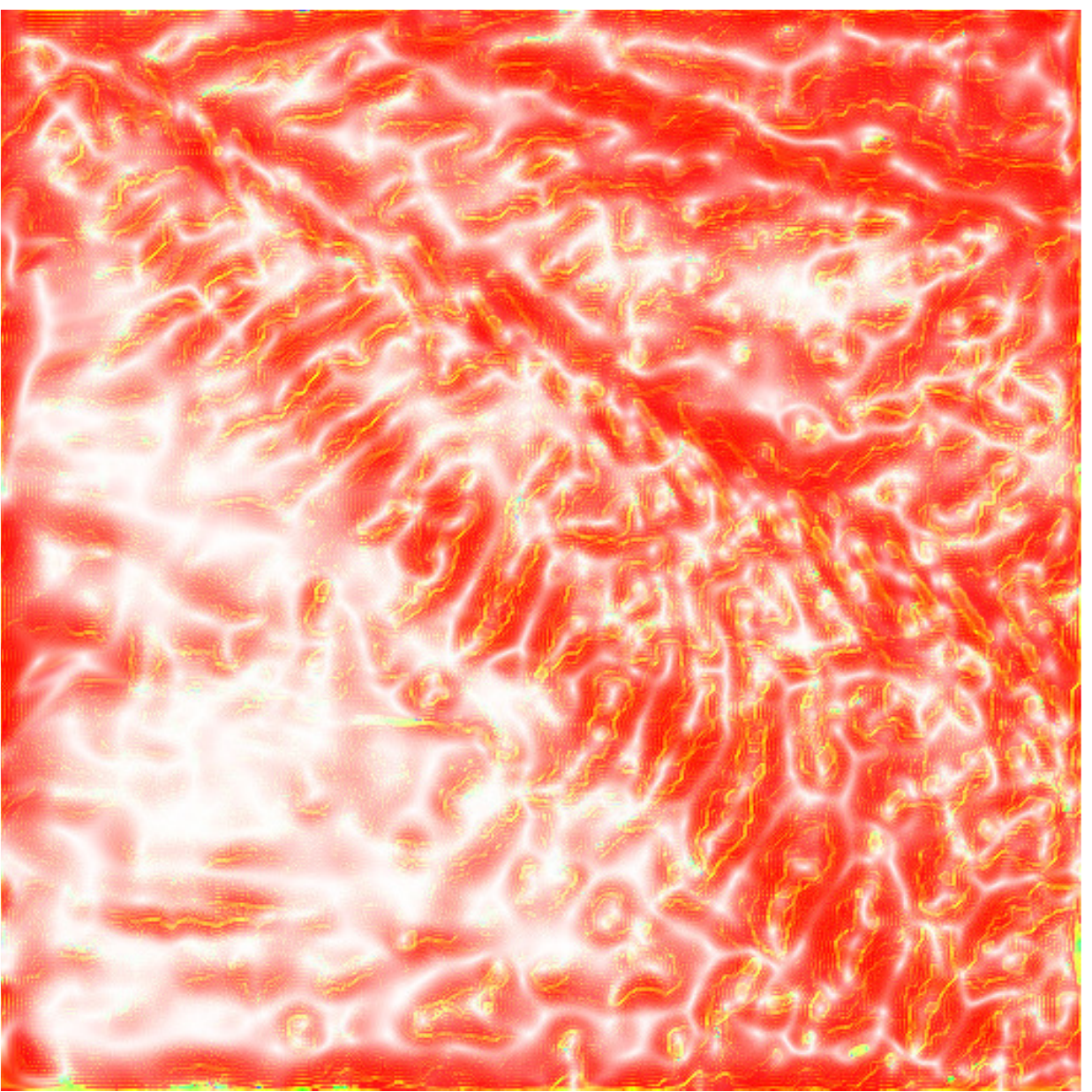} &
\includegraphics[width=0.17in]{figures/colorbar_freq-eps-converted-to.pdf} &
\includegraphics[width=1.1in]{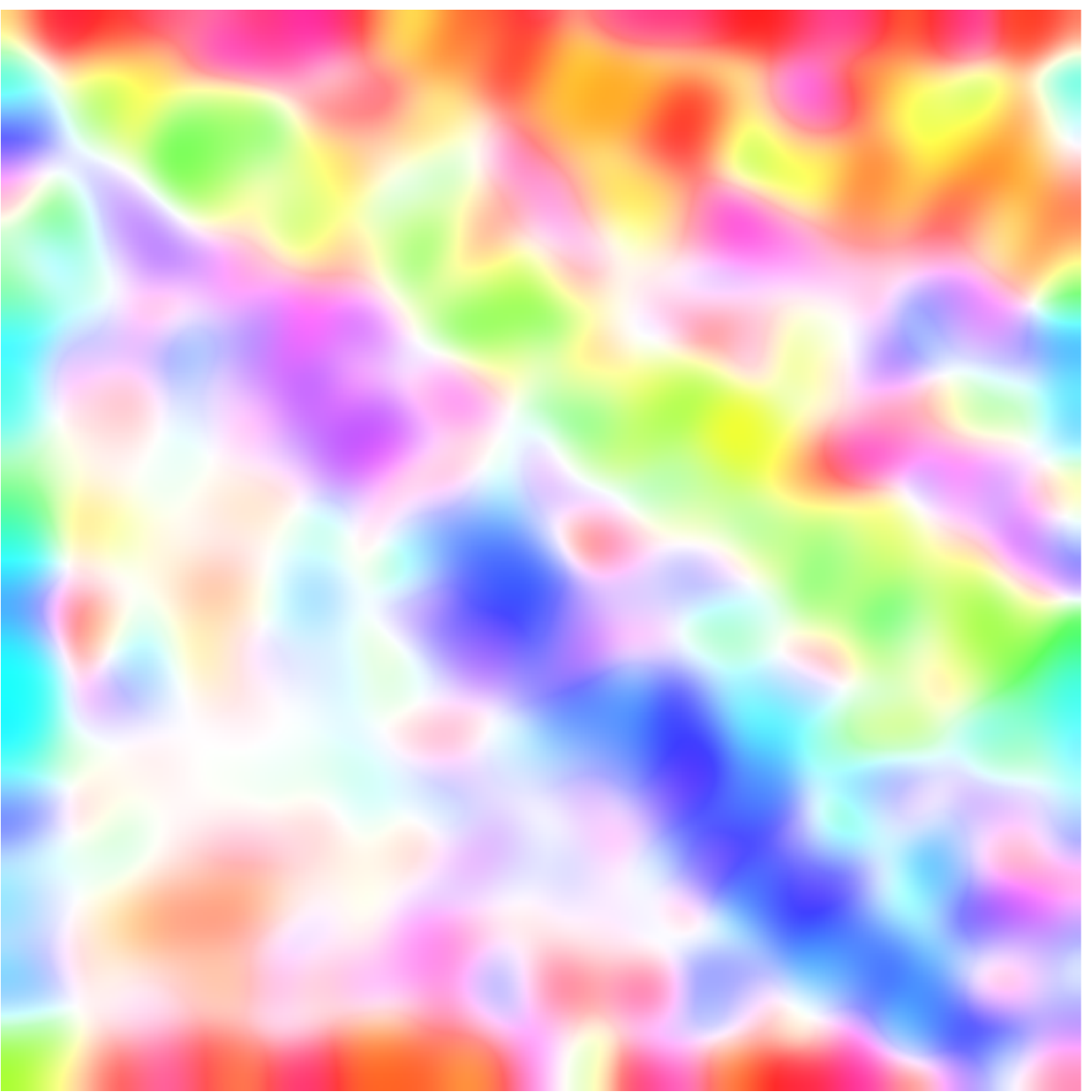} &
\includegraphics[width=0.2in]{figures/colorbar_orientation-eps-converted-to.pdf}
 \\
 
 
 \rotatebox{90}{\hspace{0.6cm}P2D--PHT} &
 \rotatebox{90}{\hspace{0.3cm}Denoised IMF 3} &
 \includegraphics[width=1.1in]{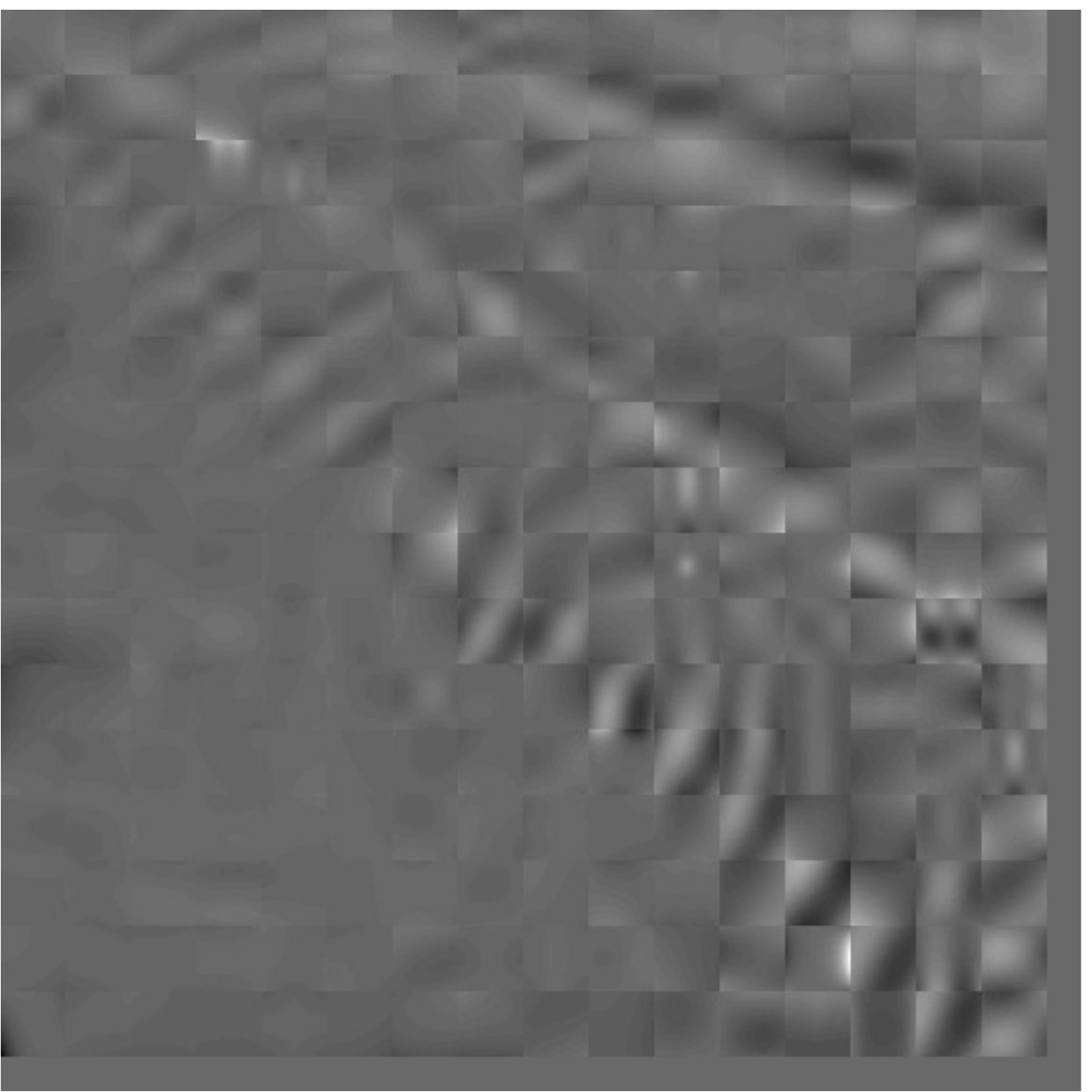} &
 \includegraphics[width=1.1in]{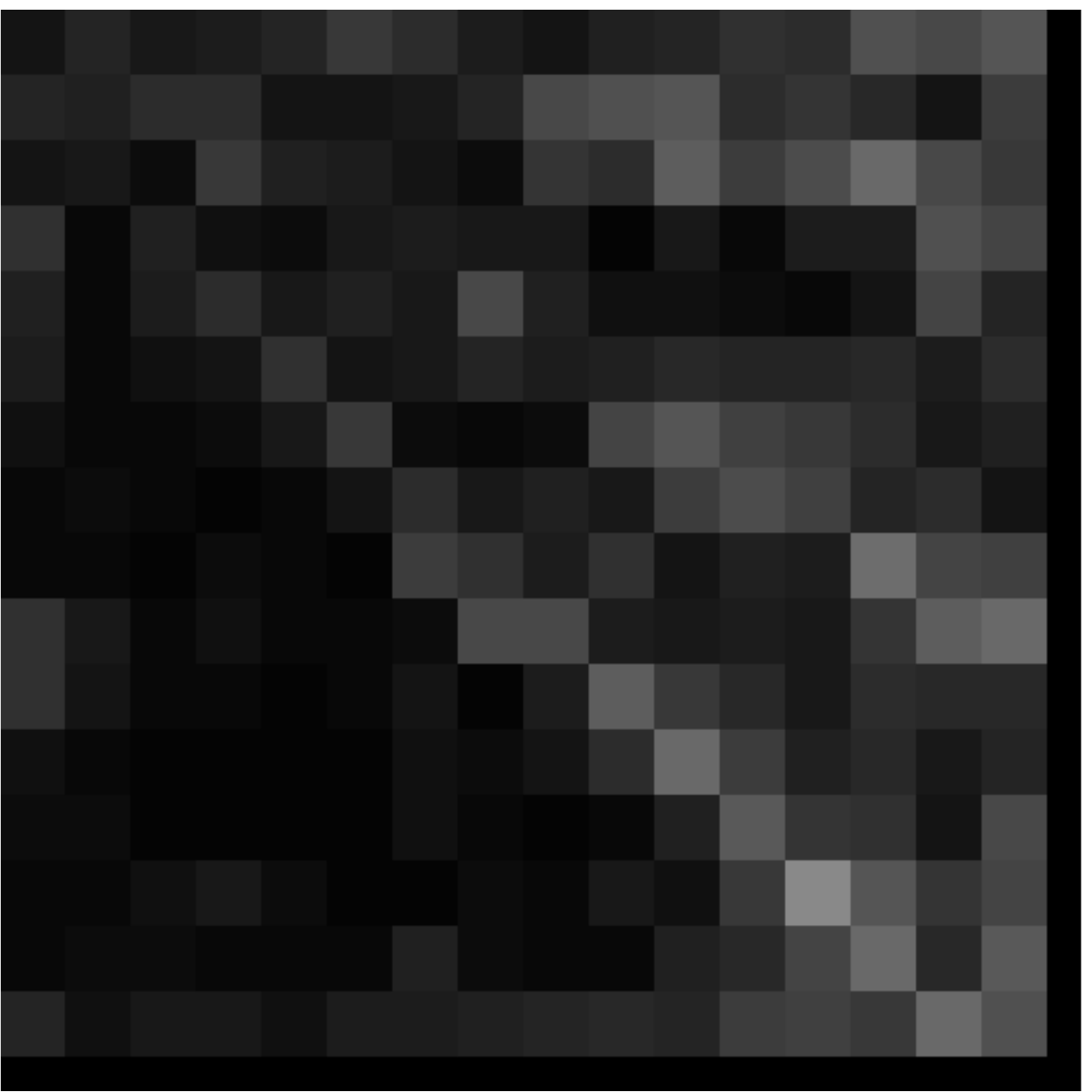} &
 \includegraphics[width=0.14in]{figures/colorbar_amplitude_psamtik_20-eps-converted-to.pdf} &
 \includegraphics[width=1.1in]{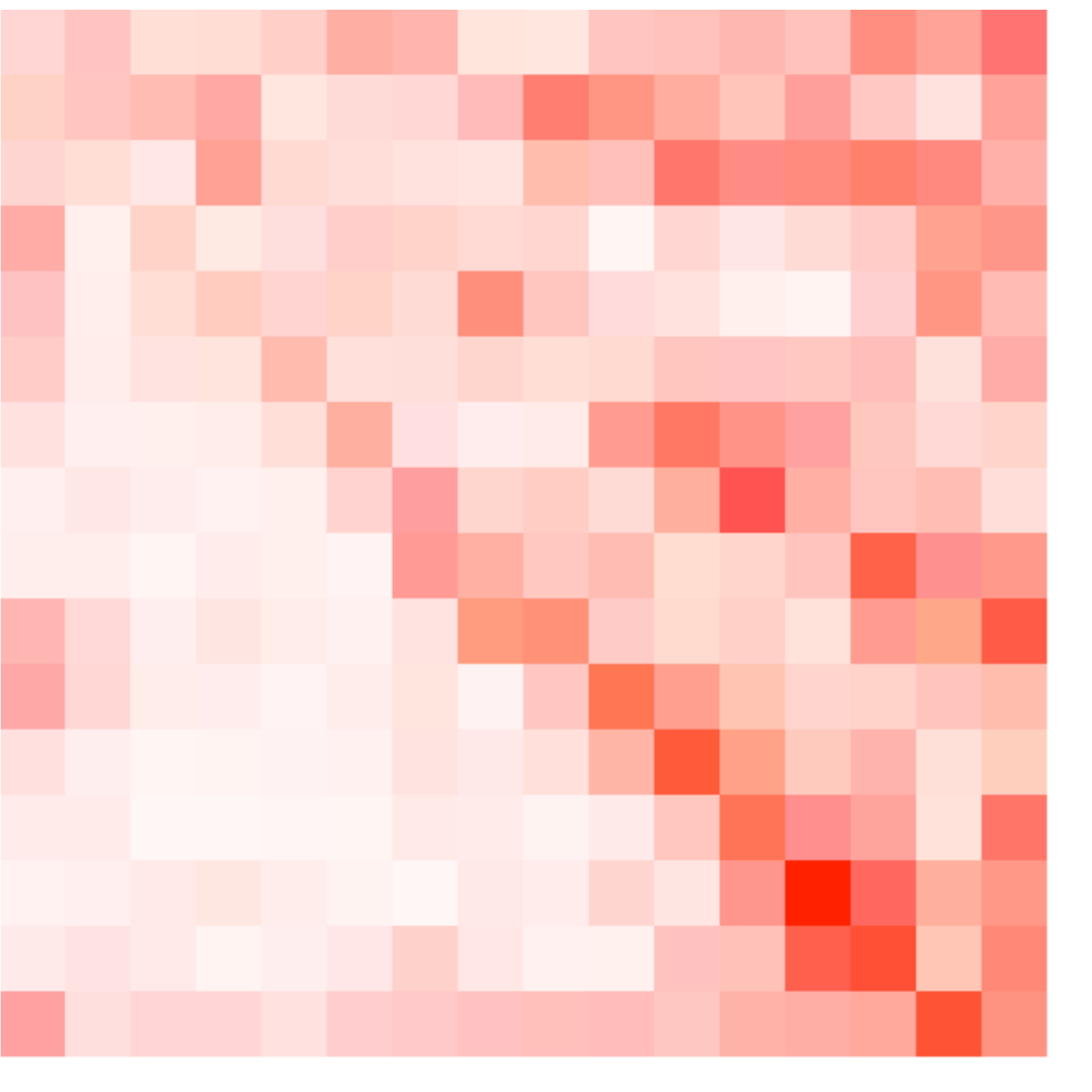} &
 \includegraphics[width=0.17in]{figures/colorbar_freq-eps-converted-to.pdf} &
\includegraphics[width=1.1in]{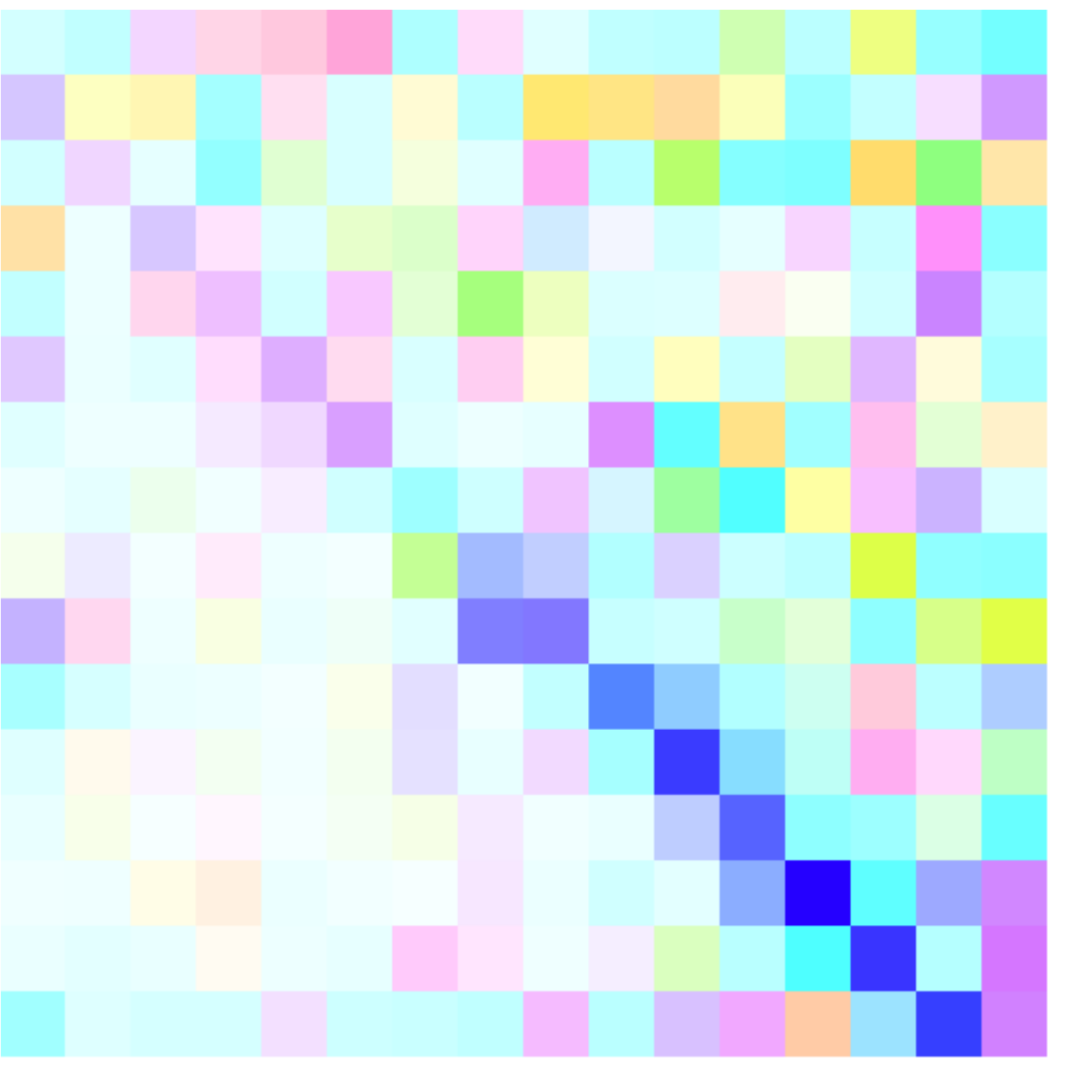} &
\includegraphics[width=0.2in]{figures/colorbar_orientation-eps-converted-to.pdf}
 \\

 \end{tabular}
 \normalsize
\caption{Spectral analysis on 3rd mode of wake image. 1st row: 4rd scale of Riesz-Laplace wavelet transform. From left to right: mode $\mathbf{d}^{(4)}$, amplitude $\boldsymbol{\alpha}^{(4)}$, frequency $\boldsymbol{\eta}^{(4)}$ and orientation $\boldsymbol{\theta}^{(4)}$.  2nd row: 3rd IMF P2D--HHT. From left to right: mode $\mathbf{d}^{(3)}$, amplitude $\boldsymbol{\alpha}^{(3)}$, frequency $\boldsymbol{\eta}^{(3)}$ and orientation $\boldsymbol{\theta}^{(3)}$. 3rd row: 3rd IMF P2D--PHT ($\overline{N}^{(3)} = 31$).  From left to right: denoised mode, amplitude $\boldsymbol{\alpha}^{(3)}$, frequency $\boldsymbol{\eta}^{(3)}$ and orientation $\boldsymbol{\theta}^{(3)}$.}
\label{fig:psamtikimf3}
\end{center}
\end{figure*}

The second experiment is performed on a boat wake image\footnote{\url{http://www.123rf.com/photo_17188220_fast-boat-in-the-far-blue-sea.html}}. Regarding the performance of P2D--HHT compared to G2D--HHT on simulated data, we have decided to focus on this method in our experiments on real data. We compare the results obtained with the proposed P2D--PHT method with P2D--HHT and Riesz-Laplace wavelet transform. The results obtained with the proposed P2D--EMD are shown in Figure~\ref{fig:psamtikemd}. We have used the following optimal parameters: $\rho^{(1)} = 50$, $\nu^{(1)} = 50$, $\rho^{(2)} = 20$, $\nu^{(2)} = 5$, $\rho^{(3)} = 20$, $\nu^{(3)} = 1$. The first IMF contains the fastest small waves. The second IMF contains a slower wave as well as some salt-and-pepper noise, while the third IMF contains the slowest waves. The trend results in the illumination map. The spectral analysis  is performed for the three IMFs and the results are displayed in Figures~\ref{fig:psamtikimf1}, \ref{fig:psamtikimf2} and \ref{fig:psamtikimf3}. For the P2D--PHT, the size of patches has to be chosen so as to be adapted to the frequency of the waves contained in each IMF: a single patch should contain at least one complete period of the wave. Consequently, we have chosen $\overline{N}^{(1)} = 14$, $\overline{N}^{(2)} = 21$ and $\overline{N}^{(3)} = 31$. 
The Riesz-Laplace transform provides good results but suffers from redundancy between the different scales. For example, the second and third wavelet scales contain the same component. Moreover, the methods based on monogenic analysis (Riesz-Laplace and P2D--HHT) give good results for the orientation estimation but are less performant for frequency estimation. Indeed, the orientation is obtained using a robust neighborhood based estimation method, while the frequency is computed pixel-by-pixel from the monogenic signal, which makes the frequency estimation very sensitive to noise. The proposed P2D--PHT method appears to be more robust and consequently gives better results for the frequency estimation. The main drawback of the P2D--PHT is the loss of resolution, especially for the coarsest IMF which requires to deal with a large size for patches.


%
%
%
%
%

\pagebreak

\section{Conclusion}
\label{sec:ccl}

This paper presents a complete method for spectral analysis of nonstationary images. This method is based on a 2-D variational mode decomposition combined with a local spectral analysis method based on Prony annihilation property of cosine functions. This method has been tested on simulated and real data. For the decomposition step, our variational 2-D--EMD proved to be more adaptive than other decomposition approaches like Riesz-Laplace wavelets and texture-geometry methods, and more efficient than existing 2-D--EMD methods in addition of having more robustness and stronger convergence guarantees. Regarding the spectral analysis step, Prony's annihilation-based method proved to be more efficient for frequency estimation and more robust with respect to noise and decomposition errors than monogenic analysis. The main drawback of the method is the loss of resolution due to its patch based approach. Further works should improve the resolution by introducing patch overlapping.


\bibliographystyle{plain}
\bibliography{abbr,EMD_IEEE}




\end{document}